\documentclass[11pt]{article}
\pdfoutput=1
\usepackage{amsmath,amssymb,epsfig,amsfonts,mathrsfs}
\usepackage{graphicx,subfigure}
\usepackage[usenames, dvipsnames]{color}
\usepackage{multirow}
\usepackage{hyperref}
\usepackage[nosort]{cite}
\usepackage{lscape}
\usepackage{rotating}
\usepackage[normalem]{ulem} 
\usepackage{slashed}

\usepackage{extarrows}

\usepackage{verbatim} 

\makeatletter

\newcommand{\be}{\begin{equation}}
\newcommand{\ee}{\end{equation}}
\newcommand{\ba}{\begin{aligned}}
\newcommand{\ea}{\end{aligned}}

\newcommand{\C}{\mathbb{C}}

\newcommand{\bea}{\begin{eqnarray}}
\newcommand{\eea}{\end{eqnarray}}

\newcommand{\R}{{\mathbb R}}
\newcommand{\Z}{{\mathbb Z}}

\def\diag{\mathop{\mathrm{diag}}\nolimits}
\def\Im{\mathop{\mathrm{Im}}\nolimits}

\def\Re{\mathop{\mathrm{Re}}\nolimits}

\def\Tr{\mathop{\mathrm{Tr}}\nolimits}

\def\half{{\frac{1}{2}}}

\def\p{\partial}

\def\unit{{1\kern-.65ex {\rm l}}}
\def\1{{1\kern-.65ex {\rm l}}}

\def\ConnV{\mathfrak{A}}
\def\DeformSpace{\mathcal{H}_{\slashed{\mathcal{D}}}}

\renewcommand{\arraystretch}{1.3}

\newcount\hour \newcount\minute
\hour=\time \divide \hour by 60
\minute=\time
\def\now{%
\ifnum \hour<13
  \ifnum \hour=0 \advance \hour by 12 \number\hour:\else \number\hour:\fi%
     \ifnum \minute<10 0\fi%
     \number\minute%
\ A.M.%
\else \advance \hour by -12 \number\hour:%
  \ifnum \minute<10 0\fi%
  \number\minute%
  \ P.M.%
\fi%
}

\makeatother

\begin{document}

\baselineskip=18pt  
\numberwithin{equation}{section}  
\allowdisplaybreaks  


\thispagestyle{empty}

\vspace*{-2cm} 
\begin{flushright}
{\tt IPMU-18-0057} 
\end{flushright}

\vspace*{0.8cm} 
\begin{center}
  {\huge  An $\mathcal{N}=1$ 3d--3d Correspondence}

 \vspace*{1.5cm}
{Julius Eckhard$\,^1$,  Sakura Sch\"afer-Nameki$\,^1$, and Jin-Mann Wong$\,^2$}\\

 \vspace*{1.1cm} 
{\it ${}^1$ Mathematical Institute, University of Oxford \\
 Woodstock Road, Oxford, OX2 6GG, UK}\\
\smallskip

{\it ${}^2$ Kavli Institute for the Physics and Mathematics of the Universe (WPI),\\
University of Tokyo, Kashiwa, Chiba 277-8583, Japan}\\
\smallskip

 {\tt {gmail:$\,$juliuseckhard, sakura.schafer.nameki, jinmannwong}}\\

\end{center}
\vspace*{.8cm}

\noindent
M5-branes on an associative three-cycle $M_3$ in a $G_2$-holonomy manifold give rise to a 
3d $\mathcal{N}=1$ supersymmetric gauge theory, $T_{\mathcal{N}=1} [M_3]$. 
We propose an $\mathcal{N}=1$ 3d--3d correspondence, based on two observables of these theories: the Witten index and the $S^3$-partition function. 
The Witten index of a  3d $\mathcal{N}=1$ theory $T_{\mathcal{N}=1} [M_3]$ is shown to be computed in terms of the partition function of a topological field theory, a super-BF-model coupled to a spinorial hypermultiplet (BFH), on $M_3$. The BFH-model localizes on solutions to a 
generalized set of 3d Seiberg-Witten equations on $M_3$. Evidence to support this correspondence is provided in the abelian case, as well as in terms of a  direct derivation of the topological field theory by twisted dimensional reduction of the 6d $(2,0)$ theory. We also consider a correspondence for the $S^3$-partition function of the $T_{\mathcal{N}=1} [M_3]$ theories, by determining the dimensional reduction of the M5-brane theory on $S^3$. The resulting topological theory is Chern-Simons-Dirac theory, for a gauge field and a twisted harmonic spinor on $M_3$, whose equations of motion are the generalized 3d Seiberg-Witten equations.
For generic $G_2$-manifolds the theory reduces to real Chern-Simons theory, in which case we conjecture that the $S^3$-partition function of $T_{\mathcal{N}=1}[M_3]$ is given by the Witten-Reshetikhin-Turaev invariant of $M_3$.

\newpage

\tableofcontents


\section{Introduction}

Starting with the Alday-Gaiotto-Tachikawa (AGT) correspondence \cite{Alday:2009aq, Wyllard:2009hg}, a series of conjectures were put forward, relating 
$d$-dimensional $\mathcal{N}=2$ supersymmetric theories $T_{\mathcal{N}=2}[M_{6-d}]$, labeled by $(6-d)$-dimensional manifolds $M_{6-d}$, 
 and topological or conformal theories in $6-d$ dimensions on the manifold $M_{6-d}$. Each of these correspondences is established in terms of the agreement between observables, e.g. sphere partition functions (or indices)  of the supersymmetric theories and partition functions/correlation functions of the topological/conformal theories  on $M_{6-d}$. 
 
 One way to motivate these conjectures is to consider M5-branes wrapped on $M_{6-d}$ with a topological twist which preserves $\mathcal{N}=2$ supersymmetry in $d$ dimensions. 
 For $d=4$ it was shown that the sphere partition function of the class $\mathcal{S}$ theories $T_{\mathcal{N}=2}[M_2]$ \cite{Gaiotto:2009we} agree with correlation functions of 2d Toda theory on the Riemann surface $M_2$ \cite{Alday:2009aq, Wyllard:2009hg}. 
 The so-called 3d--3d correspondence \cite{Dimofte:2011ju,Terashima:2011qi} similarly relates the $S^3$-partition function of 3d $\mathcal{N}=2$ theories $T_{\mathcal{N}=2}[M_3]$ to the partition function of complex Chern-Simons theory on $M_3$. Finally, for $d=2$ there is a correspondence for 2d $(0,2)$ theories labeled by four-manifolds $M_4$. The half-topologically twisted sphere partition function is conjectured to be computed by a topological sigma-model into the monopole moduli space \cite{Assel:2016lad}, whereas the Witten index is identified with the Vafa-Witten (VW) \cite{Vafa:1994tf} partition function of 4d $\mathcal{N}=4$ Super-Yang Mills theory on $M_4$ \cite{Gadde:2013sca}. 

Much of the progress in establishing these conjectures relies on the computational advances that were made for sphere partition functions of $\mathcal{N}=2$ supersymmetric theories thanks to localization techniques \cite{Nekrasov:2002qd,Pestun:2007rz} -- for a recent review see \cite{Teschner:2016yzf}. For less supersymmetry, many of these tools are not quite as well developed thus far. 
One may hope that formulating similar correspondences for $\mathcal{N}=1$ theories could give further insight into their structure. An initial step towards developing an $\mathcal{N}=1$ version of the AGT correspondence has been made in \cite{Mitev:2017jqj}, with the goal to relate the sphere partition function of the class $\mathcal{S}_k$ theories \cite{Bah:2012dg, Gaiotto:2015usa} to 2d conformal blocks.

The goal of this paper is to develop an $\mathcal{N}=1$ version of the 3d--3d correspondence, motivating it from first principles by starting with the 6d $(2,0)$ theory. As is well-known the $T_{\mathcal{N}=2}[M_3]$ are obtained by wrapping M5-branes on special Lagrangian (sLag) three-cycles in Calabi-Yau  three-folds. To retain $\mathcal{N}=1$ in 3d, we will show that the natural setup for $T_{\mathcal{N}=1}[M_3]$ is to realize $M_3$ as an associative three-cycle in a $G_2$-holonomy manifold\footnote{$G_2$-holonomy manifolds have two sets of supersymmetric, i.e. calibrated cycles: associative three-cycles calibrated with the $G_2$-three-form $\Phi$ and co-associative four-cycles, calibrated with $\star \Phi$.  M5-branes wrapping co-associative four-cycles results in the VW twist along the four-manifold, i.e. 4d-2d duality studied in \cite{Gadde:2013sca, Assel:2016lad}.}. 

A priori we do not know what the topological theories are, which would complement the 3d $\mathcal{N}=1$ theories in such a 3d--3d correspondence. To determine these, it is useful to recall the approach applied in the $\mathcal{N}=2$ setting: the topological theory, whose partition function computes the sphere partition function of the $T_{\mathcal{N}=2}[M_{6-d}]$ theories can be determined from the sphere reduction of the 6d $(2,0)$ in an $\mathcal{N}=2$ preserving conformal supergravity background \cite{Cordova:2013cea, Assel:2016lad, Cordova:2016cmu}. We will employ this approach in the following to determine the topological theories, which compute the following two observables of $T_{\mathcal{N}=1}[M_3]$: the $T^3$-partition function, i.e. Witten index, and the $S^3$-partition function. 

For 3d $\mathcal{N}=1$ the Witten index \cite{Witten:1999ds} is a well explored observable, much more so than the $S^3$-partition function. 
For this reason we will focus much of our attention on this observable and provide non-trivial checks of the proposed correspondence.  We will  derive the `dual' topological theory by considering the 6d $(2,0)$ theory first on $ T^3$, which gives 3d $\mathcal{N}=8$ SYM, which we then topologically twist along $M_3$, while preserving two topological supercharges. This twist corresponds to the embedding of $M_3$ as an associative cycle in a $G_2$-manifold. 
In summary, the theory, whose partition function computes the Witten index of the $T_{\mathcal{N}=1} [M_3]$ will be shown to be a supersymmetric BF-theory coupled to a spinorial hypermultiplet (BFH), which is a section of the normal bundle $N_{M_3}$ of $M_3$ inside the $G_2$-manifold. 

We should at this point elaborate briefly on the geometry of associative three-cycles in $G_2$-holonomy manifolds \cite{McLean, Joyce}, which will play an important role in the behavior of the Witten index. The normal bundle of an associative three-cycle is $N_{M_3} = \mathbb{S} \otimes V$, where $\mathbb{S}$ is the spin bundle and $V$ an $SU(2)$-bundle, in particular sections of the normal bundle are twisted harmonic spinors on $M_3$, satisfying a twisted Dirac equation. {On an odd-dimensional manifold the Dirac operator has vanishing index, which implies that the dimension of the kernel (infinitesimal deformations) equals that of its co-kernel (obstructions to these deformations), however the index does not reveal any information about the non-triviality of each of these spaces. 
For non-generic choices of $G_2$-structure, there can be twisted harmonic spinors, which are accompanied with non-trivial obstructions of the deformations, which they parametrize. This fact will reflect itself in the discontinuity/wall-crossing of the Witten index of $T_{\mathcal{N}=1}[M_3]$.}

The 3d--3d correspondence that we propose for $\mathcal{N}=1$ results in an identification of the partition function of the BFH-model on $M_3$ with the Witten index of $T_{\mathcal{N}=1}[M_3]$. To compute the partition function of the BFH-model, we show that its action is minimized on solutions to a non-abelian generalization of the 3d Seiberg-Witten (gSW) equations. These differ from the standard SW equations in that the spinor transforms in the adjoint of a gauge group $G$ as well as under an additional $SU(2)$-bundle $V$ (that appears in the normal bundle of the associative cycle). The partition function of the BFH-model is computed by the Euler characteristic of the moduli space $\mathcal{M}_{\text{gSW}}$ of solutions to the gSW equations\footnote{The analog for $\mathcal{N}=2$ is the moduli space of complex flat connections.}. In particular for gauge group $G=U(1)$ we derive the partition function explicitly and match it with the index of $T_{\mathcal{N}=1}[M_3, U(1)]$.

We find that already in the abelian case the index is discontinuous under metric deformations, and jumps depending on the existence of twisted harmonic spinors. The fact that the partition function of the BFH-model is only topological up to wall-crossing can, as noted earlier, be traced back to the deformation theory of associative three-cycles within $G_2$-manifolds.
At the location of the walls the normal deformations, appearing in the gSW equations, are obstructed. This means that $\mathcal{M}_{\text{gSW}}$ can become singular and its Euler characteristic can jump.

A second less-explored observable for 3d $\mathcal{N}=1$ theories is the $S^3$-partition function (for a discussion of this observable for  SCFTs  see \cite{Gerchkovitz:2014gta}). Whereas the $\mathcal{N}=2$ 3d--3d correspondence is studied for the $S^3$-partition function, and many computational results are available thanks  to localization methods  \cite{Kapustin:2009kz} (and see \cite{Teschner:2016yzf} for a recent review), the situation for $\mathcal{N}=1$ is much less explored. In particular localization will not be applicable for computing the sphere partition functions with 3d $\mathcal{N}=1$ supersymmetry. 
Here we will nevertheless determine what the `dual' topological field theory is, whose partition function on $M_3$  would provide a  conjecture for the $S^3$-partition function of $T_{\mathcal{N}=1}[M_3]$. To do so, we determine the conformal supergravity background similar to \cite{Cordova:2013cea, Assel:2016lad, Cordova:2016cmu} and perform the reduction of the 6d $(2,0)$ theory on a three-sphere, first to 5d SYM and then on an $S^2$ to 3d, whilst preserving $\mathcal{N}=1$ supersymmetry. The resulting topological theory is shown to be real Chern-Simons gauge theory on $M_3$ coupled to a twisted harmonic spinor  $\phi$, i.e. a Chern-Simons-Dirac theory whose equations of motion are the generalized Seiberg-Witten equations (for a review see \cite{kronheimer2007monopoles}). For generic associatives in $G_2$-manifolds there will be no twisted harmonic spinors, and the theory reduces to real Chern-Simons theory. In this case the topological partition function is given in terms of the Witten-Reshetikhin-Turaev invariant \cite{Witten:1988hf,MR1091619}, which we conclude must compute the $S^3$-partition function of $T_{\mathcal{N}=1}[M_3]$.

The most interesting physical application arises when viewing the M5-branes as domain walls in the 4d $\mathcal{N}=1$ theory obtained by M-theory on the $G_2$-holonomy manifold. For Lens spaces this case has been studied in 
\cite{Acharya:2001dz} and we will connect these results, when discussing concrete examples. 
This may in particular be of interest in recent constructions of new $G_2$-holonomy manifolds in \cite{Kovalev:2001zr,  CHNP1, Corti:2012kd} and singular limits thereof \cite{Braun:2017uku} that realize non-abelian gauge groups.

Finally, we should remark that in the case that there is a non-trivial IR fixed point, the M5-branes on associatives have a holographic dual description in terms of $AdS_4$-solutions, where $M_3$ is a hyperbolic three-manifold \cite{Acharya:2000mu}. This means the metric has constant sectional curvature $-1$, and by Schur's lemma the metric is Einstein. Examples of such associatives exist in the Bryant-Salamon $G_2$-manifolds \cite{BryantSalamon}, which are the total space of the spin bundle over $M_3$, with $M_3$ of constant sectional curvature $\pm 1$. For metrics on $M_3$ with negative scalar curvature the associatives can indeed have obstructions, which are determined by zero modes of the Dirac operator. It would be interesting to explore this from a holographic point of view.  

The plan of the paper is as follows: section \ref{sec:Overview} serves as an overview and summary of background material, starting with a concise statement of the  proposed $\mathcal{N}=1$ 3d--3d correspondence in section \ref{sec:3d3d}. In section \ref{sec:DeformationTheory} we provide some background on $G_2$-holonomy manifolds, calibrated cycles and their deformation theory. We conclude this section with a discussion of 3d Seiberg-Witten equations and the non-abelian generalization that we encounter. In section \ref{sec:T[M,U(1)]} we derive the abelian theory $T_{\mathcal{N}=1}[M_3, U(1)]$ and its Witten index.
The generalization to non-abelian gauge groups is discussed in section \ref{sec:NonAbelianTheory} by either considering a specialization of the three-manifold to Lens spaces, or by considering a reduction to 5d SYM and subsequently a derivation of the circle-reduction of $T_{\mathcal{N}=1}[M_3]$ to a 2d sigma-model into the moduli space of generalized 3d Seiberg-Witten equations. In section \ref{sec:BFH} we derive the topological field theory side of the 3d--3d correspondence in the case of the Witten index, and provide several checks.  
In section \ref{sec:S3} we determine the topological theory whose partition function computes the $S^3$-partition function of the 3d $\mathcal{N}=1$ theory. 
We conclude in section \ref{sec:DO} with a discussion and outlook. Various appendices summarize our notation and provide further computational details.


\section{Overview and Background}
\label{sec:Overview}

\subsection{An $\mathcal{N}=1$ 3d--3d Correspondence}
\label{sec:3d3d}

Consider 6d $(2,0)$ theory with gauge group $G$ on a three-manifold $M_3$. Depending on the topological twist, the resulting 3d theory after compactification along $M_3$, can preserve either $\mathcal{N}=2$ or $\mathcal{N}=1$ supersymmetry. We denote the Lorentz and R-symmetry of the 6d theory by $SO(1,5)_L \times Sp(4)_R$. 
With the space-time decomposition $\mathbb{R}^{1,2}\times M_3$, the two twists are realized as follows:
\be\label{N1N2}
\ba
SO(1,5)_L &\quad \to \quad SO(1,2)_L \times \underline{SO(3)_M}\cr 
Sp(4)_R & \quad \to \quad\left\{ 
\ba
\underline{SU(2)_R} \times U(1)_R  & \qquad \hbox{3d} \  \mathcal{N}=2; \ M_3= \hbox{ sLag in CY$_3$} \cr  
\underline{SU(2)_r} \times  SU(2)_{\ell}  & \qquad \hbox{3d}  \ \mathcal{N}=1; \ M_3 = \hbox{Associative in $G_2$} \,.\cr 
\ea\right.
\ea
\ee
Twisting the underlined R-symmetry groups with the local Lorentz group $SO(3)_M$ of $M_3$, results in two types of supersymmetric 3d theories:
The $\mathcal{N}=2$ case has been studied extensively in the standard 3d--3d correspondence \cite{Dimofte:2011ju, Terashima:2011qi} and the twist is realized geometrically in terms of embedding $M_3$ as a {special} Lagrangian (sLag) cycle in a Calabi-Yau three-fold. The correspondence states the equivalence between the $S^3$-partition function $\mathcal{Z}_{S^3}$ of the  $\mathcal{N}=2$ theories $T_{\mathcal{N}=2}[M_3]$, and the complex Chern-Simons partition function on $M_3$
\be\label{3d3dN2}
Z_{{\rm CS}_{\mathbb{C}}, G} (M_3) = \mathcal{Z}_{S^3} (T_{\mathcal{N}=2} [M_3, G]) \,,
\ee
for general gauge group $G$.
This correspondence is by now not only supported by computational evidence in terms of examples of $S^3$-partition functions, but in \cite{Cordova:2013cea} the complex Chern-Simons theory was derived by a dimensional reduction of the 6d $(2,0)$ theory on the three-sphere, coupled to a suitable conformal supergravity background.

\begin{figure}
\begin{center}
  \includegraphics[height=3.5cm]{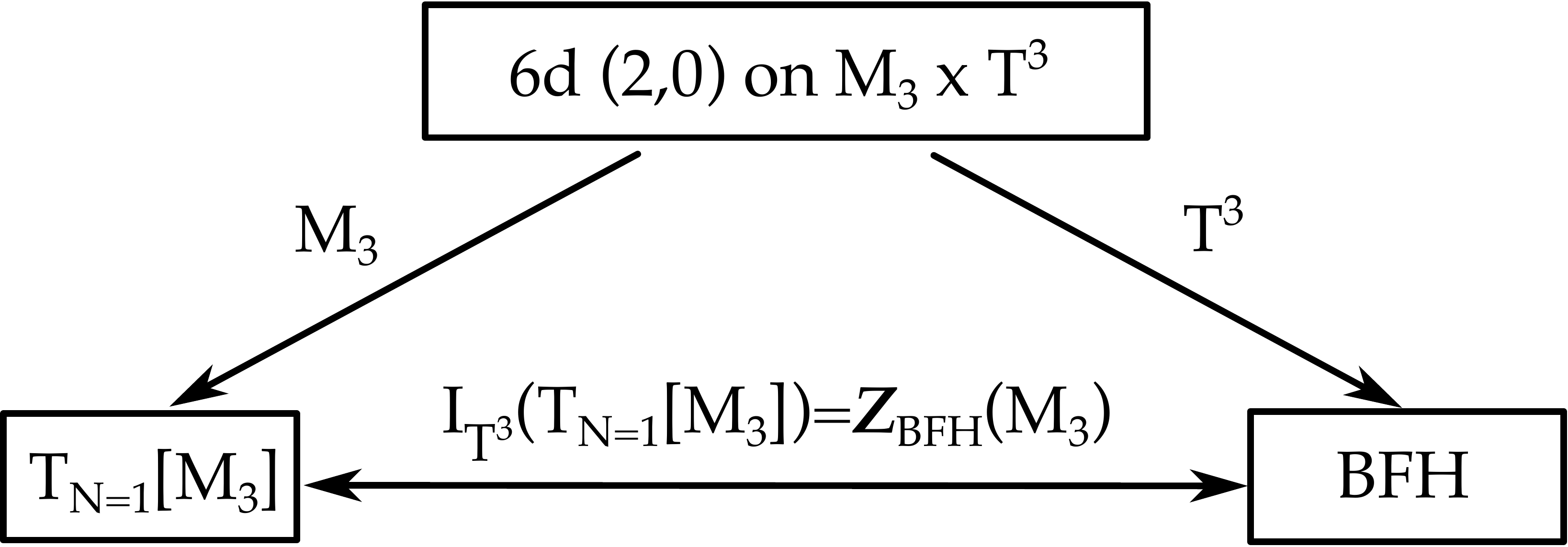}
 \caption{\label{fig:3d3dN1Duality} 
The setup for the  $\mathcal{N}=1$ 3d--3d correspondence is the 6d $(2,0)$ theory on $M_3\times T^3$. 
The topologically twisted reduction along $M_3$ preserves 3d $\mathcal{N}=1$ and geometrically corresponds to an embedding of $M_3$ as an  associative three-cycle in a $G_2$-holonomy manifold. The correspondence states that the Witten index of the resulting theory $T_{\mathcal{N}=1} [M_3]$ can alternatively be computed from the partition function on $M_3$ of a topological twist of the $T^3$-reduction of the 6d theory, a BF-theory coupled to a spinorial hypermultiplet (BFH).}
\end{center}
\end{figure}

The twist in (\ref{N1N2}) that preserves $\mathcal{N}=1$ in 3d will be the subject of this paper and we denote the corresponding 3d theories by $T_{\mathcal{N}=1}[M_3]$\footnote{We will usually consider gauge group $G= U(N)$, and in case of ambiguity specify $G$ by $T_{\mathcal{N}=1}[M_3, G]$.}. Geometrically this twist is realized by embedding $M_3$ as an associative three-cycle (i.e. calibrated and thus supersymmetric cycle) in a $G_2$-holonomy manifold. Note that the unbroken $SU(2)_\ell$ is not an R-symmetry in agreement with $\mathcal{N}=1$ supersymmetry. Instead it becomes manifest as a flavour symmetry associated to the normal bundle of $M_3$ discussed in section \ref{sec:DeformationTheory}. The observable we consider first  is the $T^3$-partition function, or Witten index \cite{Witten:1999ds}. We determine the 3d topological field theory, whose partition function on $M_3$ will be conjectured to compute the Witten index of the $T_{\mathcal{N}=1}[M_3]$ theory: first we dimensionally reduce the 6d $(2,0)$ theory on $T^3$ to 3d $\mathcal{N}=8$ SYM, which is then topologically twisted 
along $M_3$ to preserve two scalar supercharges. 
This results in a super-{\bf BF}-model coupled to a {\bf H}ypermultiplet, denoted by BFH-model in the following, and we propose the following identification of observables
\be \label{Correspondence1}
Z_{{\rm BFH}} (M_3) = I_{T^3} \left(T_{\mathcal{N}=1}[M_3]\right) \,,
\ee
where $Z$ denotes the partition function of the BFH-model on $M_3$ and $I_{T^3}$ is the Witten index or $T^3$-partition function of the $T_{\mathcal{N}=1}[M_3]$ theory -- see figure \ref{fig:3d3dN1Duality}.

The partition function of the BFH-model is computed in terms of the Euler characteristic of the moduli space of solutions to the BPS equations. These take the form of a generalized set of 3d Seiberg-Witten equations on $M_3$ for the pair $(\phi^{\alpha \hat \alpha}, A)$
\be \label{gSW}
(\hbox{gSW}_{M_3}):\qquad 
\ba
(\slashed{\mathcal{D}} \phi)^{\alpha \hat{\alpha}} &=0 \cr 
\varepsilon_{abc}F^{bc} - {i\over 2} [\phi_{\alpha \hat \alpha}, (\sigma_a)^\alpha{}_{\beta}\phi^{\beta \hat \alpha}] &=0  \,.
\ea
\ee
Here $\phi^{\alpha\hat\alpha}$ is a section of the normal bundle of the associative cycle, i.e. a section of the spin bundle twisted with an $SU(2)$-bundle $V$, and $A$ is a gauge connection for the gauge group $U(N)$, with $F = dA + [A,A]$. Furthermore,  $\slashed{\mathcal{D}}$ is the twisted Dirac operator of the covariant derivative $\mathcal{D}$, with respect to the gauge connection $A$, the $SU(2)$-connection $\ConnV$ on $V$, and the spin connection on $M_3$. The left hand side of the correspondence in \eqref{Correspondence1} can therefore be computed in terms of
\be  \label{Correspondence2}
Z_{{\rm BFH}} (M_3) = \chi\left(\mathcal{M}_{\text{gSW}_{M_3}}\right)\,,
\ee
where $\chi$ denotes the Euler characteristic and $\mathcal{M}_{\text{gSW}_{M_3}}$ is the moduli space of solutions to the generalized Seiberg-Witten equations (\ref{gSW}). 
The solutions to the gSW equations depend on the metric of the associative three-cycle, induced by the ambient $G_2$-metric, which determines the number of twisted harmonic spinors $\phi$ on $M_3$, i.e.  solutions to $(\slashed{\mathcal{D}} \phi)^{\alpha \hat{\alpha}} =0$. As a consequence of this we find that under deformations of the metric the  Witten index for the abelian theory $I_{T^3} (T_{\mathcal{N}=1}[M_3, U(1)])$ can jump, depending on whether there are such spinors.

To substantiate the conjectured correspondence \eqref{Correspondence1}, several checks are performed in the course of this paper: 
\begin{enumerate}
\item In the abelian case, the theory $T_{\mathcal{N}=1}[M_3,U(1)]$ and its Witten index can be computed explicitly by dimensional reduction of the 6d  abelian $(2,0)$ tensor multiplet. On the topological field theory side the BPS equations decouple into 
 flat $U(1)$-connections and twisted harmonic spinors, and the partition function can be computed to verify \eqref{Correspondence1}.
\item For non-abelian theories, we specialize the three-cycle $M_3$ to a Lens space $L(p,q)$ with the standard metric (which does not admit twisted harmonic spinors). The theory $T_{\mathcal{N}=1}[L(p,q), U(N)]$ was determined in \cite{Acharya:2001dz} and its index is known. Due to the absence of twisted harmonic spinors, the solutions to the gSW equations reduce to flat connections, and \eqref{Correspondence1} can again be checked explicitly.
\end{enumerate}

Note that for 3d $\mathcal{N} = 2$ theories on compact three-manifolds\footnote{The requirement of $M_3$ to be compact is necessary for the theory to have finitely many vacua \cite{Gadde:2013sca}. In the non-compact case the $T^3$ observable is difficult to compute  due to the presence of many fermionic zero modes and has only been computed for non-flat metrics on $T^3$, which preserve less supersymmetry \cite{Benini:2016hjo,Closset:2016arn}.}, obtained from the 6d $(2,0)$ theory on special Lagrangian three-manifolds inside a Calabi-Yau three-fold, it was argued in  \cite{Dimofte:2010tz} that the vacua of  $T_{\mathcal{N}=2}[M_3, G]$ on $S^1 \times \mathbb{R}^2$ are in one-to-one correspondence with \emph{complex} flat $G$-connections on $M_3$
\be
\# \hbox{vacua} (T_{\mathcal{N}=2}[M_3, G]) = \hbox{flat $G_{\mathbb{C}}$-connections on $M_3$} \,.
\ee
The complex flat connections arise from the gauge field coupling to sections of the normal bundle of the special Lagrangian three-cycle given by the cotangent bundle.
In the present case, we obtain the generalized Seiberg-Witten equations \eqref{gSW}, which couple  sections $\phi^{\alpha \hat \alpha}$ of the normal bundle of the associative cycle to real connections on $M_3$. If $M_3$ does not admit any twisted harmonic spinors, i.e. $d_{\slashed{\mathcal{D}}}(M_3,g)=0$, the solutions to $\text{gSW}_{M_3, G}$ are real flat $G$-connections, and the $\mathcal{N} = 1$ correspondence equates vacua of $T_{\mathcal{N}=1}[M_3, G]$ with \emph{real} flat $G$-connections on $M_3$, i.e. 
\be
d_{\slashed{\mathcal{D}}}(M_3,g)=0 :\qquad 
\# \hbox{vacua} (T_{\mathcal{N}=1}[M_3, G]) = \hbox{flat $G_{\mathbb{R}}$-connections on $M_3$} \,.
\ee
 In the next subsections we will first give a lightning summary of the geometry of 
associative three-cycles in $G_2$-holonomy manifolds. Secondly, we discuss generalizations of Seiberg-Witten equations and put the equations that we find in \eqref{gSW} into context. To our knowledge this generalization has not been proposed thus far and we contrast them to known generalizations of the SW equations. 

\begin{figure}
\begin{center}
  \includegraphics[height=3.5cm]{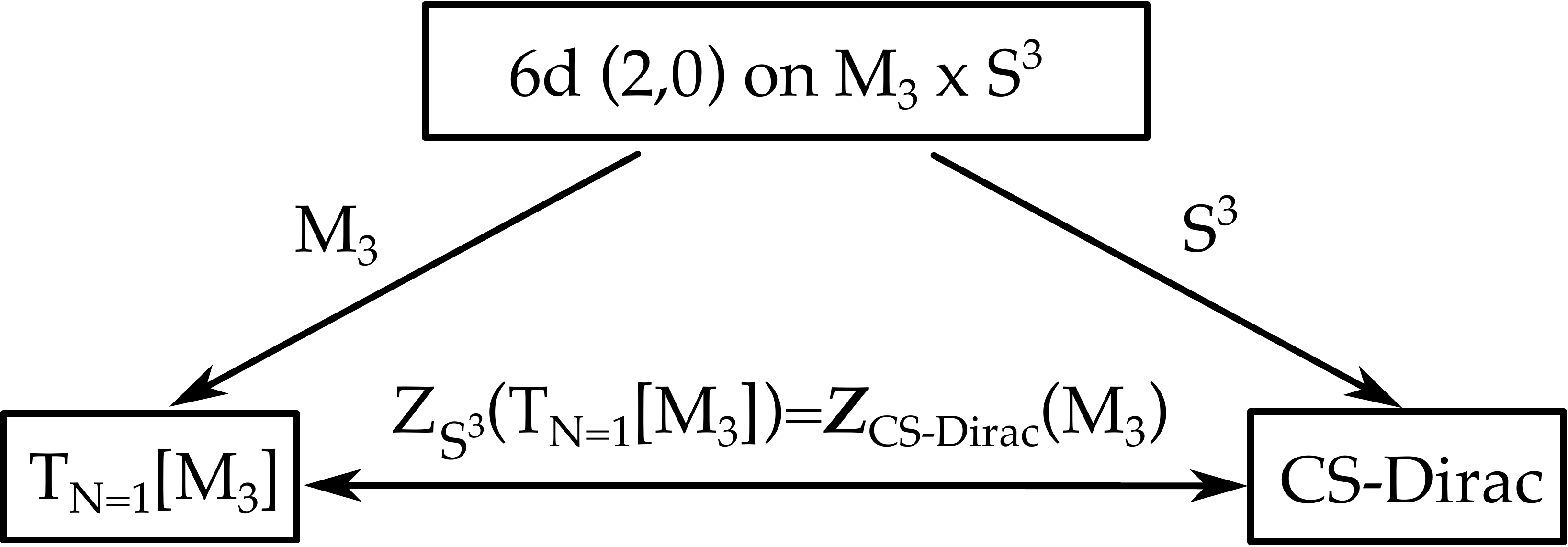}
 \caption{\label{fig:3d3dN1S3Duality} 
The setup for the  $\mathcal{N}=1$ 3d--3d correspondence is the 6d $(2,0)$ theory on $M_3\times S^3$. 
The topologically twisted reduction along $M_3$ preserves 3d $\mathcal{N}=1$ and geometrically corresponds to an embedding of $M_3$ as an  associative three-cycle in a $G_2$-holonomy manifold. 
In this variant the $S^3$-partition function of $T_{\mathcal{N}=1} [M_3]$ can alternatively be computed from the partition function of real Chern-Simons gauge theory on $M_3$ coupled to a twisted harmonic spinor, i.e. Chern-Simons-Dirac theory. In the case when there no twisted harmonic spinors, the theory reduces to real Chern-Simons theory and the partition function is given by the Witten-Reshetikhin-Turaev (WRT) invariant.}
\end{center}
\end{figure}

Before turning to this, we also summarize our findings with regards to another, less explored observable, the $S^3$-partition function of $T_{\mathcal{N}=1}[M_3]$. Very little is known about this observable in 3d $\mathcal{N}=1$ theories. We will nevertheless propose a variant of the 3d--3d correspondence for this observable, by determining  the topological field theory, whose partition function computes the $S^3$-partition function of the $\mathcal{N}=1$ theory. The strategy is again to start with the 6d $(2,0)$ theory and dimensionally reduce on $S^3$ in a  suitable conformal supergravity background -- the setup is shown in figure \ref{fig:3d3dN1S3Duality}. We find that the topological theory is a Chern-Simons-Dirac (CS-Dirac) theory, i.e. real Chern-Simons gauge theory at level 1 coupled to a twisted harmonic spinor $\phi$, whose equations of motion are precisely the generalized Seiberg-Witten equations (\ref{gSW}). We discuss the generalization to higher level as well, which correspond to the partition function on the Lens space $L(p,1)$. This leads us to conjecture that the sphere partition function of $T_{\mathcal{N}=1}[M_3, G]$ is computed by the $M_3$-partition function of this CS-Dirac theory
\be
Z_{S^3} \left(T_{\mathcal{N}=1} [M_3, G]\right)=  \mathcal{Z}_{{\rm CS}_{\mathbb{R}} {\rm-Dirac}, G} (M_3)  \,.
\ee
As we will discuss, generically in $G_2$-manifolds associatives will not have twisted harmonic spinors in which case the right hand side simply reduces to the partition function of real CS-theory on $M_3$, which computes the Witten-Reshetikhin-Turaev (WRT) invariant
\be
d_{\slashed{\mathcal{D}}}(M_3,g)=0 :\qquad  
Z_{S^3} \left(T_{\mathcal{N}=1} [M_3, G]\right)= \rm{WRT} (M_3) \,.
\ee
We will also argue that this correspondence can be further generalized by replacing the $S^3$ by a Lens space $L(p,1)$. The reduction on the Lens space yields Chern-Simons-Dirac theory at level $p$. Thus, we furthermore propose
\be
Z_{L(p,1)} \left(T_{\mathcal{N}=1} [M_3, G]\right)= \mathcal{Z}_{{\rm CSD}_p, G} (M_3)  \,.
\ee

This variant of the 3d--3d correspondence will be derived in section \ref{sec:S3}.

\subsection{$G_2$-holonomy, Associatives and Deformation Theory}
\label{sec:DeformationTheory}

This section gives a brief review of manifolds with holonomy group $G_2$ and calibrated, so-called associative, three-cycles, that play a key role in this paper. For more in depth reviews on these geometries in mathematics and string theory see \cite{Joyce, Acharya:2004qe}.
A $G_2$-holonomy manifold is a pair $(X_7,\Phi)$, where $X_7$ is an oriented manifold of real dimension seven and $\Phi$ is a harmonic three-form, the $G_2$-form. This manifold admits a metric $g_\Phi$, constructed from $\Phi$, with reduced holonomy group $G_2\subset SO(7)$. Key to the present discussion are calibrated or supersymmetric cycles. Associative three-cycles $M_3$ are calibrated by $\Phi$,\footnote{There are also co-associative four-cycles, which are calibrated by $\star \Phi$.} i.e.
\be
\text{vol}(M_3)=\Phi|_{M_3}\,,
\ee
and their properties including deformation theory have been discussed in \cite{McLean,Harvey:1982xk}. Let us fix a $G_2$-holonomy manifold $(X_7,\Phi)$ and an associative submanifold $M_3$. The cotangent space of $X_7$ can be identified with $\Im\mathbb{H}\oplus\mathbb{H}$ containing elements $(u,v)$. On $M_3$, we now identify $u$ as a one-form and $v$ as a section of the conormal bundle. In this language the $G_2$-form is
\be
\Phi=-{1 \over 3!}u_0\wedge u_0\wedge u_0-\frac{1}{2}\Re u_0\wedge \bar{v}_0\wedge v_0 \,,
\ee
for some initial element $(u_0,v_0)$. The space of deformations, which leave the structure invariant can be described by an $SO(4)$-action on $\Im\mathbb{H}\oplus\mathbb{H}$. Note that $SO(4)$ appears naturally as it is the largest group that does not mix $u$ and $v$, which is necessary to preserve the associative.  
We can write $SO(4)  \cong SU(2)_M\times SU(2)_V \ni (g,h)$, so that  $g$ acts as the rotations on $M_3$. To recover the usual transformation of one-forms we expect $u_0\rightarrow gu_0\bar{g}=u$ under the group action, where $\bar{g}=g^{-1}$. In order for $\Phi$ to transform as a three-form (i.e. in the same way as $u$) the group action on $v_0$ is $v_0\mapsto hv_0\bar{g}=v$. From these transformations one can deduce that the structure of the normal bundle consists of  two parts.
The group action $v\rightarrow  v\bar{g}$ identifies $v$ as transforming in the ${\bf 2}$ of $SU(2)_M$, and similarly also in the ${\bf 2}$ of $SU(2)_V$.
From these two transformations of $v$ one finds that the normal bundle of $M_3$ is 
\be \label{NM3}
N_{M_3}=\mathbb{S}\otimes V\,,
\ee
where $\mathbb{S}$ is the spinor bundle of $M_3$ and $V$ is the principle bundle associated to $SU(2)_V$.

The space of linear deformations is given by sections of the normal bundle, i.e. the kernel of $\slashed{\mathcal{D}}$, the twisted Dirac operator on $M_3$ coupled to the $SU(2)$-bundle $V$. 
We denote its dimension by 
\be
d_{\slashed{\mathcal{D}}}(M_3,g)= \dim \ker \slashed{\mathcal{D}} \,.
\ee
Note that $d_{\slashed{\mathcal{D}}}(M_3,g)$ depends on the metric $g$ \cite{MR0358873}. {An example of an associative that admits twisted harmonic spinors, given by an associative $T^3$ inside $T^7$, was discussed in \cite{joyce2003riemannian}. Although over a three-manifold every $SU(2)$-bundle is trivial, i.e. it admits a trivial connection, this does not imply that we can simply set the connection $\ConnV$ on $V$ to zero, as the connection relevant here is the one determined by the embedding of the associative cycle into the $G_2$-manifold.  The connection is therefore fixed, and not necessarily equal to the trivial connection. 
In the examples studied by Bryant-Salamon \cite{BryantSalamon}, the $G_2$-holonomy manifolds  are constructed as the total space of the spin bundle over a constant sectional curvature three-manifold, so that $\ConnV =0$.} In fact, whenever $\mathfrak{A}=0$ there are three distinct cases depending on the scalar curvature $\mathcal{R}$: 
\begin{enumerate}
\item $\mathcal{R}>0$: For positive curvature there are no harmonic spinors, i.e. $d_{\slashed{\mathcal{D}}}(M_3,g)=0$ \cite{MR0156292}, which implies that the associative three-cycle is rigid.
\item $\mathcal{R}=0$: The associative cycle is $T^3$ and the harmonic spinors coincide with the parallel spinors.
\item $\mathcal{R}<0$:  In this case the associative three-cycle can have non-trivial linear deformations depending on the induced metric on $M_3$. It was shown in \cite{MR1421872} that in dimensions $d = 3 \text{ mod } 4$ every closed spin manifold (in particular all closed orientable three-manifolds admit spin structures), with a given fixed spin structure, admits a Riemannian metric with $d_{\slashed{\mathcal{D}}}(M_3,g)\geq 1$. 
However, this metric need not coincide with the metric induced on the associative by the Ricci flat metric on the ambient $G_2$-manifold. 
\end{enumerate}

The index of the Dirac operator vanishes on a closed three-manifold, so that its obstruction space is of the same dimension as the deformation space \cite{McLean, joyce2003riemannian}. This implies that the virtual dimension of the moduli space of associative deformations is always zero. It is believed \cite{Joyce,joyce2003riemannian} that for generic $M_3$ there is no obstruction space, so there are also no linear deformations and $d_{\slashed{\mathcal{D}}}(M_3,g)=0$. This structure is generically preserved under small deformations of the metric, or equivalently the $G_2$-form $\Phi$. However for non-generic $G_2$-manifolds, the obstruction and deformation spaces can be non-trivial.

\begin{figure}
\begin{center}
  \includegraphics[height=4cm]{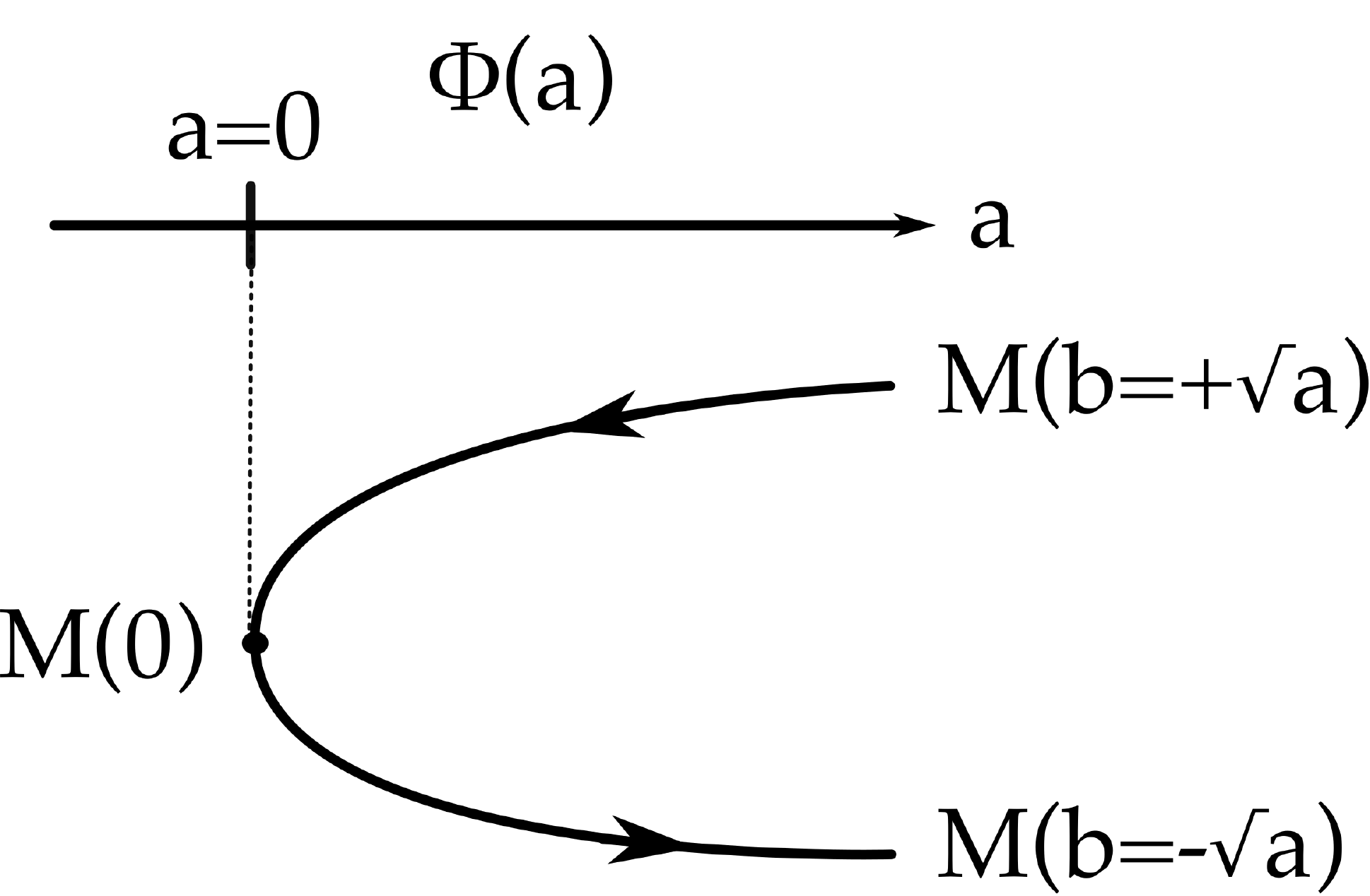}
 \caption{\label{fig:AssosOb} 
{The variation of an associative three-cycle $M(b)$ inside a $G_2$-manifold, with $G_2$-form $\Phi(a)$. The variation is shown with respect to the parameter $a$ under the constraint $b^2 = a$, required for $M(b)$ to be an associative.}}
\end{center}
\end{figure}

An example for how deformations can lead to obstructed associative cycles was given in \cite{Joyce:2016fij}. Consider families of $G_2$-manifolds $(X_7,\Phi(a))\equiv X(a)$ and embeddings $f(b): Y \to X(a)$, where $Y$ is some three-manifold, that vary smoothly with independent parameters $a,b\in \R$. It was argued in \cite{Joyce:2016fij} that one can choose the embeddings such that the $M_3(b)\equiv f(b)(Y)$ are associative inside $X(a)$ for $a=b^2$. Since they are embeddings of the same three-manifold $Y$ all the $M_3(b)$ are in the same homology class.
This setup is depicted in figure \ref{fig:AssosOb}.
Let us look at the linear deformation space of the associative $M_3(b)$ inside $X(b^2)$, i.e. consider $b\to b+\varepsilon$ and expand to linear order in $\varepsilon$. Consequently, $M_3(b+\varepsilon)$ is an associative inside $X(b^2+2b\varepsilon +\mathcal{O}(\varepsilon^2))$. For $b\neq0$ this variation does not describe a linear deformation of the associative inside a fixed $G_2$-manifold, as the latter changes. However, for $b=0$ it implies that the $M_3(\varepsilon)$ describe a family of associatives inside $X(0)$ (to leading order), so $M_3(0)$ has a non-zero linear deformation space, $d_{\slashed{\mathcal{D}}}(M_3,g)=1$. Clearly, this logic breaks down at higher order in $\varepsilon$ so the deformations are obstructed.
{Throughout this paper we will discuss variations of associative three-cycles  $M_3(b)$ in a fixed homology class and retain a description after wall-crossing, which occurs at the points where $d_{\slashed{\mathcal{D}}}(M_3,g)\neq0$.
}


\subsection{Generalized Seiberg-Witten Equations}
\label{sec:GenSWeq}

The topological field theories that will play a role in the 3d--3d correspondence for both $T^3$- and $S^3$-partition functions, are closely related to 
a set of generalized Seiberg-Witten equations on the  three-manifold $M_3$ 
\be  \label{GenSWeq}\ba
(\slashed{\mathcal{D}} \phi)^{ \alpha \hat \alpha} &=0\\
\varepsilon_{abc}F^{bc}-\frac{i}{2}[\phi_{\alpha\hat{\alpha}},(\sigma_a)^\alpha\,_\beta\phi^{\beta\hat{\alpha}}]  &=0 \,.
\ea \ee
Here $\phi^{\alpha\hat\alpha}$ is a section of the normal bundle of the three-cycle $M_3$ in the $G_2$-manifold $N_{M_3}  = \mathbb{S}\otimes V$ and thus a `bispinor' with respect to the $SU(2)_M$ and $SU(2)_V$. Both the gauge connection $A$ and $\phi$ are in the adjoint of the gauge group $G$, which we take to be $U(N)$, with $N>1$.\footnote{For the abelian theory we obtain flat connections.} The Dirac operator acting on the spinor $\phi$ is covariantized to the twisted Dirac operator, including a connection on $V$
\be
\mathcal{D} = \nabla + A + \mathfrak{A}\,,
\ee
where $\nabla$ is the covariant derivative with respect to the spin connection on $M_3$.

The above equations are non-abelian generalizations of the well-known 3d SW equations. Let us briefly recapitulate their origin and then contrast other generalizations to the one we encounter here (\ref{GenSWeq}).
The 4d Seiberg-Witten or monopole equations \cite{Witten:1994cg, Seiberg:1994rs,Seiberg:1994aj} 
consist of a $U(1)$ gauge field $A$ coupled to a positive chirality spinor $\Phi$ and its complex conjugate 
\be \label{4dSW}
F_{ij}^+-\frac{i}{2} \Phi^\dagger \Gamma_{ij} \Phi=0\,,\qquad \slashed{D}_A \Phi=0\,,
\ee
where $F^+$ is the self-dual part of the field strength $F = dA$ and $\slashed{D}_A$ is the Dirac operator in the presence of the gauge field.
The moduli space associated to these equations was shown by Witten to define a topological invariant, the Seiberg-Witten invariant, related to Donaldson invariants of four-manifolds \cite{donaldson1983}. A generalization incorporating multiple spinors transforming under an $SU(n)$ flavour symmetry has recently been studied in the context of M5-branes reduced on coassociative four-cycles \cite{dedushenko2017vertex}.
The reduction of the 4d equations on a circle yield the abelian 3d Seiberg-Witten equations, supplemented by equations involving the gauge field along the circle direction \cite{kronheimer1996embedded}.

The equations in \eqref{GenSWeq} are a generalization of the Seiberg-Witten equations in three dimensions to non-abelian gauge symmetry, and to spinors that are also sections of the $SU(2)_V$-bundle. To our knowledge these equations have not appeared as a generalization of the SW equations thus far. 
Various other generalizations of the Seiberg-Witten equations, which have an overlap with \eqref{GenSWeq}, have however arisen in the study of $G_2$-manifolds and $G_2$-instantons, which we briefly summarize below. 

A generalization of the 3d abelian Seiberg-Witten equations to incorporating $n$ spinors $\Psi$ transforming under an $SU(n)$-bundle, was recently considered in the context of counting $G_2$-instantons in \cite{MR3432158}
 \be \label{Walpu}
F = \Psi \Psi^* - \half |\Psi|^2 \,, \qquad  \slashed{\mathcal{D}}_{A \otimes B} \Psi =0\,, 
\ee
where the Dirac operator is twisted by the connection $B$ of the $SU(n)$-bundle. We note that these equations can also be obtained via dimensional reduction of the 4d abelian equations discussed in \cite{dedushenko2017vertex}.
The virtual dimension of the moduli space associated to \eqref{Walpu}, given by the difference of the dimension of the space of deformations and obstructions, is zero \cite{joyce2003riemannian}, however the count of solutions is not necessarily independent of the metric on the three-manifold. {This is an intrinsic property of the moduli space of twisted harmonic spinors, which are the zero modes of $\slashed{\mathcal{D}}_{A \otimes B}$. Consequently, the solutions to \eqref{GenSWeq}, which correspond to a non-abelian generalization of \eqref{Walpu} in the case of $n=2$, are also expected to be metric-dependent.}
  
The abelian 3d Seiberg-Witten equations with $n=2$ have appeared in relation to deformations of associative three-cycles in manifolds with (not necessarily torsion-free) $G_2$-structure in \cite{akbulut2004associative,akbulut2008deformations,akbulut2008calibrated}, where the twisted harmonic spinor condition of \cite{McLean} is supplemented with an additional Seiberg-Witten-like equation, which couples a $U(1)$ gauge field to sections of the normal bundle \eqref{NM3}, in order to make the space of deformations compact and zero-dimensional. In the context of $G_2$-strings, a non-abelian version was shown to arise as the equations of motion of the world-volume theory of topological 3-branes wrapped on an associative three-cycle in a $G_2$-manifold \cite{deBoer:2006bp}. In this context the spinor $\phi^{\alpha \hat\alpha}$ arises exactly from considering the normal modes of the associative, and the equations derived there, have some resemblance to  \eqref{GenSWeq}. It would be interesting to understand the precise relation between these two setups.

\section{$T_{\mathcal{N}=1}[M_3, U(1)]$} 
\label{sec:T[M,U(1)]}

We now turn to deriving the 3d $\mathcal{N}=1$ theory $T_{\mathcal{N}=1}[M_3, U(1)]$ from a topological twist of the 6d $\mathcal{N} = (2,0)$ theory, which geometrically corresponds to a single M5-brane wrapped on an associative three-cycle $M_3$ in a $G_2$-holonomy manifold. We reduce the equations of motion of the 6d abelian tensor multiplet to the ones of the 3d theory $T_{\mathcal{N}=1}[M_3, U(1)]$, and compute the Witten index for this theory for general $M_3$.

\subsection{Topological Twist for Associative Three-Cycles}
\label{sec:TopologicalTwist}

We consider the 6d $\mathcal{N} = (2,0)$ theory on 
\be\label{6dSpace}
\R^{1,2} \times M_3 \,,
\ee
where $M_3$ is an associative three-cycle in a $G_2$-holonomy manifold. 
The local model for this configuration is the total space of the normal bundle 
\eqref{NM3}. The relevant topological twist was first discussed in \cite{Blau:1996bx}. We will here determine the dimensional reduction to the $T_{\mathcal{N}=1}[M_3, U(1)]$ theory and its non-abelian generalization.
The Lorentz and R-symmetry group of the 6d theory is given by $SO(1,5)_L\times Sp(4)_R \subset OSp(2,6|4)$, the superconformal group in 6d. 
We consider the decompositions 
\be
\ba
SO(1,5)_L&\quad \to \quad SO(1,2)_L \times SO(3)_M\\
Sp(4)_R&\quad \to \quad SU(2)_{\ell} \times SU(2)_r \,,
\ea
\ee
where $SO(1,2)_L$ and $SO(3)_M$ are the local Lorentz groups acting on $\mathbb{R}^{1,2}$ and $M_3$, respectively.
The supersymmetry parameter transforms in the $({\bf 4}, {\bf 4})$ which decomposes as 
\be \ba 
SO(1,5)_L\times Sp(4)_R & \quad \rightarrow \quad SO(1,2)_L \times SU(2)_M \times SU(2)_\ell \times SU(2)_r  \cr 
({\bf 4}, {\bf 4}) &\quad \rightarrow \quad ({\bf 2}, {\bf 2},{\bf 2},{\bf 1}) \oplus ({\bf 2},{\bf 2},{\bf 1},{\bf 2}) \,.
\ea \ee
As explained in \eqref{N1N2} we twist the local Lorentz group of $M_3$ with the R-symmetry $SU(2)_r$
\be\label{Twist}
SU(2)_{\text{twist}} = \text{diag} (SU(2)_M , SU(2)_r) \,, 
\ee
under which the supersymmetry parameters transform as
\be\label{DecompBla}
\ba
SO(1,5)_L\times Sp(4)_R 
& \quad \rightarrow \quad  SO(1,2)_L  \times SU(2)_{\rm twist} \times  SU(2)_\ell \cr 
({\bf 4}, {\bf 4}) &\quad \rightarrow \quad ({\bf 2}, {\bf 2},{\bf 2}) \oplus ({\bf 2}, {\bf 1},{\bf 1}) \oplus ({\bf 2},{\bf 3},{\bf 1})\,.
\ea 
\ee
There are two supercharges that transform trivially under the twisted Lorentz symmetry on $M_3$, and thus $\mathcal{N}=1$ supersymmetry is preserved in the transverse 3d space-time. The $SU(2)_\ell$ is thus not an R-symmetry but rather a flavour symmetry.

To substantiate this, let us now consider the reduction of the 6d abelian tensor multiplet, 
consisting of five scalars $\Phi^{\underline{\hat{m}\hat{n}}}$, a self-dual three-form $H=dB$, and  fermions $\varrho^{\underline{\alpha\hat  m}}$ {satisfying the symplectic-Majorana-Weyl condition \eqref{SMW}}. A summary of our conventions can be found in appendix \ref{app:Conventions}, where $\underline{\hat{m}},\underline{\hat{n}} = 1, \cdots, 4$ denotes the fundamental of $Sp(4)_R$.\footnote{We take the convention of non-hatted indices and hatted indices to denote  Lorentz and R-symmetry indices, respectively.} Under \eqref{DecompBla} these become
\be\label{Decomposition6dFields}
\ba
SO(1,5)_L\times Sp(4)_R & \quad \rightarrow \quad  SO(1,2)_L  \times SU(2)_{\rm twist} \times  SU(2)_\ell\cr 
\Phi^{\underline{\hat m \hat n}}: \quad    ({\bf 1}, {\bf 5}) &\quad \rightarrow \quad  ({\bf 1}, {\bf 2},{\bf 2}) \oplus ({\bf 1}, {\bf 1},{\bf 1})\equiv(\phi^{\alpha\hat{\alpha}},\varphi) \cr 
H_{\underline{abc}}:\   ({\bf 10}, {\bf 1}) &\quad \rightarrow \quad  ({\bf 1}, {\bf 1},{\bf 1}) \oplus ({\bf 3}, {\bf 3},{\bf 1}) \equiv(h,H_{axy}) \cr
\varrho^{\underline{\alpha \hat m}}:\quad  (\bar{{\bf 4}}, {\bf 4})  &\quad \rightarrow \quad ({\bf 2}, {\bf 2},{\bf 2}) \oplus ({\bf 2}, {\bf 1},{\bf 1}) \oplus ({\bf 2},{\bf 3},{\bf 1})\equiv(\rho^{\sigma\alpha\hat{\alpha}},\lambda^\sigma,\xi_a^\sigma)\,.
\ea 
\ee
 Underlined indices denote transformations under the 6d symmetries where $\underline{a},\underline{b},\underline{c} = 0, \cdots, 5$ are flat space-time indices and $\underline{\alpha}= 1, \cdots, 8$ is a spinor index.
Under the decomposition $\sigma, \alpha, \hat\alpha$ are two component spinor indices of the groups $SO(1,2)_L$, $SU(2)_{\text{twist}}$ and $SU(2)_\ell$, respectively. Finally, $x,y$ and $a,b$ are flat space-time indices on $\mathbb{R}^{1,2}$, and $M_3$, respectively.
The key point to observe here is that from the 6d scalars we obtain a field $\phi^{\alpha \hat{\alpha}}$ transforming as a spinor under both the twisted Lorentz $SU(2)_{\rm twist}$ and the remnant flavour symmetry $SU(2)_\ell$. We therefore identify the $SU(2)_{\ell}$  with the $SU(2)$-bundle $V$ in \eqref{NM3}, which makes $\phi^{\alpha \hat \alpha}$ a section of $N_{M_3}$.\footnote{Note that this identification can also be justified more rigorously on the group level. The twist identifies the preserved R-symmetry $SO(4)_R \subset Sp(4)_R$ as the group preserving the $G_2$-structure discussed in section \ref{sec:DeformationTheory}.}

\subsection{From 6d (2,0) to 3d $\mathcal{N}=1$ $T_{\mathcal{N} = 1}[M_3, U(1)]$}
\label{sec:AbelianReduxM3}

For a single M5-brane, the dimensional reduction follows from the abelian tensor multiplet \cite{Howe:1997fb}, and we can determine  $T_{\mathcal{N} = 1}[M_3, U(1)]$ explicitly. In appendix \ref{app:AbelianDR}  we carry out the reduction and show  $T_{\mathcal{N} = 1}[M_3, U(1)]$ to be a supersymmetric Chern-Simons theory with free scalar multiplets.
The spectrum of $T_{\mathcal{N} = 1}[M_3, U(1)]$ depends on the first integral homology of $M_3$ 
\be \label{IntHomology}
H_1(M_3 , \mathbb{Z}) \cong \mathbb{Z}^{b_1(M_3)} \oplus \mathbb{Z}_{p_1} \oplus \cdots \oplus \mathbb{Z}_{p_r} \,,
\ee
as well as the number of twisted harmonic spinors $d_{\slashed{\mathcal{D}}}(M_3,g)$ on $M_3$.

These fields organize into 3d $\mathcal{N}=1$ multiplets (some basic properties of such theories are summarized in appendix \ref{app:3dN=1SUSY}). The supersymmetry transformations of the abelian 6d $\mathcal{N} = (2,0)$ theory are 
\be  \label{6dSUSY}
\ba
\delta B_{\underline{ab}}&=\epsilon_{\underline{\hat{m}}} \Gamma_{\underline{ab}} \varrho^{\underline{\hat{m}}}\\
\delta \Phi^{\underline{\hat{m}\hat{n}}}&=-4\epsilon^{[\underline{\hat{m}}}\varrho^{\underline{\hat{n}}]} -\Omega^{\underline{\hat{m}\hat{n}}} \epsilon^{\underline{\hat{r}}}\varrho_{\underline{\hat{r}}}\\
\delta \varrho^{\underline{\hat{m}}}&=\frac{1}{48} H^+_{\underline{abc}}\Gamma^{\underline{abc}} \epsilon^{\underline{\hat{m}}} +\frac{1}{4} \slashed{\p} \Phi^{\underline{\hat{m}\hat{n}}} \epsilon_{\underline{\hat{n}}}\,.
\ea
\ee 
The topological twist is implemented by requiring that the invariant supersymmetry parameter satisfies
\be \label{SUSYCondition}
\left(\Sigma_a\right)^{\underline{\alpha}}\,_{\underline{\beta}} \epsilon^{\underline{\beta\hat{m}}} + \delta_{a\hat{a}} \left(\Sigma_{\hat{a}}\right)^{\underline{\hat{m}}}\,_{\underline{\hat{n}}}\epsilon^{\underline{\alpha{\hat{n}}}}=0\,,
\ee
where $\Sigma_{a}$ and $\Sigma_{\hat a}$ are the generators of  $SU(2)_M$ and $SU(2)_r$, respectively, as defined in appendix \ref{app:6dGammas}. The solution to \eqref{SUSYCondition} is given by
\be \label{SUSYSinglet}
\epsilon^{\sigma\alpha\hat{m}}=\epsilon^\sigma \varepsilon^{\alpha\hat{m}}\qquad \hbox{and}\qquad 
\epsilon^{\sigma\alpha\hat{\alpha}} =0\,,
\ee
where $\varepsilon^{\alpha\hat m}$ is the anti-symmetric two-tensor.
The 6d symplectic-Majorana-Weyl condition implies the 3d reality condition 
\be \label{SMW3d3depsilon}
\epsilon^\sigma=
\begin{pmatrix}
\epsilon^1\\ i \epsilon^2
\end{pmatrix}
\,,
\ee
with  $\epsilon^1$ and $\epsilon^2$  real.
The dimensional reduction of the supersymmetry variations yields
\be\label{Our3dSUSY}
\ba
\delta A_x^m &=-2 \epsilon_\sigma {\gamma_x}^\sigma\,_\tau \xi^{\tau m}\\
\delta\alpha^I&=-2 \epsilon_\sigma \xi^{\sigma I}\cr 
\delta \varphi&=-2\epsilon_\sigma \lambda^\sigma\cr 
\delta h&=-2\epsilon_\sigma {\slashed{\p}_{\R^{1,2}}}^\sigma\,_\tau \lambda^\tau\cr
\delta\phi^{ i}&=-2\epsilon_\sigma \rho^{\sigma i}
\ea
\qquad \qquad 
\ba 
\delta \xi^{\sigma m}&=-\frac{1}{8}\varepsilon^{xyz} F^m_{xy} {\gamma_z}^\sigma\,_\tau \epsilon^\tau \cr 
\delta \xi^{\sigma I} &=-\frac{1}{4} {\slashed{\p}_{\R^{1,2}}}^\sigma\,_\tau \alpha^I  \epsilon^\tau \cr 
\delta \lambda^\sigma&=-\frac{1}{4}{\slashed{\p}_{\R^{1,2}}}^\sigma\,_\tau \varphi \epsilon^\tau-\frac{1}{4}h \epsilon^\sigma\cr 
\delta \rho^{\sigma i}&=-\frac{1}{4} {\slashed{\p}_{\R^{1,2}}}^\sigma\,_\tau \phi^{ i}  \epsilon^\tau\,.
\ea
\ee
Note that it is not possible to determine the supersymmetry variation of the auxiliary scalar $h$ by direct reduction as it only appears in the decomposition of the self-dual part of the field strength $H^+$, not of $B$ itself. It is straightforward to determine $\delta H_{abc}=3\p_{[a}\delta B_{bc]}$ and from this the variations of $H^\pm$. In order for $\delta H^-$ to vanish one has to impose the first equation of motion in \eqref{SpinorEOMFlat} by hand. This ensures that the 3d supersymmetry algebra closes off-shell. Then, the variation of $\delta H^+_{xyz}$ reduces to $\delta h$ in \eqref{Our3dSUSY}.

We can now compare this to the 3d $\mathcal{N}=1$ multiplet structure, which is summarized in appendix \ref{app:3dN=1SUSY}.\footnote{By an appropriate rescaling of the fermions
$\epsilon\to\sqrt{2}\epsilon$ and  $\{\lambda,\xi,\rho\}\to -\frac{1}{\sqrt{8}}\{\lambda,\xi,\rho\}$,
the supersymmetry variations can be brought into the standard form.}
Note that for the scalar multiplet with leading component $\varphi$ we obtain the full off-shell supersymmetry transformations. The field content of $T_{\mathcal{N}=1}[M_3, U(1)]$ can be interpreted as consisting of the following free 3d $\mathcal{N}=1$ multiplets:
\begin{enumerate}
\item
A single scalar multiplet $\mathcal{A}_\varphi \ni \{\varphi,\lambda^\sigma,h\}$. If we view $T_{\mathcal{N}=1}[M_3,U(1)]$ as a domain wall in the 4d $\mathcal{N}=1$ bulk theory, obtained by compactifying M-theory on  the $G_2$-holonomy manifold, this multiplet describes the center of mass.
\item
$b_1(M_3)$ massless scalar multiplets $\mathcal{A}_\alpha^I \ni \{\alpha^I,\xi^{\sigma I}\}$ coming from the free part of the first homology group of $M_3$.
\item
$d_{\slashed{\mathcal{D}}}(M_3,g)$ massless scalar multiplets $\mathcal{A}_\phi^{i}\ni \{\phi^{i},\rho^{\sigma i}\}$ which describe the deformations of the associative three-cycle $M_3$ inside the  $G_2$-holonomy manifold. These explicitly depend on the $G_2$-holonomy metric $g$ restricted to the associative cycle $M_3$.
\item
A set of $r$ massive gauge multiplets $\mathcal{V}_A^m\ni \{A^m,\xi^{\sigma m}\}$ whose masses are generated by Chern-Simons terms at levels $p_m$. Each multiplet $\mathcal{V}_A^m$ is induced by a factor in the torsion part of the first homology group \eqref{IntHomology} of $M_3$ .
\end{enumerate}
This gives the dictionary between the field content of $T_{\mathcal{N}=1}[M_3,U(1)]$ and the geometry of the associative three-cycle $M_3$.


\subsection{Witten Index of $T_{\mathcal{N} = 1}[M_3, U(1)]$}
\label{sec:AbelianWittenIndex}

{The observable for the proposed 3d--3d correspondence is the $T^3$-partition function, or Witten index, of $T_{\mathcal{N}=1}[M_3]$, which is defined by  $I=\Tr (-1)^F$ \cite{Witten:1982df}, and for $\mathcal{N}=1$ theories is the most  natural and well-explored observable to consider. We will discuss the much less explored $S^3$-partition function in section \ref{sec:S3}. We now compute the Witten index for the abelian Chern-Simons theories $T_{\mathcal{N}=1}[M_3,U(1)]$ for general three-manifolds $M_3$. }

The key geometric input from $M_3$ are its first Betti number $b_1(M_3)$, the torsion numbers $p_m$ in \eqref{IntHomology} and the number $d_{\slashed{\mathcal{D}}}(M_3, g)$ of zero-modes of the twisted Dirac operator, with $g$ being the metric arising from the restriction of the $G_2$-metric onto $M_3$. Since the abelian theory is non-interacting the multiplets decouple and the vacua can be written as tensor products. Thus, the Witten index is a product with the following contributions:
\begin{enumerate}
\item Independent of the details of the compact $M_3$ the theory includes a free scalar multiplet consisting of a bosonic and a fermionic state. Thus, the ground states always come in pairs with opposite fermion number and the full Witten index vanishes. Since this does not yield any information about the associative three-cycle we will exclude this center of mass multiplet from the computation of $I'=\Tr'(-1)^F$ in the same fashion as in \cite{Acharya:2001dz}.
\item 
There are $b_1(M_3)+d_{\slashed{\mathcal{D}}}(M_3,g)$ scalar multiplets, and as argued above, the Witten index of a free scalar multiplet vanishes. Thus, the Witten index vanishes unless $b_1(M_3) = d_{\slashed{\mathcal{D}}}(M_3,g) = 0$.
It is believed, \cite{Joyce,joyce2003riemannian}, that generic associative three-cycles have no deformation space, i.e. $d_{\slashed{\mathcal{D}}}(M_3,g) = 0$. For many examples, such as three-spheres and simple modifications thereof, furthermore $b_1(M_3)=0$.
\item The final piece is the set of $r$ vector multiplets with Chern-Simons self-interactions at levels $p_m$ given by the torsion numbers \eqref{IntHomology}. For a single $U(1)$ gauge field with Chern-Simons level $k$ the Witten index is $k$ \cite{Witten:1988hf}. Since the $r$ gauge fields are independent of each other the total contribution of this sector is $I_{\text{CS}}=\prod_{m=1}^r p_m$.
\end{enumerate}
The total Witten index $I'$, excluding the center of mass contribution, is thus given by
\be \label{AbelianWittenIndex}
I'(b_1 (M_3),d_{\slashed{\mathcal{D}}}(M_3,g),p_m)=\left\{
\begin{array}{c l}
\prod_{m=1}^r p_m &\quad \text{if }b_1(M_3)=d_{\slashed{\mathcal{D}}}(M_3,g)=0\\
0 & \quad \text{else\,.}
\end{array}\right.
\ee
Thus the Witten index is non-zero if and only if $M_3$ has trivial first rational homology ($b_1(M_3)=0$) and a vanishing obstruction space ($d_{\slashed{\mathcal{D}}}(M_3,g)=0$, see the discussion in section \ref{sec:DeformationTheory}). These manifolds are known as unobstructed rational homology 
three-spheres.\footnote{Note that if $M_3=S^3$ the index is one although there are no non-trivial cycles. This can be seen by an explicit reduction, where the Hopf fiber induces a Chern-Simons term at level one. This is consistent with \eqref{AbelianWittenIndex} as $I'$ reduces to the empty product.}
Under smooth deformations of the metric of the $G_2$-holonomy manifold, or equivalently the $G_2$-form, this index is discontinuous, whenever the kernel of $\slashed{\mathcal{D}}$ is non-vanishing. There is no guarantee that the index is invariant after passing through these loci, as the deformation space of an obstructed associative three-cycle can be singular. This phenomenon is known as wall-crossing.

We should comment on how the wall-crossing that we observe compares with the one proposed in \cite{Joyce:2016fij} for associatives in $G_2$-manifolds. 
We are interested only in variations of the $G_2$-form, which result in metric deformations on the associative three-cycle, while keeping its class in the third homology fixed. In particular we do not consider deformations, where the topology changes or the associative ceases to exists, or splits. The type of wall-crossing considered by Joyce in  \cite{Joyce:2016fij} is more closely connected to the M2-instanton partition function, as recently discussed in  \cite{Braun:2018fdp}.


\section{Non-abelian Generalization}
\label{sec:NonAbelianTheory}

Ideally at this point we would provide a generalization to non-abelian theories $T_{\mathcal{N}=1}[M_3, U(N)]$ for general, compact three-manifolds, however since the non-abelian 6d theory is unknown this is hard to come by. 
Thus, a precise dictionary between the associatives and the non-abelian theories in general is beyond the scope of the current paper. However, we provide two alternate ways to obtain some information about the non-abelian generalizations. One approach is to first reduce from 6d to 5d on a circle and study the 5d SYM theory on $M_3$. This results in a 2d sigma-model with $(1,1)$ supersymmetry, whose target space is the moduli space of the gSW equations. The alternative is to use a specialization of $M_3$, when the reduction is known, such as for $M_3$ a Lens space. More generally one can consider circle-fibrations and use methods such as 
\cite{Cordova:2013cea} to compute the dimensional reduction in those cases. We leave this for future work and focus here on the Lens spaces.  

\subsection{Circle Reduction to 2d $\mathcal{N} = (1,1)$ Sigma-Model }

\label{sec:2dSigmaModel}

\begin{figure}
\begin{center}
  \includegraphics[height=6cm]{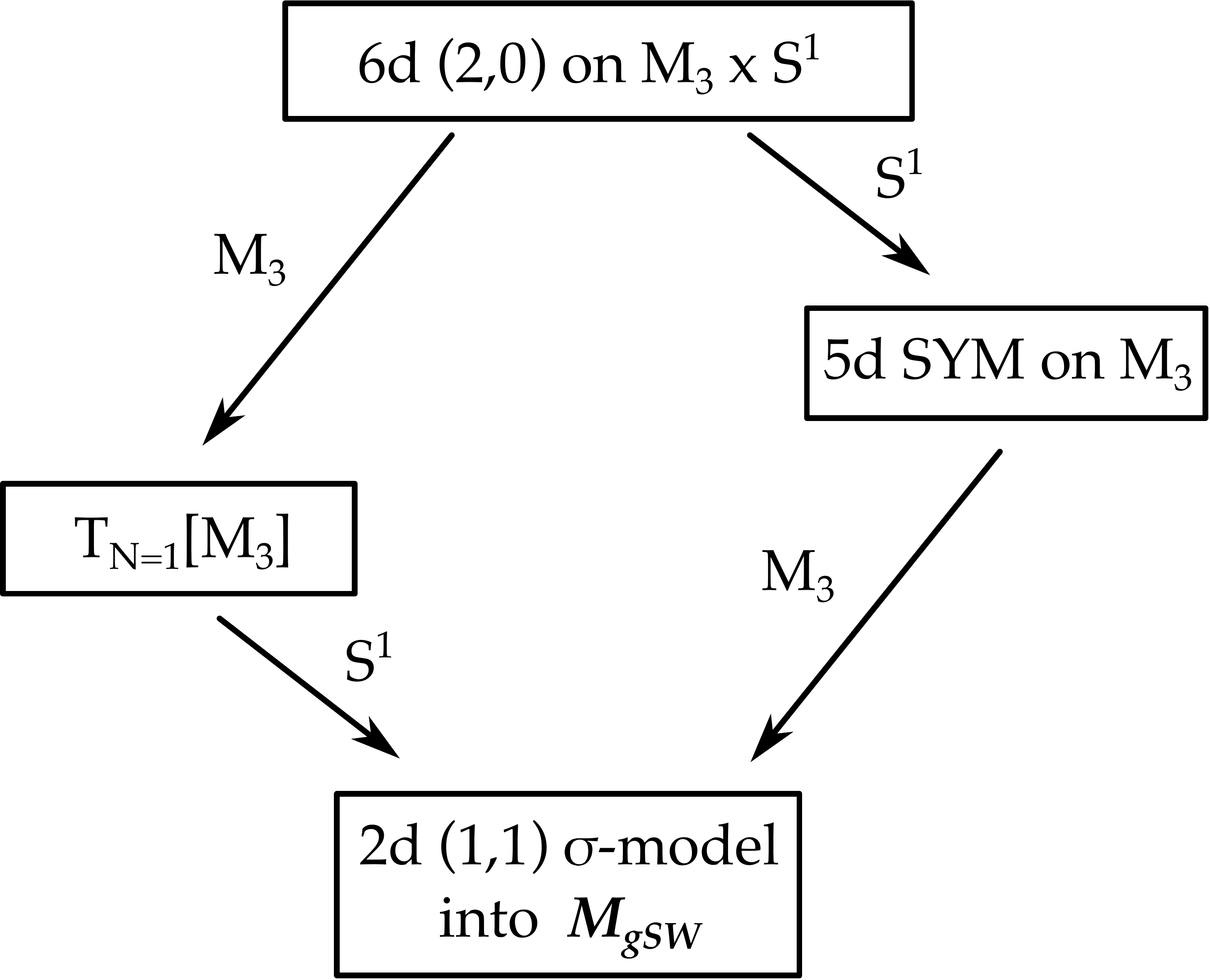}
 \caption{\label{fig:2d}  To study the non-abelian theory we consider first the dimensional reduction to 5d SYM and then the reduction on $M_3$ to 2d. This should correspond to the circle-reduction of the 3d 
 $T_{\mathcal{N}=1} [M_3]$ to a 2d $\mathcal{N}= (1,1)$ sigma-model, whose target space are the generalized Seiberg-Witten equations (gSW) in \eqref{BPSeq2d}.}
\end{center}
\end{figure}

To generalize the correspondence to non-abelian theories, without specializing the three-manifold, we first dimensionally reduce the 6d theory to 5d, and consider the non-abelian 5d SYM theory on $M_3$. 
This will not immediately reveal the $T_{\mathcal{N} = 1}[M_3]$ theory, however we will be able to generalize our results to the non-abelian version of this theory dimensionally reduced on a circle to a 2d $\mathcal{N}=(1,1)$ sigma-model. 
The target of this sigma-model is the moduli space of a particular generalization of 3d Seiberg-Witten equations. The setup is sketched in figure \ref{fig:2d}. 
A similar approach was applied by Gaiotto-Moore-Neitzke \cite{Gaiotto:2009hg} in studying the class $\mathcal{S}$ theories, obtained from M5-branes on a Riemann surface $\Sigma$, in terms of Hitchin equations, that arise in the compactification of the 5d SYM theory on $\Sigma$. The role of the Hitchin equations is in our context played by the generalized 3d Seiberg-Witten equations on $M_3$. 

The circle reduction of the 6d $\mathcal{N}=(2,0)$ theory results in 5d $\mathcal{N}=2$ SYM, i.e.  the worldvolume theory of D4-branes in type IIA string theory, which of course is known for any gauge algebra. We can then reduce further on $M_3$, which encodes information about the circle reduction of $T_{\mathcal{N}=1}[M_3]$ to two dimensions. 
The supersymmetry variations of the 5d theory are given by
\be \label{5dSUSY}
\ba
\delta A_{b'}&=-\frac{i}{4}\epsilon_{\underline{\hat{m}}}\Gamma_{b'}\rho^{\underline{\hat{m}}}\,, \qquad
\delta \Phi^{\underline{\hat{m}\hat{n}}}=-\epsilon^{[\underline{\hat{m}}}\rho^{\underline{\hat{n}}]}-\frac{1}{4}\Omega^{\underline{\hat{m}\hat{n}}}\epsilon^{\underline{\hat{r}}}\rho_{\underline{\hat{r}}}\\
\delta\rho^{\underline{\hat{m}}}&=-i\slashed{D}\Phi^{\underline{\hat{m}\hat{n}}}\Omega_{\underline{\hat{r}\hat{n}}}\epsilon^{\hat{r}}+\half F^{a'b'}\Gamma_{a'b'}\epsilon^{\underline{\hat{m}}}-\half\Omega_{\underline{\hat{n}\hat{r}}}[\Phi^{\underline{\hat{m}\hat{n}}},\Phi^{\underline{\hat{r}\hat{s}}}]\epsilon_{\underline{\hat{s}}}\,,
\ea
\ee
where $F_{a'b'}$ is the field strength of the 5d gauge field $A_{b'}$ and $\underline{\hat{m}}=1,\cdots,4$ is the R-symmetry index.
Consider the topological twist discussed in section \ref{sec:TopologicalTwist}, which is unaltered by the reduction to 5d. {The resulting field content is given by 
\be \label{5dFieldDecomposition}
\ba
SO(1,4)_L\times Sp(4)_R & \quad \rightarrow \quad  U(1)_L  \times SU(2)_{\rm twist} \times  SU(2)_\ell\cr 
A:\quad  ({\bf 5}, {\bf 1}) &\quad \rightarrow \quad ( {\bf 1},{\bf 1})_\pm \oplus ( {\bf 3},{\bf 1})_0  \equiv(A_{\pm},A_b) \cr
\rho:\quad  ({\bf 4}, {\bf 4})  &\quad \rightarrow \quad   ( {\bf 1},{\bf 1})_\pm \oplus ({\bf 3},{\bf 1})_\pm \oplus ( {\bf 2},{\bf 2})_\pm \equiv(\lambda^{\pm},\xi_a^{\pm},\rho^{{\pm}\alpha\hat{\alpha}})\cr
\Phi:\quad    ({\bf 1}, {\bf 5}) &\quad \rightarrow \quad   ({\bf 1},{\bf 1})_0 \oplus ({\bf 2},{\bf 2})_0 \equiv(\varphi,\phi^{\alpha\hat{\alpha}}) \,.
\ea 
\ee}
The twist of the 5d $\mathcal{N}=2$ vector yields gauge fields $A_b$ on $M_3$ and $A_\pm$ on $\R^{1,1}$ respectively. The decomposition of the spinor and scalar sectors is similar to \eqref{Decomposition6dFields}.
 This preserves two supercharges of opposite chirality corresponding to 2d $\mathcal{N}=(1,1)$. In order to put this theory on a general associative $M_3$ the Dirac operator has to be covariantized with respect to the spin connection on $M_3$ as well as the $SU(2)_V$ connection $\ConnV$ as discussed in appendix \ref{app:AbelianDR}.\footnote{Note that this introduces a mass term for the $\phi^{\alpha\hat{\alpha}}$ proportional to the Ricci scalar $\mathcal{R}(M_3)$ as well as a term involving the field strength $\mathfrak{F}$ of $\ConnV$. These terms arise naturally by requiring invariance under the covariantized supersymmetry transformations. This is discussed in greater detail in section \ref{sec:S3}.}
To dimensionally reduce on $M_3$ we introduce a length scale $s$ of the three-manifold and consider the theory in the limit $s\to 0$. To this end we rescale the fields and coordinates on $M_3$ to make their $s$-dependence explicit. The reduction proceeds by solving the BPS equations, which are the most divergent terms in the $s$-expansion of the supersymmetry variations
\be \label{BPSeq2d}
\ba
\varepsilon_{abc}F^{bc}-\frac{i}{2}[\phi_{\alpha\hat{\alpha}},\phi^{\beta\hat{\alpha}}](\sigma_a)^\alpha\,_\beta  &= 0\\
(\slashed{\mathcal{D}}\phi)^{ \alpha \hat \alpha} &= 0\\
D_a\varphi =[\varphi,\phi^{\alpha\hat{\alpha}}] &= 0 \,.
\ea
\ee
The first two BPS equations are the generalized Seiberg-Witten equations \eqref{gSW}. For field configurations satisfying \eqref{BPSeq2d} the leading order terms in 5d vanish and the action is minimized. The reduction to two dimensions is performed {in a standard way} and leads to a 2d $\mathcal{N} = (1,1)$ sigma model into the moduli space of solutions to \eqref{BPSeq2d}. These equations generalize the usual abelian 3d Seiberg-Witten equations in that the scalars $\phi^{\alpha\hat{\alpha}}$ transform in the adjoint of a non-abelian gauge group and in the fundamental of an additional group $SU(2)_V$. The Dirac operator $\slashed{\mathcal{D}}$ is therefore twisted by the $SU(2)$-bundle $V$.

To consider the 2d $\mathcal{N} = (1,1)$ sigma model on $T^2$ one needs to choose boundary conditions for the fermions. As described in \cite{Witten:1986bf}, choosing Ramond boundary conditions for both positive and negative chirality fermions corresponds to computing the Witten index \cite{Witten:1982df}, which is equivalent to the Euler characteristic of the moduli space of solutions to \eqref{BPSeq2d}. On the other hand, by choosing Neveu-Schwarz boundary conditions for the negative chirality fermions and Ramond boundary conditions for positive chirality one computes the elliptic genus. To make a connection to the Witten index of the theories $T_{\mathcal{N} = 1}[M_3]$ the case of interest therefore corresponds to Ramond boundary conditions for all fermions. As will be discussed in section \ref{sec:M5onT3}, the topologically twisted theory on the other side of the $\mathcal{N} = 1$ 3d--3d correspondence computes the Euler characeristic of the moduli space of gSW equations, which corresponds to solutions of \eqref{BPSeq2d} with $\varphi = 0$. The Euler characteristic of the moduli space for the abelian case, corresponding to a single M5-brane, is computed in section \ref{sec:BFHAbelian}, and shown to match the Witten index of the abelian theory found in section \ref{sec:AbelianWittenIndex}.

\subsection{Specializing $M_3$: $T_{\mathcal{N} = 1}[L(p,q), U(N)]$}
\label{sec:LensSpaceTheory}
 
Another approach to obtaining further insight into the non-abelian generalizations is to consider special cases of three-cycles $M_3$. In fact one of the most common associative cycles in compact $G_2$-manifolds that are known are three-spheres or simple modifications thereof -- see e.g. the twisted connected sum constructions in \cite{Kovalev:2001zr,  CHNP1, Corti:2012kd}, where associatives are either $S^3$ or diffeomorphic to $S^2\times S^1$, or more recently the conjecture for an infinite family of associative three-cycles with topology $S^3$ in these geometries \cite{Braun:2018fdp}.

It is thus useful to consider specializations of the three-cycles $M_3$ and determine the non-abelian theories using special properties of the geometries. This strategy has already been successfully applied for the $\mathcal{N}=2$ version of the 3d--3d correspondence. In particular we focus on Lens spaces $L(p,q)$, defined as $\mathbb{Z}_p$ quotients of $S^3$, where the action of the quotient on $S^3$, embedded as a unit sphere in $\mathbb{C}^2$ with complex coordinates $(z_1, z_2)$, is given by
\be 
(z_1, z_2) \rightarrow (e^{2\pi i / p} z_1, e^{2\pi i q/p} z_2)\,.
\ee
For $(p,q)$ coprime this action has no fixed points, and is therefore free. Lens spaces appear as associative three-cycles in the $G_2$-manifolds 
\be 
X_7 = (\mathbb{R}^4 \times S^3)/\Z_p\,,
\ee
considered in \cite{Acharya:2001dz}.
In these cases the embedding of the associative three-cycle is trivial and there are no twisted harmonic spinors. For  $q=1$,  the Lens space is a Hopf fibration $S^1\hookrightarrow L(p,1) \to S^2$. The reduction of the 6d $\mathcal{N}=(2,0)$ theory along the fiber direction yields 5d $\mathcal{N}=2$ SYM in the presence of $p$ units of graviphoton flux. This theory possesses a Lagrangian description for general gauge group and the subsequent reduction on $S^2$ can be performed explicitly. This yields a 3d SYM theory with a Chern-Simons term at level $p$ coupled to a scalar multiplet in the adjoint. The bosonic part of the action is given by
\be \label{T[L(p,q)]}
S=\frac{1}{g^2}\int d^3x \Tr\left(-\half F_{ij}  F^{ij} + \half D_i \varphi D^i \varphi \right)+\frac{p}{4\pi}\int \text{CS}(A)\,,
\ee
where the scale of the gauge coupling is given by the radius of the Lens space $g^{-2}\sim r$. In the abelian case $G=U(1)$ this matches the description in section \ref{sec:AbelianReduxM3} as $H_1(L(p,q),\Z)=\Z_p$. Since the theory only depends on the topological data of the associative this can be generalized to all $q$.

For these 3d $\mathcal{N} = 1$ Chern-Simons theories $T_{\mathcal{N} = 1}[L(p,q)]$ the Witten index was computed in \cite{Witten:1999ds,Acharya:2001dz} to be 
\be  \label{LensWittenIndex}
I_{T^3}(T_{\mathcal{N} = 1}[L(p,q),U(N)]) = {p \choose N}\,.
\ee
To compute this one uses the fact that the vacua of the Chern-Simons theory \eqref{T[L(p,q)]} are given by flat $U(N)$-connections. This moduli space consists of a set of points, each representing an $N$-dimensional representation of the fundamental group $\pi_1(L(p,q)) = \mathbb{Z}_p$, and the {Witten index} is given by the sum over all such points. The abelian flat connections are in one-to-one correspondence with the irreducible representations of $\mathbb{Z}_p$, and can be represented by the $p$ nodes on the associated affine Dynkin diagram for $A_{p-1}$ \cite{Gopakumar:1997dv}. The generalization to flat $U(N)$-connections follows by considering the number of ways in which $N$-dimensional representations of $\mathbb{Z}_p$ can be built from the 1-dimensional irreducible representations, which is given by ${p \choose N}$.
In section \ref{sec:M5onT3} we shall show that in this case the $\mathcal{N} = 1$ 3d--3d correspondence holds by determining the partition function of the BFH-model.


\section{BFH-Model on $M_3$}
\label{sec:BFH}

We now turn to the topological field theory side of the correspondence, i.e. the right hand side of figure \ref{fig:3d3dN1Duality}, whose partition function is conjectured to compute the Witten index of the 3d $T_{\mathcal{N} = 1}[M_3]$ theory.
The TQFT can be obtained from the $T^3$-reduction of the 6d (2,0) theory, to a maximally supersymmetric 3d theory, which we then topologically twist along $M_3$ in the same fashion as  in section \ref{sec:TopologicalTwist}. Following the correspondence \eqref{Correspondence1} the partition function of this topological theory computes the Witten index $I_{T^3}(T_{\mathcal{N}=1}[M_3])$ of the 3d $\mathcal{N}=1$ theory, and we verify this in examples.  
  
\subsection{Topological Twist of 3d $\mathcal{N} = 8$ SYM}
\label{sec:M5onT3}

The reduction of the 6d $\mathcal{N}=(2,0)$ theory  on $T^3$ yields maximally supersymmetric 3d $\mathcal{N}=8$ SYM theory on $M_3$. The topological twist preserves two scalar supercharges $Q^{\dot{\sigma}}$ and  the field content of the theory is:
\be \label{BFHFieldContent}
\begin{array}{c|c|c} 
& \text{Field} & SU(2)_{\text{twist}} \times SU(2)_\ell \times SU(2)_I \\ \hline
\multirow{3}{*}{\text{Bosonic}} & A_a & (\bold{3}, \bold{1}, \bold{1}) \\
& \varsigma_{\dot{a}} & (\bold{1}, \bold{1}, \bold{3}) \\
& \phi_{\alpha \hat{\alpha}} & (\bold{2}, \bold{2}, \bold{1}) \\\hline
\multirow{3}{*}{\text{Fermionic}} & \xi_a^{\dot{\sigma}} & (\bold{3}, \bold{1}, \bold{2}) \\
&  \lambda^{\dot{\sigma}} & (\bold{1}, \bold{1}, \bold{2}) \\ 
& \rho_{\alpha \hat{\alpha}}^{\dot{\sigma}} & (\bold{2}, \bold{2}, \bold{2})
\end{array}
\ee
Here, the hatted and dotted indices denote the representations under the internal symmetries $SU(2)_\ell$ and $SU(2)_I$, respectively. This 3d topological action is given by the 3d super-BF-model \cite{Witten:1989sx,Birmingham:1989is} coupled to a spinorial hypermultiplet, which  first appeared in \cite{Geyer:2001yc}. In addition, we will gauge the $SU(2)_\ell$ flavour symmetry, which is identified with the bundle $V$ in \eqref{NM3}, {to ensure that $M_3$ is embedded as an associative three-cycle.}

This theory can also be obtained as the $T^2$-reduction of the twisted 5d $\mathcal{N}=2$ theory introduced in section \ref{sec:2dSigmaModel}. After the reduction to three dimensions, the $U(1)$ symmetry on the torus gets enhanced to $SU(2)_I$, and the transformations of the fields can be repackaged into the content in \eqref{BFHFieldContent}.
Inspired by the description in \cite{Geyer:2001yc} we find that the reduction of the supersymmetry variations can be written in an off-shell form with the introduction of two auxiliary fields, $B_a$ and $W^{\alpha \hat \alpha}$,
\be \ba 
Q^{\dot{\sigma}} A_a &=\half \xi^{\dot{\sigma}}_{a} \,, & \qquad   Q^{\dot{\sigma}} \varsigma^{\dot{a}} &= -\frac{1}{2} (\sigma^{\dot{a}})^{\dot{\sigma}}{}_{\dot{\tau}} \lambda^{\dot{\tau}} \\
Q^{\dot{\sigma}}\phi^{\alpha \hat \alpha}& =\half \rho^{\dot{\sigma} \alpha \hat{\alpha}} \,,& \qquad Q^{\dot{\sigma}} \lambda^{\dot{\tau}} &=- \frac{i}{2} \varepsilon_{\dot{a} \dot{b} \dot{c}} (\sigma^{\dot{a}})^{\dot{\sigma} \dot{\tau}}[\varsigma^{\dot{b}}, \varsigma^{\dot{c}}] \\
Q^{\dot{\sigma}} \xi^{\dot{\tau}}_{a} &= (\sigma^{\dot{a}})^{\dot{\sigma} \dot{\tau}} D_a \varsigma_{\dot{a}} + \varepsilon^{\dot{\sigma} \dot{\tau}} B_a \, & \qquad Q^{\dot{\sigma}}\rho^{\dot{\tau} \alpha \hat \alpha} &=- [ (\sigma^{\dot{a}})^{\dot{\sigma} \dot{\tau}} \varsigma_{\dot{a}}, \phi^{\alpha \hat \alpha}] + \varepsilon^{\dot{\sigma} \dot{\tau}} W^{\alpha \hat \alpha} \\
Q^{\dot{\sigma}}B_a &= -\half D_a \lambda^{\dot{\sigma}} - \half (\sigma^{\dot{a}})^{\dot{\sigma}}{}_{\dot{\tau}}[\varsigma_{\dot{a}}, \xi_a^{\dot{\tau}}] \,, & \qquad Q^{\dot{\sigma}} W^{\alpha \hat {\alpha}} &= \half [\lambda^{\dot{\sigma}}, \phi^{\alpha \hat \alpha}] - \half (\sigma^{\dot{a}})^{\dot{\sigma}}{}_{\dot{\tau}}[\varsigma_{\dot{a}}, \rho^{\dot{\tau} \alpha \hat \alpha}] \,.
\ea \ee
The supersymmetry algebra closes up to gauge transformations induced by the $\varsigma_{\dot{a}}$, i.e.
\be
\{Q^{\dot{\sigma}},Q^{\dot{\tau}}\}=-(\sigma^{\dot{a}})^{\dot{\sigma}\dot{\tau}}\delta_I (\varsigma_{\dot{a}})\,,
\ee
where the gauge transformation acts as $\delta_I (\varsigma_{\dot{a}})A_b=-D_b  \varsigma_{\dot{a}}$ and $\delta_I (\varsigma_{\dot{a}})= [\varsigma_{\dot{a}},\cdot]$ on all other fields. 
{The action invariant under this set of supersymmetry transformations has a BF-type coupling and is given by 
\be \label{BFH}
\ba
S_{\text{BFH}} &= \frac{1}{e^2}\int d^3 x  \Tr \left[-B^a\left(B_a-\varepsilon_{abc}F^{bc}+\frac{i}{2}[\phi_{\alpha\hat{\alpha}},(\sigma_a)^\alpha\,_\beta\phi^{\beta\hat{\alpha}}]\right)-\half W_{\alpha\hat{\alpha}}\left(W^{\alpha\hat{\alpha}}-2i\slashed{D}^\alpha\,_\beta \phi^{\beta\hat{\alpha}}\right) \right. \\
& \left. \qquad \qquad+\half[\varsigma_{\dot{a}},\varsigma_{\dot{b}}][\varsigma^{\dot{a}},\varsigma^{\dot{b}}]+D_a\varsigma_{\dot{a}}D^a\varsigma^{\dot{a}}+\half [\varsigma_{\dot{a}},\phi_{\alpha\hat{\alpha}}][\varsigma^{\dot{a}},\phi^{\alpha\hat{\alpha}}] + \xi_{a\dot{\sigma}}D^a \lambda^{\dot{\sigma}}+\half\varepsilon^{abc}\xi_{a\dot{\sigma}}D_b\xi_c^{\dot{\sigma}}\right]\\
& \left. \qquad \qquad  +\frac{i}{4}\rho_{\dot{\sigma}\alpha\hat{\alpha}}\slashed{D}^\alpha\,_\beta \rho^{\dot{\sigma}\beta\hat{\alpha}} +\frac{1}{4}\rho_{\dot{\sigma}\alpha\hat{\alpha}}(\sigma^{\dot{a}})^{\dot{\sigma}}\,_{\dot{\tau}}[\varsigma_{\dot{a}},\rho^{\dot{\tau}\alpha\hat{\alpha}}] +\half \xi_{a\dot{\sigma}}(\sigma^{\dot{a}})^{\dot{\sigma}}\,_{\dot{\tau}}[\varsigma_{\dot{a}},\xi^{a\dot{\tau}}]\right. \\
& \left. \qquad \qquad +\half \lambda_{\dot{\sigma}}(\sigma^{\dot{a}})^{\dot{\sigma}}\,_{\dot{\tau}}[\varsigma_{\dot{a}},\lambda^{\dot{\tau}}] +\half\rho_{\dot{\sigma}\alpha\hat{\alpha}}[\phi^{\alpha\hat{\alpha}},\lambda^{\dot{\sigma}}]-\frac{i}{2}\xi_{a\dot{\sigma}}[\rho^{\dot{\sigma}}_{\alpha\hat{\alpha}},(\sigma^a)^\alpha\,_\beta\phi^{\beta\hat{\alpha}}]
\right]\,,
\ea
\ee}
which can be written in a $Q$-exact form
\be
S_{\text{BFH}} = \frac{1}{2e^2} \varepsilon_{\dot{\sigma} \dot{\tau}} Q^{\dot{\sigma}} Q^{\dot{\tau}} V_{\text{BFH}}\,,
\ee
where 
\be \label{BFHV} \ba 
V_{\text{BFH}} &= \int d^3 x \Tr \left(2 \varepsilon^{abc}( A_{a} \partial_{b} A_c + \frac{2}{3} A_a A_b A_c) +\half ( \xi_{a\dot{\sigma}} \xi^{a\dot{\sigma}} - \lambda_{\dot{\sigma}}\lambda^{\dot{\sigma}}) \right. \\
& \left. \qquad \qquad \phantom{\frac{1}{1}} +i\phi_{\alpha \hat \alpha}(\sigma^a)^{\alpha}\,_{\beta}D_a \phi^{\beta \hat \alpha}+\frac{1}{4} \rho_{\dot{\sigma}\alpha \hat \alpha} \rho^{\dot{\sigma} \alpha \hat{\alpha}}\right)\,.
\ea \ee
We identify the gauge coupling constant $e$ as 
\be
\frac{1}{e^2}=\frac{v_{T^2}}{8\pi^2 r}\,,
\ee
where $v_{T^2}$ is the $T^2$-volume and $r$ is the radius of the M-theory circle. Note that the radii of the three circles enter the gauge coupling differently, as we first reduce to  5d, where the fields are rescaled by the corresponding radius $r$ to ensure canonical scaling dimension.\footnote{This is very similar to the arguments in the geometric interpretation of S-duality in $4d$ $\mathcal{N}=4$ SYM obtained by a $T^2$-reduction of the $6d$ $\mathcal{N}=(2,0)$ theory.}

The auxiliary fields can be integrated out to give
\be
\ba
B_a&=\half\left(\varepsilon_{abc}F^{bc}-\frac{i}{2}[\phi_{\alpha\hat{\alpha}},\phi^{\beta\hat{\alpha}}](\sigma_a)^\alpha\,_\beta\right)\\
W^{\alpha\hat{\alpha}}&=i\slashed{D}^\alpha\,_\beta\phi^{\beta\hat{\alpha}}\,,
\ea
\ee
such that the bosonic action can be written as
\be \label{T3ActionBosonic}
\ba
S_{\text{BFH}}|_{\text{bosonic}}&=\frac{1}{e^2}\int d^3 x \Tr \left(\half F_{ab} F^{ab}+\half[\varsigma_{\dot{a}},\varsigma_{\dot{b}}][\varsigma^{\dot{a}},\varsigma^{\dot{b}}]+D_a\varsigma_{\dot{a}}D^a\varsigma^{\dot{a}}-\half  \phi_{\alpha\hat{\alpha}} (\slashed{D}^2 \phi)^{\alpha\hat{\alpha}} \right. \\
& \left. \qquad \qquad +\half [\varsigma_{\dot{a}},\phi_{\alpha\hat{\alpha}}][\varsigma^{\dot{a}},\phi^{\alpha\hat{\alpha}}]+\frac{1}{8}[\phi_{\alpha\hat{\alpha}},\phi_{\beta\hat{\beta}}][\phi^{\alpha\hat{\alpha}},\phi^{\beta\hat{\beta}}]\right)\,.
\ea
\ee
Note that the above derivation holds for curved $M_3$ by covariantizing the derivatives with respect to the spin connection on $M_3$ and the connection $\ConnV$ of the $SU(2)$ bundle $V$.
Indeed, the energy momentum tensor $T_{ab}$, the variation of $S_{\text{BFH}}$ with respect to the metric on $M_3$, can be written in a $Q$-exact form  
\be\ba
T_{ab} &= \varepsilon_{\dot{\sigma}\dot{\tau}} Q^{\dot{\sigma}}Q^{\dot{\tau}} \left( \varepsilon^{cd}{}_{(a}A_{b)} F_{cd} - 2 A_c \varepsilon^{cd}{}_{(a} F_{b)d}- \varepsilon^{cd}{}_{(a}A_{b)}[A_c,A_d]+ \half \xi_{a \dot{\rho}} \xi_{b}^{\dot{\rho}} +i\phi_{\alpha \hat \alpha} (\sigma_{(a})^{\alpha}{}_{\beta} D_{b)} \phi^{\beta \hat \alpha} \right. \\
&\left.  \qquad \qquad \qquad \phantom{\half} - g_{ab} \mathcal{V}_{\text{BFH}} \right)\,,
\ea \ee
where $\mathcal{V}_{\text{BFH}}$ is the Lagrangian density in \eqref{BFHV}.
The BFH-model is therefore topological in the sense of Witten, however this does not forbid the observables to exhibit wall-crossing. As discussed in section \ref{sec:BFHAbelian}, the partition function of the BFH-model, under the proposed correspondence \eqref{Correspondence1}, is expected to undergo wall-crossing for metrics admitting twisted harmonic spinors. We can understand this as the moduli space of vacua becoming singular due to the obstructions discussed in section \ref{sec:DeformationTheory}.

The BFH-model \eqref{BFH} preserves two topological supercharges and  the partition function is expected to compute the Euler characteristic of some moduli space \cite{Blau:1996bx,Dijkgraaf:1996tz}.\footnote{For the Euler characteristic to be well-defined the moduli space needs to be compact. This was shown to be the case for the 3d abelian Seiberg-Witten equations coupled to $n$ spinors transforming under an $SU(n)$-bundle in \cite{MR3432158}. {The moduli space of the non-abelian generalisation in 4d can be non-compact, but has a natural compactification analogous to the Uhlenbeck compactification of anti-self-dual instantons. This has been studied for the case of $PU(N)$-monopoles in \cite{PidstrigachTyurin,MR1405956}, however to our knowledge the compactification of the moduli space of 3d non-abelian Seiberg-Witten equations has not been studied.}}
For the case of the supersymmetric BF-model, where the spinorial hypermultiplet is absent, it was shown in \cite{Witten:1989sx} that for gauge group $SU(2)$ and $M_3 = S^3$ the partition function computes the Casson invariant.
For general three-manifolds the partition function computes the Euler characteristic of the moduli space of flat connections \cite{blau1993n}, which was conjectured to be the generalization of the Casson invariant to arbitrary three-manifolds. For the BFH-model \eqref{BFH} we expect the associated moduli space to be the moduli space of solutions to the generalized Seiberg-Witten equations \eqref{gSW}, which minimizes the action \eqref{T3ActionBosonic} for $D_a\varsigma_{\dot{a}}=[\varsigma_{\dot{a}},\phi^{\alpha\hat{\alpha}}]=0$. The partition function of this topological theory is then, under the $\mathcal{N}=1$ 3d--3d correspondence, conjectured to match the $T^3$-partition function of the $T_{\mathcal{N}=1}[M_3]$ theory. 

Indeed, for the simple case of Lens spaces $L(p,q)$ considered in section \ref{sec:LensSpaceTheory} there are no twisted harmonic spinors for the standard metric and the gSW equations reduce to flat $U(N)$-connections. Thus, the partition function of the BFH model reduces to the Euler characteristic of the moduli space of flat connections. 
Since the moduli space of flat connections for Lens spaces consists of a set of  points, the counting of flat connections computed in section \ref{sec:LensSpaceTheory} to determine the vacua of the theory $T_{\mathcal{N}=1}[L(p,q)]$ corresponds to the Euler characteristic of the moduli space. We thus find the partition function of the BFH-model for Lens spaces is given by 
\be 
Z_{\text{BFH},U(N)}(L(p,q)) = \chi(\mathcal{M}_{\text{flat}, U(N)}) = \begin{pmatrix}
 p \\ N
\end{pmatrix}\,,
\ee
and matches with the Witten index of $T_{\mathcal{N}=1}[L(p,q),U(N)]$.

\subsection{The Abelian BFH-Model}
\label{sec:BFHAbelian}

Let us look at the abelian case more closely, where we are able to perform  concrete computations and comparisons. The action of the abelian BFH-model is minimized by solutions to the now decoupled equations
\be \label{Flat+Harm}
F_{ab}=0\,,  \quad ({\slashed{\mathcal{D}}}\phi)^{\alpha\hat{\alpha}}=0\,, \quad \p_a\varphi=0\,.
\ee
This implies that the moduli space splits into a product where the individual pieces relate to one of the equations in \eqref{Flat+Harm}. The solutions to the last equation are constant scalars on $M_3$, which we identify with the center of mass mode introduced in section \ref{sec:AbelianReduxM3}. The two remaining contributions correspond to flat $U(1)$-connections and twisted harmonic spinors on $M_3$ respectively.
It is discussed in appendix \ref{app:AbelianDR} that the moduli space of flat $U(1)$-connections has two independent contributions coming from the free and torsion part of the first homology group \eqref{IntHomology} respectively. For the free part the moduli space is given by a $b_1(M_3)$-dimensional torus. The torsion part $\mathcal{M}_{\mathcal{T}}$ of the moduli space is a finite set of $\prod_{m=1}^r p_m$ points, where the $p_m$ are the torsion numbers. Finally, the elements of the moduli space $\DeformSpace$ of twisted harmonic spinors on $M_3$ are locally described by the sections of the normal bundle $N_{M_3}=  \mathbb{S}\otimes V$, which is the space of deformations of the associative.

The full moduli space of solutions to \eqref{Flat+Harm}, neglecting the center of mass contribution, is thus given by
\be \label{MBFHU1}
\mathcal{M}_{U(1)}= T^{b_1(M_3)}\times \mathcal{M}_{\mathcal{T}}\times\DeformSpace\,.
\ee
Under the 3d--3d correspondence the $T^3$-partition function of $T_{\mathcal{N}=1}[M_3,U(1)]$ is equivalent to the partition function of the abelian BFH-model which is computed by the Euler characteristic of the moduli space \eqref{MBFHU1},
\be
Z_{\text{BFH}}(M_3)=\chi\left(\mathcal{M}_{U(1)}\right) = \chi\left(T^{b_1(M_3)}\right) \chi \left(\mathcal{M}_{\mathcal{T}}\right) \chi\left(\DeformSpace\right)\,.
\ee
For generic metrics on $M_3$ no solutions for twisted harmonic spinors exist besides the trivial solution $\phi^{\alpha \hat \alpha}=0$. In that case the partition function computes the Euler characteristic of the moduli space of flat $U(1)$-connections on $M_3$, which is given by
\be \label{AbelianBFHIndex}
Z_{\text{BFH}}(M_3) = \left\{
\begin{array}{c l}
\prod_{m=1}^r p_m & b_1(M_3)=0\\
0 & \text{else}
\end{array}\right.\,.
\ee
For the case of $d_{\slashed{\mathcal{D}}}(M_3,g) = 0$ this result is in agreement with the Witten index computed in section \ref{sec:AbelianWittenIndex}. In order to recover the full Witten index, i.e. including $d_{\slashed{\mathcal{D}}}(M_3,g) \neq 0$, one needs to prove that the index $Z_{\text{BFH}}(M_3)$ vanishes unless $b_1(M_3) = d_{\slashed{\mathcal{D}}}(M_3,g) = 0$.
 However, the moduli space of twisted harmonic spinors is not necessarily smooth and compact for $d_{\slashed{\mathcal{D}}}(M_3,g)\geq0$, and these cases require a more careful treatment of how the partition function is defined in terms of integrals over the moduli space\footnote{A similar situation arises in Donaldson theory where the moduli space is described by the pair $(A, \phi)$ satisfying
\be \nonumber
F^+ = 0\,, \quad D_{\mu} \phi = 0\,.
\ee
The moduli space becomes non-compact for solutions with non-zero scalar field $\phi$ \cite{Witten:1990bs}, which correspond to reducible connections. From the results of Uhlenbeck, there exists a natural compactification of the anti-self-dual instanton moduli space from which one computes the Donaldson invariants.}.
 For cases when the moduli space $\mathcal{H}_{\slashed{\mathcal{D}}}$ is compact we conjecture via the $\mathcal{N} = 1$ 3d--3d correspondence that
\be \label{HEulerConjecture}
\chi(\mathcal{H}_{\slashed{\mathcal{D}}}) = 0\qquad \text{for }\ d_{\slashed{\mathcal{D}}}(M_3,g)\neq0\,.
\ee 
This has the consequence that the partition function of the abelian theory is not completely invariant under metric deformations, but can jump, whenever the metric deforms to allow for harmonic spinors. Although this is not explicitly verified in the non-abelian case we expect this type of wall-crossing to persist in the more general case as well. 

{In the case of $T^7$ expressed as the product $T^3 \times T^4$, it is noted in \cite{joyce2003riemannian} that there exist a family of flat $G_2$-structures for which $T^3$ is an associative three-cycle with a deformation space of dimension four. In fact this deformation space is exactly the transverse $T^4$, and one finds a confirmation of the conjecture \eqref{HEulerConjecture}.} {In this case, the twisted harmonic spinors simply correspond to the covariantly constant spinors on $T^3$.}

We note that in the well-studied case of 3d  $\mathcal{N} = 2$ supersymmetry the reduction of the 6d abelian theory differs from section \ref{sec:AbelianReduxM3} only in the scalar and spinor sectors. From the $\mathcal{N} = 2$ topological twist and reduction on $M_3$ we obtain instead $b_1(M_3)$ chiral multiplets in addition to the centre of mass multiplet. The Witten index in this case is given by  
\be \label{3dN=2WittenIndex}
I_{\mathcal{N} =2}=\left\{
\begin{array}{c l}
\prod_{m=1}^r p_m & b_1(M_3) = 0\\
0 & \text{else}
\end{array}\right.\,.
\ee
On the topological field theory side the BFH-model is replaced by a complex super-BF-model that localizes on complex flat $G$-connections. In the abelian case these coincide with real $U(1)$-connections and the moduli space is $T^{b_1(M_3)} \times M_{\mathcal{T}}$. Thus, its Euler characteristic coincides with the index \eqref{3dN=2WittenIndex}.


\section{$S^3$-Partition Function and Chern-Simons-Dirac Theory}
\label{sec:S3}

In this last section we will discuss a much less explored observable for 3d $\mathcal{N}=1$ theories, the $S^3$-partition function. Much progress in computing this observable has been made for $\mathcal{N}=2$ thanks to localization results, however these seem to be not applicable in the minimal supersymmetric situation. 
Nevertheless there is a well-defined question as to what a 3d--3d correspondence for this case would look like, i.e. what constitutes the topological field theory whose partition function on $M_3$ computes the $S^3$-partition function of  $T_{\mathcal{N}=1}[M_3]$. 
For 3d $\mathcal{N}=2$ this topological theory is the complex Chern-Simons theory  (\ref{3d3dN2}). 
In this section we will argue that for $\mathcal{N}=1$ the topological theory is real Chern-Simons gauge theory at level $k=1$ coupled to a twisted harmonic spinor. The equations of motion of this theory are precisely the generalized Seiberg-Witten equations in (\ref{gSW}). We thus propose that the $S^3$-partition function of $T_{\mathcal{N}=1} [M_3, G]$ is computed by 
\be \label{ZCSD=ZS3}
\mathcal{Z}_{{\rm CS}_{\mathbb{R}}-{\rm Dirac}, G} (M_3) = Z_{S^3} (T_{\mathcal{N}=1} [M_3, G]) \,.
\ee
The derivation of this proceeds by considering M5-branes on $S^3$ in a conformal supergravity background that preserves $\mathcal{N}=1$ supersymmetry. 
The circle-reduction along the fiber of the Hopf fibration $S^1\hookrightarrow S^3 \to S^2$ gives 5d SYM on 
$S^2$ with radius
\be
R_{S^2}=\frac{r}{2}\,,
\ee
where $R_{S^2}$ is the radius of the two-sphere and $r$ is the radius of the Hopf circle. 
For the $\mathcal{N}=2$ preserving background this reduction from {5d to 3d after non-abelianization} was carried out in \cite{Cordova:2013cea}.
Due to the non-trivial geometry of the $S^2$ it is necessary to couple to conformal supergravity \cite{Bergshoeff:1999db,Cordova:2013bea}. We determine the values of the background fields by solving the Killing spinor equations. The coupling to supergravity then leads to additional mass terms and interactions.  

After determining the supergravity background we perform the topological twist on the equations of motion coupled to supergravity, which preserves two scalar supercharges on $M_3$, and dimensionally reduce the theory on $S^2$, to determine the 3d topological theory. Most of the technical details can be found in appendix \ref{app:M3xS2}. In the following, we will summarize the salient features and the results. 

\subsection{Supergravity Background for 5d SYM}
\label{sec:S3BackgroundAnsatze}

The 6d $(2,0)$ theory on $M_3\times S^3$ can be described in terms of a supergravity background of 5d SYM on $M_3 \times S^2$, that is obtained after dimensional reduction along the Hopf-fiber. Here $M_3$ will be as before, an associative three-cycle in a $G_2$-manifold. 
The metric of the background for 5d SYM is 
\be  \label{5dMetric}
ds^2_{5d} = \frac{r^2}{4}(d \theta^2 + \sin^2 \theta d \phi^2) + dx_a dx^a \,,
\ee
where $(\theta, \phi)$ are the spherical polar coordinates. As the Hopf fibration is non-trivial, there is a non-vanishing graviphoton  $C = \cos^2 \frac{\theta}{2} d \phi$, in 5d with field strength
\be \label{FieldStrengthGraviphoton}
G_{xy}=-\half \left(\frac{2}{r}\right)^2\varepsilon_{xy}\,,
\ee
where $x,y$ are flat indices on $S^2$, and $\varepsilon_{xy}$ is the rank 2 antisymmetric tensor such that $G$ is proportional to the volume form of the unit two-sphere. In the following we determine the supergravity background fields, that ensure {$\mathcal{N}=1$ supersymmetry for the theory along $S^3$ (or equivalently, two scalar supercharges along $M_3$)}. 

We begin with the 5d $\mathcal{N} =2$ SYM coupled to background supergravity, which was derived in \cite{Kugo:2000hn, Cordova:2013bea} from a dimensional reduction of the  6d $\mathcal{N} = (2,0)$ tensor multiplet coupled to conformal supergravity \cite{Bergshoeff:1999db}. The bosonic 5d supergravity background fields are summarized in table \ref{tab:5dSugraFields}, where the index conventions are those in appendix \ref{app:Conventions}.
\begin{table}
\centering
{\renewcommand{\arraystretch}{1.2}
\begin{tabular}{|c|c|c|c|}
\hline
Label & Background Field & $Sp(4)_R$ & Properties \\\hline
$e^{a'}_{\mu'}$ & Frame & $\bold{1}$ & \\
$C_{\mu '}$ & Graviphoton & $\bold{1}$ & $G = dC$ \\
$V_{a'}^{\underline{\hat m \hat n}}$ & R-symmetry gauge field & $\bold{10}$ & \\
 $S^{\underline{\hat m \hat n}}$ & Auxiliary scalar & $\bold{10}$ & \\
 $T_{a'b'}^{\underline{\hat m \hat n}}$ & Auxiliary 2-form & $\bold{5}$ & $T_{a'b'}^{\underline{\hat m \hat n}} = -T_{b'a'}^{\underline{\hat m \hat n}}\,,  T_{a'b'}^{\underline{\hat m \hat n}} = -T_{a'b'}^{\underline{\hat n \hat m}}$ \\
\multirow{2}{*}{$ D^{\underline{\hat m \hat n}}{}_{\underline{ \hat r \hat s}}$} &\multirow{2}{*}{ Auxiliary scalar} & \multirow{2}{*}{$\bold{14}$} & $D^{\underline{\hat m \hat n}}{}_{\underline{ \hat r \hat s}} = -D^{\underline{\hat n \hat m}}{}_{\underline{ \hat r \hat s}}\,, D^{\underline{\hat m \hat n}}{}_{\underline{ \hat r \hat s}} = -D^{\underline{\hat m \hat n}}{}_{\underline{ \hat s \hat r}} $\\
& & & $ \Omega_{\underline{\hat m \hat n}} D^{\underline{\hat m \hat n}}{}_{\underline{ \hat r \hat s}} = \Omega^{\underline{\hat r \hat s}} D^{\underline{\hat m \hat n}}{}_{\underline{ \hat r \hat s}} =  D^{\underline{\hat m \hat n}}{}_{\underline{ \hat m \hat n}} = 0$ \\
$b_{a'}$ & Dilaton gauge field & $\bold{1}$ & \\ \hline
\end{tabular}}
\caption{The bosonic background fields for 5d $\mathcal{N}=2$ supergravity. \label{tab:5dSugraFields}}
\end{table}

To determine which fields are compatible with the topological twist along the associative three-cycle we have to work out whether there are any singlets under the full twisted symmetry group 
\be 
G_{\text{twist}} =  SO(2)_L \times SU(2)_{\text{twist}} \times SU(2)_{\ell}\,,
\ee
and we will determine the group theoretic decomposition of the background fields under
\be
SO(5)_L \times Sp(4)_R \longrightarrow SO(2)_L \times SO(3)_M \times SU(2)_\ell \times SU(2)_r \xrightarrow{\text{twist}} G_{\text{twist}}\,.
\ee
For  $b_{a'}=0$ and $S^{\underline{\hat m \hat n}}=0$ we find no singlets under $G_{\text{twist}}$.
The  R-symmetry gauge field $V_{a'}^{\underline{\hat m \hat n}}$ decomposes as 
\be
V_{a'}^{\underline{\hat m \hat n}}:\qquad ({\bf 5}, {\bf 10}) \longrightarrow \left( \underline{({\bf 1},{\bf 3})} \oplus ({\bf 2},{\bf 1}) ,\underline{({\bf 1}, {\bf 3})} \oplus ({\bf 3}, {\bf 1}) \oplus ({\bf 2}, {\bf 2})\right)
\xrightarrow{\text{twist}} \underline{({\bf 1}, {\bf 1}, {\bf 1})}  \oplus \cdots\,.
\ee
For the singlet we make an ansatz in terms of the generators of the $SU(2)_r$,  $(\Sigma_{\hat{a}})^{\underline{\hat{m}}}\,_{\underline{\hat{n}}}$, defined in appendix \ref{app:6dGammas}
\be \label{AnsatzV}
{V_a}^{\underline{\hat{m}}}\,_{\underline{\hat{n}}} = -\frac{2iv}{r} \delta_{a\hat{a}} \left(\Sigma_{\hat{a}}\right)^{\underline{\hat{m}}}\,_{\underline{\hat{n}}}\,,
\ee
where $a,\hat{a}= 1,2,3$ are the vector indices of $SO(3)_M$ and $SU(2)_r$ respectively, and the normalization is chosen for convenience. The field $D^{\underline{\hat m \hat n}}{}_{\underline{\hat r \hat s}}$ decomposes as 
\be
D^{\underline{\hat m \hat n}}{}_{\underline{\hat r \hat s}}:\qquad ({\bf 1}, {\bf 14})\longrightarrow  ({\bf 1}, {\bf 1}, {\bf 1}) \oplus  ({\bf 1}, {\bf 2}, {\bf 2}) \oplus  ({\bf 1}, {\bf 3}, {\bf 3})\,,
\ee
and a singlet exists independent of the twist.
Instead of making an ansatz in terms of the $Sp(4)_R$ description it is much more convenient to go to the $SO(5)_R$ description $D_{\underline{\hat{a}\hat{b}}}$, where $\underline{\hat{a}}=1,\dots,5$, which are related by
\be
D^{\underline{\hat{m}\hat{n}}}\,_{\underline{\hat{r}\hat{s}}}=D_{\underline{\hat{a}\hat{b}}}(\gamma_{\underline{\hat{a}}})^{\underline{\hat{m}\hat{n}}}(\gamma_{\underline{\hat{b}}})_{\underline{\hat{r}\hat{s}}}\,.
\ee
The singlet can then be written as
\be
D_{\hat{5}\hat{5}}=-4d, \qquad D_{\hat{A}\hat{B}}=d\, \delta_{\hat{A}\hat{B}}\,, \quad \hat A, \hat B = 1, \cdots, 4 \,,
\ee
where the relative prefactor is chosen such that $D$ is traceless.
The field $T_{a'b'}^{\underline{\hat m \hat n}}$ reduces as
\be
T_{a'b'}^{\underline{\hat m \hat n}}:\qquad ({\bf 10}, {\bf 5}) \longrightarrow \left( \underline{({\bf 1},{\bf 1})} \oplus ({\bf 1},{\bf 3}) \oplus ({\bf 2},{\bf 3}),\underline{({\bf 1}, {\bf 1})} \oplus ({\bf 2}, {\bf 2}) \right) 
\xrightarrow{\rm twist} \underline{({\bf 1}, {\bf 1},{\bf 1})}  \oplus \cdots\,,
\ee
such that the ansatz for the singlet can be written as
\be
T^{\underline{\hat{m}\hat{n}}}_{xy}=\frac{t}{r}\varepsilon_{xy}\gamma_{\hat{5}}^{\underline{\hat{m}\hat{n}}}\,,
\ee
where we have included a factor of $r$ for later convenience.

The Killing spinor equations are solved  in appendix \ref{app:KSE}, and we  find a one-parameter family of solutions, parametrized by $v$, of the form
\be 
t = \frac{1-v}{2}\,, \quad d = \frac{3(5v^2 -2)}{8r^2}\,.
\ee
The supersymmetry parameter is solved to be constant along $M_3$ and satisfies
\be \label{S2KSE}
\left(\sigma_3 \slashed{\mathcal{D}}_{S^2}\right)^\mu\,_\nu \epsilon^\nu=-\epsilon^\mu\,,
\ee 
along $S^2$. The resulting supergravity background preserves two scalar supercharges on $M_3$.

\subsection{5d SYM on $M_3 \times S^2$}
\label{sec:5donM3xS2}

The action for 5d SYM on $\mathbb{R}^{3}\times S^2$ in the supergravity background of section \ref{sec:S3BackgroundAnsatze} is derived in appendix \ref{app:FieldsM3S2} using the decomposition of the 5d fields as in \eqref{5dFieldDecomposition}.
We find a one-parameter family of theories, where the masses of the fields, with the exception of the 3d gauge field and its superpartner, are dependent on the background parameter $v$. 
We note that the final action  with $v = 0$ matches the action derived in \cite{Gustavsson:2011af} for M5-branes on $\mathbb{R}^{1,2} \times S^3$ from an alternative, deconstruction point of view starting with the BLG-theory. Whereas in \cite{Gustavsson:2011af} the action was argued to be unique, here we find, by  coupling to off-shell supergravity, that there is a one parameter family of solutions for this background. The key difference lies in the presence of the spinorial kinetic term for the field $\phi^{\alpha \hat \alpha}$ of the form
\be 
S_{v} = v \phi_{\alpha \hat \alpha} \slashed{D}^{\alpha}{}_{\beta} \phi^{\beta \hat \alpha}\,.
\ee
This term is absent in \cite{Gustavsson:2011af}, but can be included while preserving the same amount of supersymmetry by adding mass terms for the bosonic fields $\varphi$ and $\phi^{\alpha \hat \alpha}$ and their superpartners.

The 3d flat space action and supersymmetry variations can be generalized to curved $M_3$ by covariantizing the derivatives and determining the additional terms required in the action for supersymmetry to be preserved,  by first noting that the non-vanishing curvature on $M_3$ modifies the commutation relations satisfied by the covariant derivatives, which enters into the supersymmetry variation of the action. In addition we gauge the flavour symmetry $SU(2)_{\ell}$, which is identified with $SU(2)_V$ associated with the bundle $V$ in the normal bundle of $M_3$. 

For the spinor $\phi^{\alpha \hat \alpha}$ on $M_3$ we obtain 
\be  \label{DerivativeCurvature}
([\mathcal{D}_a, \mathcal{D}_{ b}]\phi)^{\alpha\hat{\alpha}}=F_{ab}\phi^{\alpha\hat{\alpha}}+\frac{1}{4}\mathcal{R}_{ab}\,^{cd}(\sigma_{cd})^\alpha\,_\beta\phi^{\beta\hat{\alpha}} + \mathfrak{F}_{ab}{}^{\hat \alpha}{}_{\hat \beta} \phi^{\alpha \hat \beta}\,,
\ee
where $\mathcal{D}$ is the covariant derivative on the three-manifold with respect to the metric, with Riemann tensor $\mathcal{R}_{ab}\,^{cd}$, the gauge connection $A$ and the $SU(2)_V$ connection $\ConnV$, with field strength $\mathfrak{F}$. To cancel the contribution of these terms in the supersymmetry variation of the action it is necessary to introduce the terms  
\be \label{RphiMass}
S_\Phi^{\text{curv}}=\frac{1}{16\pi^2 r}\int d^5x \Tr\left(\frac{\mathcal{R}}{4}\phi_{\alpha\hat{\alpha}}\phi^{\alpha\hat{\alpha}} - \frac{i}{2} \varepsilon^{abc} {\mathfrak{F}}_{ab}{}^{\hat \alpha}{}_{\hat \beta}(\sigma_{c})^{\alpha}{}_{\beta}\phi_{\alpha \hat \alpha}\phi^{\beta \hat \beta}\right)\,.
\ee
No additional corrections are required for the other fields. {In appendix \ref{app:CurvedM3Sugra} the  additional curvature terms are determined by turning on an R-symmetry gauge field, to cancel the spin-connection on $M_3$, in the background supergravity. We find that the terms required for preserving supersymmetry on a curved three-manifold agree, however in order to solve the supergravity Killing spinor equations $M_3$ is required to be Einstein. } 

In the action the additional terms \eqref{RphiMass} can be combined with the covariantized kinetic term for $\phi^{\alpha \hat \alpha}$ using the Lichnerowicz-Weitzenb\"ock formula for the twisted Dirac operator, derived in \cite{2017arXiv170106061M} 
\be \label{DM3^2}
\phi_{\alpha\hat{\alpha}}\left(\mathcal{D}_a\mathcal{D}^a-\frac{\mathcal{R}}{4} \right)\phi^{\alpha\hat{\alpha}}+\frac{i}{2} \varepsilon^{abc} {\mathfrak{F}}_{ab}{}^{\hat \alpha}{}_{\hat \beta}(\sigma_{c})^{\alpha}{}_{\beta}\phi_{\alpha \hat \alpha}\phi^{\beta \hat \beta}=\phi_{\alpha\hat{\alpha}}({\slashed{\mathcal{D}}^2} \phi)^{\alpha\hat{\alpha}}+ \frac{i}{2} \varepsilon_{abc} F^{ab} (\sigma^c)^{\alpha}{}_{\beta}[\phi_{\alpha \hat{\alpha}}, \phi^{\beta \hat{\alpha}}]\,.
\ee
Making use of this identity we generalize the action derived in section \ref{app:FieldsM3S2} to curved $M_3$
\be \label{5dActionM3S2}
\ba 
S&=S_A+S_\Phi+S_\rho+S_{\text{int}}\\
S_A &=\frac{r}{32\pi^2} \int \sqrt{|g|} d^3x \sin \theta d \theta d \phi \Tr\left( \half F_{ab}F^{ab}  + \frac{4}{r^2} F_{xa} F^{xa} + \frac{8}{r^4}F_{xy}F^{xy}+\frac{2i}{r} \text{ CS}(A)  \right) \\
 S_{\Phi} &=   \frac{r}{64 \pi^2 } \int \sqrt{|g|}  d^3x  \sin \theta d \theta d \phi \Tr\left( 2 \mathcal{D}_a \varphi \mathcal{D}^a \varphi  -  \phi_{\alpha \hat \alpha} (\slashed{\mathcal{D}}^2 \phi)^{\alpha \hat \alpha} -\frac{i}{2}\varepsilon_{abc} F^{ab} [\phi_{\alpha\hat{\alpha}},\phi^{\beta\hat{\alpha}}](\sigma^c)^\alpha\,_\beta  \right.  \\
& \left. \qquad \qquad  - \frac{v}{r} \phi_{\alpha \hat \alpha} (\slashed{\mathcal{D}} \phi)^{\alpha \hat \alpha} +\frac{8}{r^2}\mathcal{D}_{x} \varphi \mathcal{D}^x \varphi + \frac{4}{r^2}\mathcal{D}_{x} \phi_{\alpha \hat \alpha} \mathcal{D}^{x} \phi^{\alpha \hat \alpha} - \frac{16(1-v)}{r^3}\varepsilon^{xy} F_{xy} \varphi  \right.  \\
& \left. \qquad \qquad  + \frac{8(1-v)}{r^2}\varphi^2 + \frac{(2+v)(2-v)}{4r^2}\phi_{\alpha \hat \alpha} \phi^{\alpha \hat \alpha}   \right) \\
S_{\rho} &= \frac{r}{128\pi^2}\int\sqrt{|g|}   d^3x  \sin \theta d \theta d \phi \Tr\left(-4\xi_{a \mu}(\sigma_3)^{\mu}\,_{\nu} \mathcal{D}^{a}\lambda^{\nu} - 2 \varepsilon_{abc} \xi^a_{\mu} (\sigma_3)^{\mu}\,_{\nu} \mathcal{D}^b \xi^{c\nu} \right. \\
& \left. \qquad \qquad + i \rho_{\mu \alpha \hat \alpha} (\sigma_3)^{\mu}\,_{\nu} (\slashed{\mathcal{D}}_{M_3}\rho)^{\nu \alpha \hat \alpha} +\frac{4i}{r} \lambda_{\mu}\slashed{\mathcal{D}}_{S^2}{}^{\mu}\,_{\nu} \lambda^{\nu} +\frac{4i}{r} \xi_{a\mu} \slashed{\mathcal{D}}_{S^2}{}^{\mu}\,_{\nu} \xi^{a \nu} + \frac{2i}{r} \rho_{\mu\alpha \hat \alpha} \slashed{\mathcal{D}}_{S^2}{}^{\mu}\,_{\nu} \rho^{\nu \alpha \hat \alpha} \right. \\
& \left. \qquad \qquad -\frac{4iv}{r}\lambda_{\mu}(\sigma_3)^{\mu}\,_{\nu}\lambda^{\nu} - \frac{i (2-v)}{2r}\rho_{\mu \alpha \hat \alpha} (\sigma_3)^{\mu}\,_{\nu} \rho^{\nu \alpha\hat \alpha} \right) \\
S_{\text{int}} &= \frac{r}{128\pi^2}\int \sqrt{|g|}  d^3x  \sin \theta d \theta d \phi \Tr\left(-2 \lambda_{\mu}[\varphi, \lambda^{\mu}] -2 \xi_{a \mu}[ \varphi, \xi^{a \mu}] + \rho_{\mu \alpha \hat \alpha}[\varphi, \rho^{\mu \alpha \hat \alpha}] -2\lambda_\mu[\rho^\mu_{\alpha\hat{\alpha}},\phi^{\alpha\hat{\alpha}}]\right. \\
& \left. \qquad \qquad +2i\xi_{a\mu}[\rho^\mu_{\alpha\hat{\alpha}},\phi^{\beta\hat{\alpha}}](\sigma^a)^\alpha\,_\beta + 2[\varphi, \phi_{\alpha \hat \alpha}][\varphi, \phi^{\alpha \hat \alpha}]+ \half [ \phi_{\alpha \hat \alpha}, \phi_{\beta \hat \beta}][\phi^{\alpha \hat \alpha}, \phi^{\beta \hat \beta}]\right)\,,
\ea
\ee
where $\text{CS}(A)$ is the Chern-Simons functional on $M_3$. Note that this action is only invariant under the supersymmetry variations 
\be \label{5dSUSYS2}
\ba
\delta A_b&=-\frac{1}{2}\epsilon_\mu (\sigma_3)^\mu\,_\nu\xi_b^\nu\,, \qquad \qquad \delta A_x = -\frac{ir}{4} \epsilon_{\mu}(\sigma_x)^{\mu}\,_{\nu} \lambda^{\nu} \\
 \delta\varphi & =\frac{1}{2}\epsilon_\mu\lambda^\mu\,, \qquad \qquad \qquad \quad \, \delta\phi^{\alpha\hat{\alpha}}=-\frac{1}{2}\epsilon_\mu\rho^{\mu \alpha \hat \alpha}\\
\delta\lambda^\mu&=-\frac{2i}{r}\varphi (\sigma_3)^\mu\,_\nu \epsilon^\nu+\frac{2i}{r}\mathcal{D}_x\varphi (\sigma^x)^\mu\,_\nu \epsilon^\nu +\frac{2i}{r^2}F_{xy}\varepsilon^{xy}(\sigma_3)^{\mu}\,_{\nu} \epsilon^{\nu}\\
\delta\xi_a^\mu&=-\frac{1}{2}\varepsilon_{abc}F^{bc}\epsilon^\mu-\mathcal{D}_a\varphi(\sigma_3)^\mu\,_\nu\epsilon^\nu-\frac{2i}{r}F_{xa}(\sigma_3\sigma^x)^\mu\,_\nu \epsilon^\nu+\frac{i}{4}[\phi_{\alpha\hat{\alpha}},(\sigma_a)^\alpha\,_\beta \phi^{\beta\hat{\alpha}}]\epsilon^\mu\\
\delta\rho^{\mu\alpha\hat{\alpha}}&=\frac{i}{2r}(2-v)\phi^{\alpha\hat{\alpha}}(\sigma_3)^\mu\,_\nu\epsilon^\nu-i(\sigma_a)^\alpha\,_\beta \mathcal{D}^a\phi^{\beta\hat{\alpha}}(\sigma_3)^\mu\,_\nu\epsilon^\nu-\frac{2i}{r}\mathcal{D}_x\phi^{\alpha\hat{\alpha}}(\sigma^x)^\mu\,_\nu \epsilon^\nu+[\varphi,\phi^{\alpha\hat{\alpha}}]\epsilon^\mu\,.
\ea
\ee
 after imposing \eqref{S2KSE} on the supersymmetry parameter.

\subsection{Reduction to Chern-Simons-Dirac Theory}
\label{sec:RealCS}

The reduction of the action and supersymmetry variations proceeds by expanding the fields in terms of harmonics on $S^2$, which are detailed in appendix \ref{app:HarmonicsonS2}. For the gauge fields we first note that $S^2$ does not admit any non-trivial one-forms and therefore there are no zero modes for the gauge field $A_x$. However, we need to integrate out $F_{xy}$ which sets it to
\be 
F_{xy} = \frac{r}{2}(1-v) \varphi \varepsilon_{xy}\,,
\ee 
 leading to an additional mass term for $\varphi$. Taking the zero mode of $A_b$ on the sphere leads to the 3d action
\be \label{3dActionGauge}
S_A=\frac{1}{4\pi}\int d^3x \Tr\left(\frac{r}{2} F\wedge\star F+i \text{CS}(A)\right)\,,
\ee
where all higher modes become massive and decouple. In the limit $r\to 0$ the kinetic term is suppressed and we obtain real Chern-Simons theory at level $k=1$. 
Depending on the choice of $v$ this is coupled to additional fields.
The values of $v$ for which the remaining bosonic and fermion fields becomes massless are computed in appendix \ref{app:MasslessSpectrum}. 
For $v=0,1$ we show that the theory reduces to real Chern-Simons theory. However it is clear that if the 3d--3d correspondence is to hold, the topological theory has to be sensitive to the twisted harmonic spinors $\phi$ on $M_3$. Even for the abelian theory it is clear that non-trivial $d_{\slashed{\mathcal{D}}}(M_3,g)$ will result in additional scalar multiplets that contribute to the sphere partition function. This motivates us to consider the case $v=2$, which has a massless twisted harmonic spinor. It would indeed be very nice to have another first principle way to fix $v$ from the reduction in the supergravity background.

Recall that the 3d gauge field is massless for all values of $v$ and its action is given by \eqref{3dActionGauge}. The case $v=2$ is the simplest one for which there is also a massless field coming from $\phi^{\alpha\hat{\alpha}}$, namely the one that is constant along the $S^2$. The reduction of its kinetic term is straightforward
\be
S^{v=2}_\phi=\frac{1}{8\pi}\int d^3x \Tr\left(-\frac{r}{2} \phi_{\alpha\hat{\alpha}}({\slashed{\mathcal{D}}^2} \phi)^{\alpha\hat{\alpha}}-  \phi_{\alpha\hat{\alpha}} (\slashed{\mathcal{D}} \phi)^{\alpha\hat{\alpha}} \right)\,.
\ee
Crucially, the spinorial kinetic term is leading in the limit $r\to0$. The massless field content is completed by spinors $\lambda_{(2,j)}$ and scalars $\varphi_{(1,m)}$ as can be seen from the conditions \eqref{vConditionvarphi} and \eqref{vConditionFermions}. Since the kinetic term of $\lambda$ couples to $\xi$ in \eqref{5dActionM3S2} the massive modes $\xi^a_{(2,j)}$ have to be integrated out correctly. This same procedure was also used in \cite{Cordova:2013cea} and will be written out in detail in appendix \ref{app:v=0,1} for the (slightly simpler) case $v=1$. Crucially, the kinetic terms of the $\lambda_{(2,j)}$ become bosonic and scale with $r$. The same naturally happens for the kinetic terms of the scalars $\varphi_{(1,m)}$. Here, we will only consider terms at leading order in $r$, so these fields do not receive a kinetic term.\footnote{In \cite{Cordova:2013cea} the corresponding fields are interpreted as ghosts responsible for gauge fixing. We will find the same interpretation for the case $v=1$. In the case at hand this interpretation is rather unclear.}
Nevertheless, they still appear in the action to order $\mathcal{O}(r^0)$ {in the form of non-vanishing Yukawa couplings.}
However, these fields do not couple to the gauge field or the bispinor and are thus non-dynamical. This interaction term thus vanishes on-shell. At leading order in $r$ we are left with the action
\be
S^{v=2}=\frac{i}{4\pi} \int d^3x \Tr\left(\text{CS}(A)+\frac{i}{2}\phi_{\alpha\hat{\alpha}} (\slashed{\mathcal{D}} \phi)^{\alpha\hat{\alpha}}\right)\,.
\ee
This theory is non-abelian Chern-Simons theory coupled to twisted harmonic spinors i.e. non-abelian Chern-Simons-Dirac theory. The equations of motion are given by the generalized Seiberg-Witten equations
\be (\slashed{\mathcal{D}} \phi)^{\alpha \hat{\alpha}} =0 \,, \qquad 
\varepsilon_{abc}F^{bc} - {i\over 2} [\phi_{\alpha \hat \alpha}, (\sigma_a)^\alpha{}_{\beta}\phi^{\beta \hat \alpha}]=0\,,
\ee
that have already appeared at various points throughout this paper.

\subsection{Generalization to Lens Spaces}

In the preceding sections we have determined the reduction of the 6d $\mathcal{N}=(2,0)$ theory on a three-sphere. There is a straightforward generalization to Lens spaces $L(p,1)$ which were defined in section \ref{sec:LensSpaceTheory} as the free $\Z_p$ action
\be \label{ActionLp1}
(z_1, z_2) \rightarrow e^{2\pi i / p} (z_1, z_2)\,,
\ee
on $S^3=\left\{(z_1,z_2)\in\C^2\ : \ |z_1|^2+|z_2|^2=1\right\}$. Writing the $S^3$ as a Hopf fibration the base $S^2$ is parametrized by $z=\frac{z_1}{z_2}$, so the group action \eqref{ActionLp1} does not act on the base.\footnote{Note that this logic fails for general $L(p,q\neq1)$.} It does however act on the fiber and we can write the  Lens space as a Hopf fibration $S^1\hookrightarrow L(p,1) \to S^2$, where the radii of the spheres are given by
\be
R_{S^1}=\frac{r}{p}\,, \qquad R_{S^2}=\frac{r}{2}\,.
\ee
The length scale $r$ is chosen to have the same interpretation as in the previous sections. For $p=1$ the setup thus reduces to $L(1,1)=S^3$. The metric of the Lens space is given by
\be \label{LensMetric}
ds^2=\left(\frac{r}{2}\right)^2 \left(d\theta^2+\sin^2\theta\ d\phi^2\right)+\left(\frac{r}{p}\right)^2 \left(p C +d\psi\right)^2\,,
\ee
where $C$ is the graviphoton with field strength $G$ as in \eqref{FieldStrengthGraviphoton} and $\psi$ is the coordinate on the fiber. One can then repeat the reduction of the 6d $\mathcal{N}=(2,0)$ along the fiber in \cite{Cordova:2013bea} for general $p$. From the metric \eqref{LensMetric} we find that the dilaton $g_{\psi\psi}^{-1/2}=\frac{p}{r}$ gets rescaled by $p$ and we can interpret the resulting 5d theory as $\mathcal{N}=2$ SYM coupled to $p$ units of graviphoton flux.

The rescaling of the dilaton by $p$ has the effect of rescaling the entire 5d action as
\be
S_{p}=pS_{p=1}\,,
\ee
where $S_{p=1}$ is the action relevant for the reduction on the $S^3$. The 5d supersymmetry variations and Killing spinor equations in \cite{Cordova:2013bea} are unchanged since they only depend on the invariant combination $G\ g_{\psi\psi}^{1/2}$. This rescaling does not affect the subsequent reduction on the $S^2$ and we can thus conclude that the reduction on the Lens space yields the same one-parameter family of theories, now with Chern-Simons level $p$. Choosing $v=2$ we obtain
\be
S^{v=2}_p=\frac{ip}{4\pi} \int d^3x \Tr\left(\text{CS}(A)+\frac{i}{2}\phi_{\alpha\hat{\alpha}} (\slashed{\mathcal{D}} \phi)^{\alpha\hat{\alpha}}\right)\,,
\ee
i.e. Chern-Simons-Dirac theory at level $p$. This generalizes our proposed correspondence \eqref{ZCSD=ZS3} to
\be
\mathcal{Z}_{{\text{CSD}}_p, G} (M_3) = Z_{L(p,1)} (T_{\mathcal{N}=1} [M_3, G]) \,.
\ee

\subsection{Chern-Simons-Dirac Partition Function and WRT-Invariants}

As we have repeatedly stated, not much is known about the $S^3$-partition function for 3d $\mathcal{N}=1$ theories. In fact this motivated studying the topological theory, which would under a 3d--3d correspondence compute this quantity in terms of the partition function on $M_3$. 
Taking stock we should assess how concrete this proposal can be made. We have seen that the topological theory is a Chern-Simon-Dirac theory, whose equations of motion are the generalized Seiberg-Witten equations gSW (\ref{gSW}), the fields are a gauge field and twisted harmonic spinors solving $(\slashed{\mathcal{D}} \phi)^{\alpha \hat{\alpha}} =0$, which are twisted by an $SU(2)$ bundle. Put differently, these are sections of the normal bundle of $M_3$ inside the $G_2$-holonomy manifold. Let us re-emphasize that these gSW equations, or equivalently the CS-Dirac functional, have in this form not been studied. The CS-Dirac functional has played an important role related to the standard SW equations, see e.g. \cite{kronheimer2007monopoles}, however the generalization that we find here seems to not have appeared so far in this literature. 
However recall that in a generic $G_2$-manifold, the associatives will not have any deformations/obstructions (see the discussion in section \ref{sec:DeformationTheory}). It is thus also of interest to consider the case when $\phi=0$ and the topological theory reduces to real Chern-Simons. In this case much more is known about three-manifold partition functions, which we will briefly summarize now.

The partition function of real Chern-Simons theory is a topological invariant of three-manifolds, the  Witten-Reshetikhin-Turaev invariant \cite{Witten:1988hf,MR1091619}. In \cite{MR1091619} an oriented three-manifold invariant was constructed, which was proposed to be equal to Witten's invariant, the partition function of real Chern-Simons theory. Analytic expressions of these have been computed for $M_3=S^3$ and $L(p, q=\pm1)$ Lens spaces in \cite{Witten:1988hf, FreedGompf}.  For level $k$ and gauge group $G=SU(2)$ these are 
\be
\ba
\mathcal{Z}_{\rm{CS}, k} (S^3) &=  \sqrt{2\over k+2} \sin \left({\pi \over k+2}\right)\cr 
\mathcal{Z}_{\rm{CS}, k} (L(p, q=\pm1)) &=  {2\over k+2} e^{2 \pi i (q {3k \over 8(k+2)})} \sum_{\ell =1}^{k+1} \sin^2 \left({\ell \pi \over k+2} \right) e^{2\pi i \left({\ell^2-1 \over 4(k+2)}\right) q p}\,.
\ea
\ee
For general Lens spaces (and gauge groups) the partition function was determined in \cite{Jeffrey:1992tk}.
Futhermore for Seifert manifolds exact expressions were determined for any simply-laced gauge groups in \cite{LawrenceRozansky, Marino:2002fk, ClossetKimWillett}.
For $k=1$ our result suggests that these are the values of the $S^3$-partition functions of the $T_{\mathcal{N}=1}[M_3]$ theories for $M_3$ without any deformations. Equivalently, for $k=p$, this should reproduce the $L(p,1)$-partition function. Needless to say, this would be indeed very interesting to check. Perhaps a more accessible framework similar to the one applied for $\mathcal{N}=1$ AGT in \cite{Mitev:2017jqj} is to start with quotients of $\mathcal{N}=2$ theories for which the $S^3$-partition function has been computed. 


\section{Discussions and Outlook}
\label{sec:DO}

We proposed an $\mathcal{N} = 1$ 3d--3d correspondence relating two observables of 3d $\mathcal{N}=1$ theories $T_{\mathcal{N} = 1}[M_3]$ labeled by three-manifolds $M_3$, 
to partition functions of topological theories: 
the Witten index, which is related to the $M_3$-partition function of a BFH-model (a super-BF-model  coupled to a hypermultiplet), and the $S^3$-partition function, which is related to the partition function of  Chern-Simons-Dirac theory on $M_3$. Both topological theories are closely linked to a set of generalized Seiberg-Witten equations (\ref{gSW}) for a gauge field and a twisted harmonic spinor on $M_3$.
We have defined the theories $T_{\mathcal{N} = 1}[M_3]$ as arising from M5-branes wrapped on associative three-cycles in $G_2$-manifolds with a topological twist, and have shown that the geometry of the associative cycle, specifically its normal bundle, play a key role in the dimensional reductions.

In the case of the Witten index we were able to perform various checks of the proposed correspondence: the partition function of the BFH-model is computed in terms of the Euler characteristic of the moduli space of solutions to the generalized Seiberg-Witten equations (\ref{gSW}), and we provided evidence for this correspondence in specific examples. In particular this can be made explicit for abelian theories, where the moduli space decouples  into a product of the space of real flat connections with the space of twisted harmonic spinors on $M_3$.

Unfortunately not much is known about localization results for 3d theories preserving $\mathcal{N} = 1$ supersymmetry, unlike their higher-supersymmetric cousins, which limits the scope of checks in the case of the $S^3$-partition function. 
It would indeed be interesting to use the results in section \ref{sec:S3} and either gain insight into the computation of $S^3$-partition functions of 3d $\mathcal{N}=1$ through the correspondence with CS-theory, or to compare with direct computations by other means, e.g. from orbifolds of $\mathcal{N}=2$ theories. Thanks to some resurgence in interest in dualities in 3d theories  without \cite{Aharony:2012nh,Jain:2013gza,Seiberg:2016gmd,Karch:2016sxi} and with minimal \cite{Armoni:2009vv,Gomis:2017ixy,Bashmakov:2018wts,Benini:2018umh} supersymmetry, as well as new geometric constructions \cite{Braun:2018joh},  further progress on 3d $\mathcal{N}=1$ theories may be on the horizon.
Another interesting direction to pursue is the relation of the $\mathcal{N}=1$ 3d--3d correspondence to an $\mathcal{N}=1$ AGT type correspondence, much along the lines of \cite{Dimofte:2011ju}, where the 3d theories are defects in the 4d $\mathcal{N}=1$ theories.

We have seen that there is a wall-crossing phenomenon for the Witten index of $T_{\mathcal{N} = 1}[M_3]$ and it would be interesting to further investigate this, in tandem with a better understanding of the moduli space of generalized Seiberg-Witten equations for metrics on $M_3$, which admit twisted harmonic spinors. In the abelian theories $T_{\mathcal{N} = 1}[M_3, U(1)]$ these special metrics give rise to additional zero modes, which result in the vanishing of the Witten index, and one natural question is the non-abelian generalization of this.

One possible method of accessing this information is the following. It was conjectured in \cite{blau2000relationship} that the Euler characteristic of the moduli space of solutions to
the non-abelian 3d Seiberg-Witten equations, coupled to additional matter multiplets, is proportional to the Rozansky-Witten invariant \cite{Rozansky:1996bq} of $M_3$. The set of equations in \cite{blau2000relationship} take the form 
\be \label{BTGenSW}
\varepsilon_{abc} (F^{bc})^A - \frac{1}{2} \sum_{\text{I} = 1}^{N_f} (\phi_{\alpha})^{\phantom{A}}_\text{I} T^A_\text{I} (\sigma_a)^{\alpha}{}_{\beta} (\phi^{\beta})_\text{I} = 0 \,, \qquad \slashed{D}_A (\phi^{\alpha })_\text{I} = 0 \,,
\ee
where $T^A_\text{I}$ are the generators of $G$ in representation $R_\text{I}$, and $N_f$ is the number of flavors, and the generalized Seiberg-Witten equations \eqref{gSW} are obtained from \eqref{BTGenSW} for $N_f = 2$ and $R_I=${\bf Adj}. It is therefore natural to conjecture, when the bundle $V$ is trivial,  that the partition function of the BFH-model can be computed by the Rozansky-Witten invariant 
for the sigma-model on $M_3$ with target the Coulomb branch of a 3d $\mathcal{N} = 4$ theory. The relevant 3d $\mathcal{N} = 4$ theory is singled out, in that after the topological twist, its BPS equations should be given by the gSW equations.
To make use of the proposal in \cite{blau2000relationship} knowledge of the full non-perturbative corrections to the Coulomb branch of 3d $\mathcal{N} = 4$ $SU(N)$ SYM in the presence of matter is required, which are not known in general. However, it would be interesting to explore this further for the case of two additional adjoint hypermultiplets in the light of the correspondence proposed in this paper.

\subsection*{Acknowledgements}

We thank Benjamin Assel, Andreas Braun, Clay Cordova, Michele del Zotto, Heeyeon Kim, Dario Martelli, Itamar Yaakov and Masahito Yamazaki for discussions. 
SSN  and JE are supported by the ERC Consolidator Grant 682608 ``Higgs bundles: Supersymmetric Gauge Theories and Geometry (HIGGSBNDL)''. J-M.Wong is supported by World Premier International Research Center Initiative (WPI), MEXT, Japan.

\appendix

\section{Conventions}
\label{app:Conventions}

In this appendix we summarize our conventions for indices, gamma matrices and spinors in this paper. 

\subsection{Index Conventions}

For convenience we summarize the space-time and R-symmetry indices in tables \ref{tab:LorentzIndices} and \ref{tab:RSymmIndices}, respectively.
All 6d indices are underlined and all hatted indices are associated to an R-symmetry.
\begin{table}[h]
\centering
\begin{tabular}{c|c|c|c|c|c}
 & 6d  & 5d & $M_3$ & $\R^{1,2}$ & $S^2$ \\
 \hline 
 Vector & $\underline{a},\underline{b} = 0, \cdots, 5$ & $a',b'=0, \cdots, 4$ & $a,b= 1, \cdots, 3$& $x,y= 0, \cdots, 2$ & $x,y=1,2$  \\
 Spinor &  $\underline{\alpha},\underline{\beta}= 1, \cdots, 8$ & $\alpha',\beta'=1, \cdots, 4$  & $\alpha,\beta= 1, 2$ & $\sigma,\tau= 1,2$ & $\mu,\nu=1,2$
\end{tabular}
\caption{Space-time indices \label{tab:LorentzIndices}}
\end{table}
\begin{table}[h]
\centering
\begin{tabular}{c|c|c|c}
& $SO(5)_R$ &  $SU(2)_r$ & $SU(2)_\ell$ \\
\hline
Vector & $\underline{\hat{a}},\underline{\hat{b}}= 1, \cdots, 5$ &$\hat{a},\hat{b}= 1, \cdots, 3$&  $\hat{a}',\hat{b}'= 1, \cdots, 3$ \\
Spinor & $\underline{\hat{m}},\underline{\hat{n}}= 1, \cdots, 4$ &   $\hat{m},\hat{n}=1,2$ & $\hat{\alpha},\hat{\beta}=1,2$
\end{tabular}
\caption{R-symmetry indices\label{tab:RSymmIndices}}
\end{table}
Note that after imposing the twist $SU(2)_{\text{twist}}=\diag(SU(2)_M,SU(2)_r)$ we identify the indices $\alpha=\hat{m}$ as the spinor index of the $SU(2)_{\text{twist}}$.
Note that the spinor components denote those of a Dirac spinor and we need to impose the symplectic-Majorana-Weyl condition, which halves the degrees of freedom, see (\ref{SMW}).

\subsection{Gamma Matrix and Spinor Conventions}

\label{app:6dGammas}

There are two types of gamma matrices involved in the calculations, describing either the space-time or the R-symmetry.

\subsubsection*{Gamma matrices of $SO(1,5)_L$}

On $ \mathbb{R}^{1,2}\times M_3$ we use the gamma matrices
\be
\ba
\Gamma_0 = \bold{1}_2 \otimes i \sigma_1 \otimes \sigma_2\,,\qquad
\Gamma_1 = \bold{1}_2 \otimes \sigma_{2} \otimes \sigma_2\,, \qquad
\Gamma_2 = \bold{1}_2 \otimes \sigma_{3} \otimes \sigma_2\,, \\
\Gamma_3 = \sigma_1 \otimes \bold{1}_2 \otimes \sigma_1\,,\qquad
\Gamma_4 = \sigma_2 \otimes \bold{1}_2 \otimes \sigma_1\,, \qquad
\Gamma_5 = \sigma_3 \otimes \bold{1}_2 \otimes \sigma_1\,, \\
\ea
\ee
with the Pauli matrices
\be
\sigma_1=\begin{pmatrix} 0 & 1 \\ 1 & 0 \end{pmatrix}\,,\quad
\sigma_2=\begin{pmatrix} 0 & -i \\ i & 0 \end{pmatrix}\,,\quad
\sigma_3=\begin{pmatrix} 1 & 0 \\ 0 & -1 \end{pmatrix} \,.
\ee
The 6d gamma matrices obey $\{\Gamma_{\underline{a}},\Gamma_{\underline{b}}\}=2\eta_{\underline{ab}}$, where $\underline{a} = 0, \cdots, 5$ and $\eta=\diag(-+++++)$.
The 6d chirality matrix $\Gamma = -\prod_{\underline{a}} \Gamma_{\underline{a}}$, the charge conjugation matrix $C$ and the reality matrix $B$ are given by
\be
\ba
\Gamma &=\diag(\bold{1}_4,-\bold{1}_4) \\
C_{\underline{\alpha\beta}}&= -\sigma_2 \otimes  \sigma_2 \otimes \sigma_1=C^{\underline{\alpha\beta}}\\
B_{\underline{\alpha\beta}}&= i\sigma_2\otimes \sigma_3 \otimes \sigma_3\,,
\ea
\ee
and satisfy the relations
\be \ba 
C\Gamma_{\underline{a}}C^{-1}&=-\Gamma_{\underline{a}}^T \\
B \Gamma_{\underline{a}} B^{-1}&=\Gamma_{\underline{a}}^\ast\,.
\ea \ee
The antisymmetrized product of gamma matrices is defined as 
\be
\Gamma_{\underline{a}_1\cdots\underline{a}_n}=\frac{1}{n!}\prod_{\sigma \in S_n} (-1)^{\text{sgn}(\sigma)}\Gamma_{\underline{a}_{\sigma(1)}}\cdots \Gamma_{\underline{a}_{\sigma(n)}} \,,
\ee
where $S_n$ is the symmetric group of order $n$.
The natural index structure of the gamma matrices is $\Gamma^{\underline{\alpha}}\,_{\underline{\beta}}$ acting on the spinors $\Psi^{\underline{\beta}}$. The spinor indices are raised and lowered by the charge conjugation matrix
\be
\Psi_{\underline{\alpha}}=\Psi^{\underline{\beta}} C_{\underline{\beta\alpha}} \,,\qquad \Psi^{\underline{\alpha}}=C^{\underline{\alpha\beta}}\Psi_{\underline{\beta}}\,.
\ee
The spinors of $SO(1,5)_L$ obey a symplectic-Majorana-Weyl condition
\be\label{SMW}
\Psi_{\underline{\alpha\hat{m}}}^\ast\equiv\left(\Psi^{\underline{\alpha\hat{m}}}\right)^\ast=B_{\underline{\alpha\beta}}\Omega_{\underline{\hat{m}\hat{n}}}\Psi^{\underline{\beta\hat{n}}}\,,\ee
which can be written in terms of the Dirac conjugate spinor
$\bar{\Psi}_{\underline{\alpha\hat{m}}}=\Psi_{\underline{\beta\hat{m}}}^\ast {\Gamma_0}^{\underline{\beta}}\,_{\underline{\alpha}}$
as
\be
\bar{\Psi}^{\underline{\hat{m}}}=\Psi^{\underline{\hat{m}}}\,.
\ee
This reduces the number of independent components by half. We will point out the appropriate reality condition on the spinors explicitly when carrying out the decomposition.
Lastly, we define a set of $SO(3)$ generators
\be
\ba
SO(3)_M:&\quad \Sigma_a=-\frac{i}{2}\sigma_a \otimes \bold{1}_4\,,
\ea
\ee
that act along $M_3$.

\subsubsection*{Gamma matrices of $SO(1,4)_L$ and $SO(5)_L$}

Our conventions for the 5d gamma matrices $\Gamma_{a'}$, where $a' = 0, \cdots, 4$, on $\R^{1,1}\times M_3$ are
\be \label{5dLGammas}
\Gamma_0=i\sigma_1\otimes\bold{1}_2\quad
\Gamma_1=\sigma_2\otimes\bold{1}_2\quad
\Gamma_2=\sigma_3\otimes\sigma_1\quad
\Gamma_3=\sigma_3\otimes\sigma_2\quad
\Gamma_4=\sigma_3\otimes\sigma_3\,.
\ee
In the Euclidean case on $S^2 \times M_3$ we take the same gamma matrices \eqref{5dLGammas} but with a Wick rotation on $\Gamma_0$ such that $\Gamma^{\text{Eucl}}_0=\sigma_1\otimes\bold{1}_2$. 
For both the Minkowski and Euclidean gamma matrices the charge conjugation matrix is given by
\be
C_{\alpha'\beta'}=\sigma_1 \otimes i\sigma_2=C^{\alpha'\beta'}\,,
\ee
raising and lowering the spinor indices as
\be \label{Raising5d}
\Psi_{\alpha'}=\Psi^{\beta'} C_{\beta'\alpha'} \,,\qquad \Psi^{\alpha'}=C^{\alpha'\beta'}\Psi_{\beta'}\,.
\ee

\subsubsection*{Gamma matrices of $Sp(4)_R$}

\label{app:RSymmetry}

The R-symmetry of the full 6d theory is $Sp(4)_R\cong SO(5)_R$. The gamma matrices in the $SO(5)_R$ representation are
\be
\gamma_{\hat{1}}=\sigma_2\otimes\sigma_1\,,\quad \gamma_{\hat{2}}=\sigma_2\otimes\sigma_2\,, \quad \gamma_{\hat{3}}=\sigma_2\otimes\sigma_3\,, \quad \gamma_{\hat{4}}=\sigma_1\otimes\bold{1}_2\,, \quad \gamma_{\hat{5}}=\sigma_3\otimes\bold{1}_2\,.
\ee
The index structure is $\left(\gamma_{\underline{\hat{a}}}\right)^{\underline{\hat{m}}}\,_{\underline{\hat{n}}}$ where $\underline{\hat{a}}=\hat{1},\cdots,\hat{5}$ is the vector index of $SO(5)_R$. The symplectic structure is given by
\be
\Omega_{\underline{\hat{m}\hat{n}}}=\bold{1}_2\otimes i\sigma_2=\Omega^{\underline{\hat{m}\hat{n}}}\,,\qquad \Omega \gamma_{\underline{\hat{a}}} \Omega=-\left(\gamma_{\underline{\hat{a}}}\right)^T \,, \qquad\Omega^{\underline{\hat{m}\hat{n}}}=-(\Omega^{-1})^{\underline{\hat{m}\hat{n}}}\,,
\ee
which raises and lowers the R-symmetry indices
\be
\Psi_{\underline{\hat{m}}}=\Psi^{\underline{\hat{n}}}\Omega_{\underline{\hat{n}\hat{m}}}\,,\quad\Psi^{\underline{\hat{m}}}=\Omega^{\underline{\hat{m}\hat{n}}}\Psi_{\underline{\hat{n}}}\,.
\ee
In order to break the R-symmetry $SO(5)_R\to SO(4)_R\cong SU(2)_r\times SU(2)_\ell$ we regard $\gamma_{\hat{5}}$ as the chirality matrix. The two identical copies of $SU(2)$ are generated by the (anti-)self-dual part of the $SO(4)_R$ and the generators can be written as
\be \label{SU2lrGenerators}
SU(2)_r:\quad \Sigma_{\hat{a}}=-\frac{i}{2}\begin{pmatrix} 1 & 0 \\ 0 & 0 \end{pmatrix} \otimes \sigma_{a} 
\,,\qquad
SU(2)_\ell:\quad \Sigma_{\hat{a}'}=-\frac{i}{2}\begin{pmatrix} 0 & 0 \\ 0 & 1 \end{pmatrix} \otimes \sigma_{a} \,,
\ee
with $\hat{a}=1,2,3$ and $\hat{a}'=1,2,3$ the adjoint indices of $SU(2)_r$ and $SU(2)_\ell$, respectively.

\subsection{Spinor Decomposition}
\label{app:SpinorDecomp}

In the following we will give the spinor decompositions used throughout this paper.

\subsubsection*{6d $\to$ 3d + 3d {Conventions for Section \ref{sec:T[M,U(1)]}}}

On $\R^{1,2}\times M_3$ the 6d spinor decomposes as
\be
\Psi^{\underline{\alpha}}\to
\begin{pmatrix}
\Psi_+^{\sigma\alpha}\\ i\Psi_-^{\sigma\alpha}
\end{pmatrix}\,,
\ee
where $\sigma=1,2$ and $\alpha=1,2$ are the spinor indices of $SO(1,2)_L$ and $SO(3)_M$, respectively. The subscript $\pm$ denotes the 6d chirality and the additional factor of $i$ is included for convenience. The symplectic-Majorana-Weyl condition can then be imposed by setting $\Psi^{\sigma}=\begin{pmatrix}
\Psi^1\\ i\Psi^2
\end{pmatrix}$, where the fields $\Psi^1$ and $\Psi^2$ are real. The spinor indices are raised and lowered by
\be \label{3dRL}
\ba
C_{\sigma\tau}=-C^{\sigma\tau}=\begin{pmatrix} 0 & -i \\ i & 0 \end{pmatrix} = C_{\alpha\beta}=-C^{\alpha\beta}\\
\Psi_{\sigma\alpha}=\Psi^{\tau\beta}C_{\tau\sigma}C_{\beta\alpha}\,,\qquad \Psi^{\sigma\alpha}=C^{\sigma\tau}C^{\alpha\beta}\Psi_{\tau\beta}\,.
\ea
\ee
The gamma matrices reduce in the obvious way such that on the two three-dimensional spaces they are given by
\be
\ba
\R^{1,2}:\quad (\gamma_x)^\sigma\,_\tau&=\{i\sigma_1,\sigma_2,\sigma_3\}^\sigma\,_\tau\\
M_3:\quad (\gamma_a)^\alpha\,_\beta&=(\sigma_a)^\alpha\,_\beta\,.
\ea
\ee
We will also use that the 6d Dirac operator acting on an (anti-)chiral spinor $\Psi^{\underline{\alpha}}_\pm$ decomposes as
\be
\ba
\slashed{\p}^{\underline{\alpha}}\,_{\underline{\beta}} \Psi_\pm^{\underline{\beta}}\to\left(\slashed{\p}^\alpha\,_\beta \delta^\sigma_\tau\pm i \slashed{\p}^\sigma\,_\tau \delta^\alpha_\beta\right)\Psi_\pm^{\tau\beta}\,.
\ea
\ee

%
%
%
\subsubsection*{5d $\to$ 3d + 2d  Conventions for Section \ref{sec:S3}}
Under the splitting $SO(5)_L\to SO(2)_L\times SO(3)_M$ of the 5d Lorentz symmetry the spinor representation decomposes as $\bold{4}\to(\bold{2},\bold{2})$. A spinor thus reduces as
\be
\Psi^{\alpha'}\to\Psi^{\mu\alpha}
\,.
\ee
where $\mu$ and $\alpha$ represent the spinor indices of $SO(2)_L$ and $SO(3)_M$, respectively. Note that all components are strictly real. The index $\mu$ is raised and lowered by
\be
\ba
C_{\mu\nu}=C^{\mu\nu}=\begin{pmatrix} 0 & 1 \\ 1 & 0 \end{pmatrix}\,, \qquad C_{\alpha\beta}=C^{\alpha\beta}=\begin{pmatrix} 0 & 1 \\ -1 & 0 \end{pmatrix}\\
\Psi_{\mu\alpha}=\Psi^{\nu\beta}C_{\nu\mu}C_{\beta\alpha}\,,\qquad \Psi^{\mu\alpha}=C^{\mu\nu}C^{\alpha\beta}\Psi_{\nu\beta}\,.
\ea
\ee

\subsubsection*{R-symmetry Conventions}

Under the decomposition of the R-symmetry $Sp(4)_R\to SU(2)_r\times SU(2)_\ell$ the fundamental representation reduces as $\bold{4}\to(\bold{2},\bold{1})\oplus (\bold{1},\bold{2})$, and the R-symmetry index splits as $\underline{\hat{m}}\to(\hat{m},\hat{\alpha})$, where $\hat{m}$ and $\hat{\alpha}$ represent the spinor indices of $SU(2)_r$ and $SU(2)_\ell$, respectively. We denote this decomposition by
\be
\Psi^{\underline{\hat{m}}}\to
\begin{pmatrix}
\Psi^{\hat{m}}\\ \Psi^{\hat{\alpha}}
\end{pmatrix}
\,,
\ee
where we can view $\gamma_{\hat{5}}$ as the chirality matrix.
The symplectic structure $\Omega^{\underline{\hat m \hat n}}$ does not mix the left and right $SU(2)$ indices, and the spinor indices after the decomposition are raised and lowered as
\be
\ba
\Psi_{\hat{m}}&=\Psi^{\hat{n}}\varepsilon_{\hat{n}\hat{m}}\,,\qquad \Psi^{\hat{m}}=\varepsilon^{\hat{m}\hat{n}}\Psi_{\hat{n}}\\
\Psi_{\hat{\alpha}}&=\Psi^{\hat{\beta}}\varepsilon_{\hat{\beta}\hat{\alpha}}\,,\qquad \Psi^{\hat{\alpha}}=\varepsilon^{\hat{\alpha}\hat{\beta}}\Psi_{\hat{\beta}}\,.
\ea
\ee

\section{Dimensional Reduction of the  Tensor Multiplet}
\label{app:AbelianDR}

In this appendix we dimensionally reduce the abelian tensor multiplet from 6d on $M_3$ with the topological twist discussed in section \ref{sec:TopologicalTwist}. The tensor multiplet consists of a self-dual tensor field $H=dB$, five scalars $\Phi^{\underline{\hat{m}\hat{n}}}$ and four symplectic-Majorana-Weyl spinors $\varrho^{\underline{\alpha\hat{m}}}$ of negative chirality. We will look at them in turn.
The three-form $H$ satisfies the equations of motion
\be \label{ThreeFormEoM} \ba
dH=0 \,,\qquad 
H_{\underline{abc}} =\frac{1}{3!}\varepsilon_{\underline{abc}}\,^{\underline{def}}H_{\underline{def}}\,.
\ea \ee
We can make an ansatz for $H$
\be \label{DecompositionH6d}
H=\frac{1}{3!} H_{xyz} e^x\wedge e^y \wedge e^z + \frac{1}{2!} H_{axy} e^a\wedge e^x \wedge e^y +  \frac{1}{2!} H_{abx} e^a\wedge e^b \wedge e^x+\frac{1}{3!}H_{abc} e^a\wedge e^b \wedge e^c \,,
\ee
where the $e^a$ are local coordinates of $T^*M_3$ and $x,y,z=0,1,2$ label coordinates along $\R^{1,2}$. Self-duality then reads
\be
H_{abc}=-\frac{1}{6}\varepsilon_{abc}\varepsilon^{xyz}H_{xyz}\,,\qquad H_{abx}=\frac{1}{2} \varepsilon_{ab}\,^c\varepsilon_x\,^{yz} H_{cyz}\,,
\ee
where we normalize $\varepsilon_{012345}=\varepsilon_{012}=\varepsilon_{345}=1$. Inserting this into $dH=0$ we obtain
\be\label{EOMGaugeField6d}
\ba 
-\frac{1}{6} \varepsilon_{abc} \partial_w \varepsilon^{xyz} H_{xyz} -\half \varepsilon_w\,^{xy} \partial_{[a}\varepsilon_{bc]}\,^d H_{d xy} &=0 \\
\frac{1}{2}\p_{[x}\varepsilon_{y]}\,^{wz} \varepsilon_{ab}\,^c H_{cwz} -\partial_{[a} H_{b]xy} &=0 \\
 \partial_{[x} H_{yz]a}-\partial_a H_{xyz}&=0\,.
\ea 
\ee
To obtain the massless spectrum in 3d one expands $H_{axy}$ in a basis of harmonic one-forms 
$\omega_{I,a}$ that depend on the local coordinates on $M_3$
\be
H_{axy}=\sum_{I=1}^{b_1(M_3)} F^{I}_{xy}\ \omega_{I,a}\,,
\ee
where $b_1(M_3)$ is the first Betti number of $M_3$.
The $F^{I}_{xy}$ then only depend on the coordinates of $\R^{1,2}$. After taking $H_{xyz}$ to be a constant on $M_3$ the equations of motion take the form
\be
dF^I =0\,,\qquad 
d\star F^I  =0  \,,
\ee
where $\star$ is the Hodge star on $\mathbb{R}^{1,2}$. The first equation of motion implies that locally we can write $F^I=dA^I$ and from the second equation we obtain the usual equation of motion for free gauge fields. The massless spectrum therefore consists of $b_1(M_3)$ gauge fields. Since in three dimensions a massless gauge field is dual to a massless scalar we can introduce a set of scalars $\alpha^I$ with
\be \label{3dDuality}
\p_x \alpha^I=\half \varepsilon_x\,^{yz} F^I_{yz}\,,
\ee
which are periodic due to the gauge symmetry of the $F^I$. In terms of these scalars the 3d action can be written as
\be
S=\int_{\R^{1,2}} d \alpha_I \wedge \star \, d \alpha^I \,.
\ee
Now consider the equation of motion for the three-form $H_{xyz}$ which is given by  
\be 
d \star H =0\,,
\ee
where $H_{xyz}$ is constant on the $M_3$. This equation of motion can be obtained from the 3d action 
\be 
\int_{\R^{1,2}} H \wedge \star\, H\,,
\ee
where $H = dB$, since $H$ closes trivially. In terms of the scalar $H_{xyz}=-\varepsilon_{xyz} h$ dual to the three-form this action gives rise to the equation of motion \cite{Kaloper:1993fg}
$h = 0$, and we obtain a scalar field which is set to zero by its equation of motion. We will see from the supersymmetry variations that this scalar can be identified with the auxiliary field in the center of mass scalar multiplet.

By this direct reduction it is not possible to detect gauge fields with non-zero Chern-Simons term. The action of a single such field takes the form
\be
S=g^{-2} \int_{\R^{1,2}} F\wedge\star F + \frac{k}{4\pi}\int_{\R^{1,2}} A\wedge F\,,
\ee
where $g$ is the gauge coupling and $k$ is the integer Chern-Simons level. 
In the presence of such a Chern-Simons term the gauge field acquires a mass \cite{Schonfeld:1980kb} 
\be
m=\frac{g^2 k}{4\pi}\,.
\ee
Despite the massive nature of the gauge field the theory is not completely trivial at low energies \cite{Witten:1999ds}. It was argued in \cite{Cecotti:2011iy} that the integral first homology group of the three-manifold is presented by the matrix of Chern-Simons levels in a $U(1)^R$ theory, where $R = b_1(M_3) + r$. In other words, the Chern-Simons levels induce the relation
\be \label{kGamma=0}
k_{MN}\gamma^{N}=0\,,
\ee
where $\gamma^N$ are the generators of the full integral homology $H_1(M_3, \mathbb{Z})$ as in \eqref{IntHomology}.\footnote{This also explains why we were not able to see these fields in the reduction from 6d. In order to impose the self-duality $H$ has to be $\R$-valued so the torsion part of the homology is not detected.} Thus, given $H_1(M_3, \mathbb{Z})$, one can present it in terms of a diagonal matrix of Chern-Simons levels where the torsion numbers $p_{m}$, where $m = 1, \cdots, r$, correspond to the (non-zero) integer eigenvalues of $k_{MN}$ and the free part of $H_1(M_3,\Z)$, generated by the one-cycles $\gamma^I$, span the kernel of $k_{MN}$. The resulting theory is a $U(1)^R$ theory with $b_1(M_3)$ massless gauge fields $A_I$ and $r$ gauge fields $A_m$ with level $p_m$ Chern-Simons self-interactions.

Next, let us discuss the remaining bosonic field content coming from the five scalars $\Phi^{\underline{\hat{m}\hat{n}}}$.
Under the topological twist the scalars decompose as
\be
\ba
SO(1,5)_L\times Sp(4)_R &\to SO(1,2)_L \times SU(2)_{\text{twist}}\times SU(2)_\ell\\
(\bold{1},\bold{5}) & \to (\bold{1},\bold{1},\bold{1})\oplus (\bold{1},\bold{2},\bold{2})\,,
\ea
\ee
yielding a singlet $\varphi$ and a field $\phi^{\alpha \hat \alpha}$ that, as discussed in section \ref{sec:TopologicalTwist}, is a section of the normal bundle \eqref{NM3} of $M_3$. The decomposition of the R-symmetry naturally leads to the identification
\be
\ba
\Phi^{\alpha\beta}&=-\varphi\varepsilon^{\alpha\beta}\,, \qquad \Phi^{\hat{\alpha}\hat{\beta}}=\varphi\varepsilon^{\hat{\alpha}\hat{\beta}}\\
\Phi^{\alpha\hat{\alpha}}&=\phi^{\alpha\hat{\alpha}}\,, \qquad\quad \Phi^{\hat{\alpha}\alpha}=\phi^{\hat{\alpha}\alpha}\,,
\ea
\ee
where we have identified the $SU(2)_r$ index $\hat{m}$ with the spinor index $\alpha$ on $M_3$ as dictated by the topological twist. The conventions are chosen such that $\phi^{\alpha \hat \alpha}$ satisfies
$
\phi^{\alpha\hat{\alpha}}=-\phi^{\hat{\alpha}\alpha}
$
and the reality condition
\be \label{BispinorReality}
{\phi^{1\hat{1}}}^*=\phi^{2\hat{2}}\,,\qquad {\phi^{1\hat{2}}}^*=-\phi^{2\hat{1}}\,.
\ee
Since $\varphi$ is a singlet the 6d equation of motion $D^2 \Phi^{\underline{\hat{m}\hat{n}}}=0$, where $D$ is the 6d covariant derivative, reduces to
\be
\left(\Delta_{M_3}+\p^2_{\mathbb{R}^{1,2}}\right)\varphi=0\,,
\ee
and we obtain a single massless scalar in 3d for $\varphi$ constant on $M_3$. 
As discussed in section \ref{sec:TopologicalTwist}, the fields $\phi^{\alpha \hat \alpha}$ are identified with sections of the normal bundle of $M_3$. For general three-manifolds the additional $SU(2)_V$-bundle is twisted with the spin bundle and can act non-trivially. Thus, the derivative acting on $\phi^{\alpha \hat \alpha}$  needs to be covariantized not only with respect to the spin connection on $M_3$, but also an $SU(2)_V$ connection.
The equation of motion for $\phi^{\alpha \hat \alpha}$ therefore takes the form
\be
({{\slashed{\mathcal{D}}}^2 \phi})^{\alpha \hat \alpha}+\p_{\mathbb{R}^{1,2}}^2 \phi^{\alpha\hat{\alpha}}=0\,,
\ee
where $\slashed{\mathcal{D}}$ is the twisted Dirac operator on $M_3$ explicitly given by
\be \label{TwistedDirac}
(\mathcal{D}_a \phi)^{\alpha \hat \alpha} = \partial_a \phi^{\alpha \hat \alpha} + \frac{1}{4}{\omega_a}{}^{bc} (\sigma_{bc}){}^{\alpha}{}_{\beta} \phi^{\beta \hat \alpha} + {\ConnV_a}{}^{\hat \alpha}{}_{\hat \beta} \phi^{\alpha \hat \beta}\,,
\ee
where ${\omega_a}{}^{bc}$ is the spin connection on $M_3$ and ${\ConnV_a}{}^{\hat \alpha}{}_{\hat \beta}$ is the $SU(2)_V$ connection. Since the kernels of $\slashed{\mathcal{D}}$ and $\slashed{\mathcal{D}}^2$ are equal $\phi^{\alpha\hat{\alpha}}$ is expanded in a basis of twisted harmonic spinors $\zeta_i^{\alpha \hat \alpha}$
\be
\phi^{\alpha\hat{\alpha}}=\sum_{i=1}^{d_{\slashed{\mathcal{D}}}(M_3,g)}  \phi^{ i} \zeta_i^{\alpha \hat{\alpha}}\,.
\ee
The number of massless modes is given by the dimension, $d_{\slashed{\mathcal{D}}}(M_3,g)$, of the kernel of the twisted Dirac operator $\slashed{\mathcal{D}}$ on the associative three-cycle. As reviewed in section \ref{sec:DeformationTheory} this dimension depends on the metric on the associative three-cycle $M_3$ induced by the $G_2$-holonomy metric.

By $\mathcal{N}=1$ supersymmetry the bosonic fields are supplemented by their fermionic superpartners. The negative chirality spinor $\varrho^{\underline{\alpha\hat{m}}}$ was shown in \eqref{Decomposition6dFields} to decompose after the twist as 
\be\label{SpinorDecomposition}
\ba
\varrho^{\underline{\alpha\hat{m}}} & \quad \rightarrow \quad i\lambda^\sigma \varepsilon^{\alpha\hat{\alpha}} 
+ \xi_a^\sigma ({\sigma^a})^\alpha\,_\beta \varepsilon^{\beta\hat{\alpha}} 
- i\rho^{\sigma\alpha\hat{\alpha}}\,, 
\ea
\ee
where the prefactors have been introduced for convenience. The 6d symplectic-Majorana-Weyl condition \eqref{SMW}, implies a reality condition on a 3d spinor $\psi^{\sigma}$ such that it takes the form
\be 
\psi^{\sigma} = \begin{pmatrix}
\psi^1 \\ i \psi^2
\end{pmatrix} \,,
\ee
where $\psi^1, \psi^2$ are real. Applied to the fields $\rho^{\alpha\hat{\alpha}}$ there is an additional condition
\be \label{RhoReality}
{\rho^{1\hat{1}}}^*=\rho^{2\hat{2}}\,,\qquad {\rho^{1\hat{2}}}^*=-\rho^{2\hat{1}}\,.
\ee
Note that this is the same reality condition \eqref{BispinorReality} as for the $\phi^{\alpha\hat{\alpha}}$.
The 6d equations of motion $\slashed{\mathcal{D}}^{\underline{\alpha}}\,_{\underline{\beta}}\varrho^{\underline{\beta\hat{m}}}=0$ reduce to
\be\label{SpinorEOMFlat} 
\ba
\star d \star \xi^{\sigma} + \slashed{\partial}_{\mathbb{R}^{1,2}}{}^{\sigma}{}_{\tau}\lambda^{\tau} & = 0\cr 
d\lambda^{\sigma} +  \star d \xi^{\sigma} -{\slashed{\p}_{\mathbb{R}^{1,2}}}{}^{\sigma}{}_{\tau} \xi^{\tau} &=0   \\
i(\slashed{\mathcal{D}} \rho)^{ \sigma\alpha \hat \alpha } + \slashed{\p}_{\mathbb{R}^{1,2}}{}^{\sigma}{}_{\tau}\rho^{\tau \alpha \hat \alpha} & = 0\,,
\ea
\ee
where $\star$, $d$ and $\slashed{\mathcal{D}}$ denote the Hodge star, the exterior derivative and the twisted Dirac operator along $M_3$, respectively.
As $\lambda^\sigma$ is the superpartner of the scalar $\varphi$ it is constant along $M_3$. 
The equations then decouple leading to a single massless spinor $\lambda^\sigma$ in 3d. The massless modes from $\xi^{\sigma}_a$ are obtained by expanding
\be
\xi^{\sigma}_a =\sum_{I=1}^{b_1(M_3)} \xi^{\sigma I} \omega_{I,a}\,,
\ee
where the $\omega_{I,a}$ are a basis of harmonic one-forms on $M_3$. This yields the superpartners of the $b_1(M_3)$ massless gauge fields $A^I$.
The superpartners of the massive gauge fields $A^m$ also arise from the reduction of $\xi_a^{\sigma}$. Each of the torsion cycles $\gamma^m\in H_1(M_3,\Z)_{\text{torsion}}$ corresponds to a massive fermion $\xi^{\sigma m}$ whose mass is governed by the respective torsion number $p_m$.
Lastly, the massless superpartners of the $\phi^{i}$ arise from the expansion of $\rho^{\sigma\alpha\hat{\alpha}}$ in terms of twisted harmonic spinors on $M_3$
\be
\ba
\rho^{\sigma\alpha\hat{\alpha}}&=\sum_{i=1}^{d_{\slashed{\mathcal{D}}}(M_3,g)}  \rho^{\sigma i} \zeta^{\alpha \hat{\alpha}}_{i}\,.
\ea
\ee
Summarizing this reduction the spectrum of the abelian theory $T_{\mathcal{N}=1}[M_3,U(1)]$ is given by 
\begin{enumerate}
\item $b_1(M_3)$ massless gauge fields $A^I$ (or dually, periodic scalars $\alpha^I$) and gauginos $\xi^{\sigma I}$.
\item $r$ massive gauge fields $A^m$ with Chern-Simons levels $p_m$, and massive fermionic superpartners  $\xi^{\sigma m}$ from the torsion part of the first integral homology.
\item A single real massless scalar $\varphi$ and spinor $\lambda^\sigma$.
\item An auxiliary real scalar field $h$.
\item $d_{\slashed{\mathcal{D}}}(M_3,g)$ real scalars $\phi^{ i}$ and spinors $\rho^{\sigma i}$, where $i=1, \cdots, d_{\slashed{\mathcal{D}}}(M_3,g)$ is the number of twisted harmonic spinors on $M_3$, which depends on the metric on $M_3$. 
\end{enumerate}

\section{3d $\mathcal{N}=1$ Supersymmetry}

\label{app:3dN=1SUSY}

In this appendix we review the basic multiplet structure of 3d $\mathcal{N}=1$ in signature $(-++)$, following \cite{Gates:1983nr}. The gamma matrices are given by $\gamma_x=\{i\sigma_1,\sigma_2,\sigma_3\}$. We begin with defining coordinates $\theta^\sigma$, where $\sigma = 1,2$, on the superspace. Throughout this discussion all spinors are of the form $\Psi^\sigma=\begin{pmatrix} \Psi^1 \\ i\Psi^2 \end{pmatrix}$ with real components $\Psi^{1,2}$, where the indices are raised and lowered as in \eqref{3dRL}.
Next, define derivatives $\p^\sigma$ on the superspace with the convention
\be
\p^\sigma \theta_\tau=\delta^\sigma_\tau\,.
\ee
Note that written as a spinors $\p^\sigma$ and $\theta^\sigma$ fulfill the above reality condition on the components.

To define supersymmetry transformations the usual Poincar\'e algebra is extended by supercharges $Q^\sigma$ which obey
\be
[P_x,P_y]=0\,, \qquad \{Q^\sigma,Q_\tau\}= 2 (\gamma^x)^\sigma\,_\tau P_x\,, \qquad [Q^\sigma,P_x]=0
\ee
where $P_x=-\p_x$ is the momentum operator generating translations. The supercharges are given by
\be
Q^\sigma=\p^\sigma-(\gamma^x)^\sigma\,_\tau \theta^\tau \p_x\,.
\ee
Since the partial derivative acting on the superspace coordinates is not invariant under $Q$, one defines a covariant derivative
\be
D^\sigma=\p^\sigma+(\gamma^x)^\sigma\,_\tau \theta^\tau \p_x\,,
\ee
such that $[D^\sigma,P_x]=\{D^\sigma,Q^\tau\}=0$.
The fields invariant under supersymmetry are functions of space-time coordinates and the $\theta^\sigma$. Since the latter anticommute it is possible to expand the fields in powers of $\theta^\sigma$ terminating at $\theta^2$. There are two types of superfields that we will discuss in more detail.

\subsubsection*{Scalar Multiplet}
The scalar multiplet $\mathcal{A}_\varphi$ contains a real scalar $\varphi$, a spinor $\lambda^\sigma$ and a real auxiliary scalar $h$
\be
\mathcal{A}_\varphi=\varphi+ \theta_\sigma \lambda^\sigma + \frac{1}{2} \theta_\sigma \theta^\sigma h\,.
\ee
Its transformation under the supersymmetry is generated by the supercharges $Q^\sigma$
\be
\delta\mathcal{A}_\varphi= \epsilon_\sigma Q^\sigma \mathcal{A}_\varphi\,,
\ee
where $\epsilon^\sigma$ is an infinitesimal supersymmetry parameter. Direct calculation yields
\be\label{3dN=1ScalarSUSY}
\delta\varphi =\epsilon_\sigma \lambda^\sigma\,,\qquad 
\delta\lambda^\sigma =h\epsilon^\sigma+\slashed{\p}^\sigma\,_\tau \varphi \epsilon^\tau\,,\qquad 
\delta h=\epsilon_\sigma \slashed{\p}^\sigma\,_\tau \lambda^\tau\,.
\ee
From these scalar multiplets one can build supersymmetric actions that are invariant under \eqref{3dN=1ScalarSUSY}. Since $D^\sigma$ is invariant, every Lorentz invariant function of the superfields and its derivatives can be inserted into
\be
S_{\mathcal{A}}=\int d^3x\ d^2\theta f(\mathcal{A},D\mathcal{A},\cdots)\,.
\ee
If we choose $f(\mathcal{A},D\mathcal{A},\cdots)=-\half(D^\sigma \mathcal{A}_\varphi)^2$ then we obtain the kinetic term
\be
S_{\mathcal{A}_\varphi}^{\text{kin}}=\half\int d^3x \left(\p_x \varphi \p^x \varphi+\lambda_\sigma \slashed{\p}^\sigma\,_\tau \lambda^\tau-h^2\right)\,.
\ee
Interaction terms can be added by picking more general functions.

\subsubsection*{Vector Multiplet}

Let us now couple such a scalar multiplet to a real gauge field $A_x$. In order to preserve supersymmetry we  can again make an ansatz for a vector multiplet $\mathcal{V}_A^\sigma$ consisting of the gauge field $A_x$ and a spinor $\xi^\sigma$. Then the covariant derivative changes as
\be
D^\sigma\to D^\sigma + \mathcal{V}_A^\sigma\,,
\ee
to ensure gauge invariance using the same logic as for the non-supersymmetric case. In a particular gauge, the Wess-Zumino gauge, the vector multiplet and its associated gauge invariant field strength can be written as
\be
\ba
\mathcal{V}_A^\sigma&= A_x (\gamma^x)^\sigma\,_\tau \theta^\tau+ \theta_\tau \theta^\tau \xi^\sigma\\
\mathcal{W}_A^\sigma&=\half D^\tau D^\sigma {\mathcal{V}_A}_\tau=\xi^\sigma -\half \varepsilon^{xyz} F_{xy} (\gamma_z)^\sigma\,_\tau \theta^\tau+\half \slashed{\p}^\sigma\,_\tau \xi^\tau \theta_\upsilon\theta^\upsilon\,.
\ea
\ee
The two simplest gauge invariant actions are
\be
\ba
S_{\mathcal{V}_A}^{\text{kin}}&=-\frac{1}{g^2}\int d^3x\ d^2\theta\ \mathcal{W}_\sigma \mathcal{W}^\sigma=\frac{1}{g^2}\int d^3x \left(-\half F_{xy} F^{xy}- \xi_\sigma \slashed{\p}^\sigma\,_\tau \xi^\tau \right)\\
S_{\mathcal{V}_A}^{\text{CS}}&=\frac{k}{4\pi}\int d^3x\ d^2\theta\ \mathcal{W}_\sigma \mathcal{V}^\sigma= \frac{k}{4\pi}\int d^3x\left(\half\varepsilon^{xyz}A_x F_{yz}+ \xi_\sigma \xi^\sigma\right)\,.
\ea
\ee
The first action describes the canonical kinetic term for a gauge field with gauge coupling $g$. The Chern-Simons term is unusual as it involves the vector multiplet, which is not gauge invariant on its own. However, for integer $k$ the quantum theory is gauge invariant. This construction is unique to three dimensions and gives the field strength an effective mass.

\section{M5-branes on $S^3$}
\label{app:M3xS2}

This appendix contains details of the reduction of M5-branes on $S^3$ in section \ref{sec:S3}.

\subsection{The Killing Spinor Equations on $S^2 \times \mathbb{R}^3$}
\label{app:KSE}

In this section we solve the Killing spinor equations and determine the resulting conditions on the supersymmetry parameter and the supergravity background. 

\subsubsection{Gravitini-variation}

In our background, the non-vanishing components of the supersymmetry variation of the gravitini $\psi_{a'}^{\underline{\hat m}}$, are \cite{Cordova:2013bea}
\be \label{GravitiniKSE}
\delta\psi_{a'}^{\underline{\hat{m}}}=\mathcal{D}_{a'}\epsilon^{\underline{\hat{m}}}+\frac{ir}{2}G_{a'b'}\Gamma^{b'}\epsilon^{\underline{\hat{m}}}+\frac{i}{8}\left[rG^{b'c'}\Omega^{\underline{\hat{m}\hat{n}}}-4(T^{\underline{\hat{m}\hat{n}}})^{b'c'}\right]\Gamma_{a'b'c'}\epsilon_{\underline{\hat{n}}}\,,
\ee
with covariant derivative
\be
\mathcal{D}_{a'}\epsilon^{\underline{\hat{m}}}=\p_{a'}\epsilon^{\underline{\hat{m}}}+\frac{1}{4}\omega_{a'}\,^{b'c'}\Gamma_{b'c'}\epsilon^{\underline{\hat{m}}}-\frac{1}{2}{V_{a'}}^{\underline{\hat{m}}}\,_{\underline{\hat{n}}}\epsilon^{\underline{\hat{n}}}\,.
\ee
 We impose that the supersymmetry parameter solves the equation for the topological twist on $M_3$ \eqref{SUSYCondition}, and therefore, using the spinor decompositions detailed in appendix \ref{app:SpinorDecomp}, the non-vanishing components can be written as 
\be \label{5dSusySinglet}
\epsilon^{\alpha' \underline{\hat m}} = \epsilon^\mu \varepsilon^{\alpha\hat{m}}\,,
\ee
where $\mu = 1,2$, $ \alpha = 1,2$ and $ \hat m  = 1,2$ denote the spinor representations of $SO(2)_L$, $SO(3)_M$ and $SU(2)_r$, respectively.\footnote{Here we do not decompose the spinors on $S^2$ into one component positive and negative chirality spinors labelled by $\pm$ as in section \ref{sec:2dSigmaModel}. Instead we will leave them as two component spinors for compactness. Similarly, we keep $x=1,2$ as the flat vector index on the $S^2$.}

\subsubsection*{Flat directions}
Let us first consider the components of \eqref{GravitiniKSE} along $\R^3$. Inserting the ans\"atze for the background fields and taking the supersymmetry parameter to be constant along $\mathbb{R}^3$, the Killing spinor equation reduces to
\be
0=\delta\psi_a^{\underline{\hat{m}}}=-\frac{1}{2}{V_a}^{\underline{\hat{m}}}\,_{\underline{\hat{n}}}\epsilon^{\underline{\hat{n}}}+\frac{i}{8}\left[rG^{xy}\Omega^{\underline{\hat{m}\hat{n}}}-4(T^{\underline{\hat{m}\hat{n}}})^{xy}\right]\Gamma_{axy}\epsilon_{\underline{\hat{n}}}\,,
\ee
which is solved for
\be \label{t(v)}
t=\half(1-v)\,,
\ee
and no further restrictions on $\epsilon^\mu$ arise.

\subsubsection*{$S^2$ directions}

\label{app:KSEPsiS2}

To solve the equations along the $S^2$ we have to include an explicit dependence of the supersymmetry parameter on the coordinates $\xi=(\theta,\phi)$ . The non-vanishing contributions are
\be
0=\delta\psi_\xi^{\underline{\hat{m}}}=\mathcal{D}_\xi \epsilon^{\underline{\hat{m}}}+\frac{ir}{2}G_{\xi\zeta}\Gamma^\zeta \epsilon^{\underline{\hat{m}}}\,.
\ee
Taking the two angles individually and inserting the singlet ansatz \eqref{5dSusySinglet} for the supersymmetry parameter one obtains
\be
\ba
\p_\theta\epsilon^{\mu}&=\frac{i}{2}(\sigma_2)^\mu\,_\nu\epsilon^\nu\\
\p_\phi\epsilon^\mu&=\left(\frac{i}{2}\cos\theta(\sigma_3)^\mu\,_\nu-\frac{i}{2}\sin\theta(\sigma_1)^\mu\,_\nu\right)\epsilon^\nu\,, \label{KSES2}
\ea
\ee
which is equivalent to
\be \label{S2eps}
(\sigma_3\slashed{\mathcal{D}}_{S^2})^\mu\,_\nu\epsilon^\nu=-\epsilon^\mu\,,
\ee
where
\be \label{DiracOperatorS2}
\slashed{\mathcal{D}}_{S^2}{}^{\mu}\,_{\nu}=\left(\p_\theta +\half \cot\theta\right) (\sigma_1)^{\mu}\,_{\nu}+\frac{\p_\phi}{\sin\theta}(\sigma_2)^{\mu}\,_{\nu}\,,
\ee
is the Dirac operator on the unit two-sphere. Since the operator $(\sigma_3\slashed{\mathcal{D}}_{S^2})$ has two eigenvectors with eigenvalue $-1$ we find that the background preserves two supercharges, which transform as scalars on $M_3$.

\subsubsection{Dilatino-variation}

The non-vanishing components of the second Killing spinor equation, given by the supersymmetry variation of the dilatino $\chi^{\underline{\hat{m}\hat{n}}}\,_{\underline{\hat{r}}}$, are
\be
0=\delta\chi^{\underline{\hat{m}\hat{n}}}\,_{\underline{\hat{r}}}=\left[-{R_{b'c'}}^{[\underline{\hat{m}}}\,_{\underline{\hat{r}}}\delta^{\underline{\hat{n}}]}_{\underline{\hat{s}}}\Gamma^{b'c'}+rG^{a'b'}T_{a'b'}^{\underline{\hat{m}\hat{n}}}\Omega_{\underline{\hat{r}\hat{s}}}-2T_{a'b'}^{\underline{\hat{m}\hat{n}}}T^{a'b'}_{\underline{\hat{r}\hat{s}}}-\frac{4}{15}D^{\underline{\hat{m}\hat{n}}}\,_{\underline{\hat{r}\hat{s}}}\right]\epsilon^{\underline{\hat{s}}}-[\text{traces}]\,,
\ee
where the curvature is given by
\be \label{VCurv}
R_{a'b'}^{\underline{\hat{m}\hat{n}}}=2\p_{[a'}{V_{b']}}^{\underline{\hat{m}\hat{n}}}+{V_{[a'}}^{\underline{\hat{r}}(\underline{\hat{m}}}{V_{b']}}^{\underline{\hat{n}})}\,_{\underline{\hat{r}}}\,.
\ee
The trace terms, determined in \cite{Bergshoeff:1999db}, ensure that the variation fulfills the symmetries of the dilatino
$\Omega_{\underline{\hat{m}\hat{n}}}\chi^{\underline{\hat{m}\hat{n}}}\,_{\underline{\hat{r}}}=\chi^{\underline{\hat{m}\hat{n}}}\,_{\underline{\hat{m}}}=0$. 

Using the ans\"atze for the background fields derived in section \ref{sec:S3BackgroundAnsatze} the second Killing spinor equation is solved for
\be
d=\frac{3t(t-1)}{r^2}+\frac{9v^2}{8r^2}=\frac{3\left(5v^2-2\right)}{8r^2}\,,
\ee
where we used \eqref{t(v)} for the second equality.

\subsection{5d $\mathcal{N} =2$ SYM on $\mathbb{R}^{1,2} \times S^2$}
\label{app:FieldsM3S2}

In this section we determine the 5d action coupled to off-shell supergravity fields using the ans\"atze determined in sections \ref{sec:S3BackgroundAnsatze} and \ref{app:KSE}. The field content of the SYM after the topological twist is given in \eqref{5dFieldDecomposition}. We find a one parameter family of actions where the masses of the twisted scalars $(\varphi, \phi^{\alpha \hat \alpha})$, and their superpartners, depend on the free parameter $v$.

\subsubsection*{Gauge field}

The 5d action of the gauge field is given by
\be \label{5dActionGauge}
S_A=\frac{1}{8\pi^2 r}\int \Tr\left(F\wedge \star F +i r\ C\wedge  F \wedge F\right)\,.
\ee
The graviphoton $C$ is only non-vanishing in the $\phi$-direction on $S^2$ and therefore from the second term in the 5d action we obtain a Chern-Simons term on $\mathbb{R}^3$. The action thus reduces to
\be \ba
S_A &=\frac{r}{32\pi^2} \int d^3x \sin \theta d \theta d \phi \Tr\left( \half F_{ab}F^{ab}  
+ \frac{4}{r^2} F_{xa} F^{xa} + \frac{8}{r^4}F_{xy}F^{xy} \right. \\
& \left.  \qquad \qquad+\frac{2i}{r} \varepsilon^{abc} \left( A_a\partial_bA_c+\frac{2}{3}A_a A_b A_c\right) \right)\,,
\ea \ee
where we have taken out the factor of $\frac{r^2}{4}$ in the metric on the $S^2$ in \eqref{5dMetric}. From here on curved indices on $S^2$ are therefore raised and lowered using the metric on a unit radius two-sphere.

\subsubsection*{Scalars}

The 5d action of the scalars is given by
\be
S_\Phi=\frac{1}{32\pi^2 r}\int d^5x\ \sqrt{g} \Tr\left( \mathcal{D}_{a'} \Phi^{\underline{\hat m \hat n}} \mathcal{D}^{a'} \Phi_{\underline{\hat m \hat n}}-4 \Phi^{\underline{\hat m \hat n}} F_{a'b'}T^{a'b'}_{\underline{\hat m \hat n}} - \Phi_{\underline{\hat m \hat n}} (M^2_{\Phi})^{\underline{\hat m \hat n}}\,_{\underline{\hat r \hat s}} \Phi^{\underline{\hat r \hat s}}\right)\,.
\ee
In the 5d theory the covariant derivative and mass term for the scalars are given by
\be \label{5dCovDandMass}
\ba
\mathcal{D}_{a'}\Phi_{\underline{\hat m \hat n}}&=D_{a'} \Phi_{\underline{\hat m \hat n}}-{V_{a'}}^{\underline{\hat{r}}}\,_{[\underline{\hat{m}}}\Phi_{\underline{\hat n}]\underline{ \hat r}}\\
\mathcal{D}^2\Phi_{\underline{\hat m \hat n}}&=\left(D^{a'}+\omega_{b'}\,^{a'b'}\right)\mathcal{D}_{a'}\Phi_{\underline{\hat m \hat n}}-{V_{a'}}^{\underline{\hat{r}}}\,_{[\underline{\hat{m}}}\mathcal{D}^{a'}\Phi_{\underline{\hat n}]\underline{ \hat r}}\\
(M^2_{\Phi})^{\underline{\hat m \hat n}}\,_{\underline{\hat r \hat s}}&=\left(\frac{r^2}{20}G_{a'b'}G^{a'b'}-\frac{\mathcal{R}}{5}\right)\delta^{\underline{\hat{m}}}_{\underline{\hat{r}}}\delta^{\underline{\hat{n}}}_{\underline{\hat{s}}}-\frac{1}{15}D^{\underline{\hat{m}\hat{n}}}\,_{\underline{\hat{r}\hat{s}}}-T_{a'b'}^{\underline{\hat{m}\hat{n}}}T^{a'b'}_{\underline{\hat{r}\hat{s}}}\,,
\ea
\ee
where $D_{a'}=\p_{a'}+[A_{a'},\cdot]$.
Inserting in the form of the background fields and the field decomposition under the topological twist the mass of $\varphi$ takes the form
\be
M^2_{\varphi}=\frac{4(v-1)}{r^2}\,.
\ee
For the spinor $\phi^{\alpha \hat \alpha}$ on $M_3$ there is an additional contribution to its mass besides the explicit 5d mass term in \eqref{5dCovDandMass}, originating from the R-symmetry gauge field in the covariant derivative. This term is proportional to $V^2$ and the combination of these two mass contributions gives
\be
M_{\phi,\text{total}}^2=\frac{1}{4r^2}(v+2)(v-2)\,.
\ee
The linear coupling with the R-symmetry gauge field, arising from the covariant derivative, in the scalar equations of motion gives rise to a spinorial kinetic term for $\phi^{\alpha \hat \alpha}$. The full scalar action becomes
\be \label{5dScalars}
\ba
S_{\Phi} &= \frac{r}{64 \pi^2 } \int d^3x  \sin \theta d \theta d \phi \Tr\left( 2 D_a \varphi D^a \varphi +\frac{8}{r^2}D_x \varphi D^x \varphi - \frac{16 (1-v)}{r^3}\varepsilon^{xy}F_{xy}\varphi +\frac{8(1-v)}{r^2}\varphi^2  \right.  \\
& \left. \qquad \qquad + D_a \phi_{\alpha \hat \alpha} D^a \phi^{\alpha \hat \alpha}+ \frac{v}{r} D_a \phi_{\alpha \hat \alpha} (\sigma^a)^{\alpha}\,_{\beta} \phi^{\beta \hat \alpha}+\frac{4}{r^2} D_{x} \phi_{\alpha \hat \alpha} D^{x} \phi^{\alpha \hat \alpha} + \frac{(2+v)(2-v)}{4r^2}\phi_{\alpha \hat \alpha} \phi^{\alpha \hat \alpha} \right) \,.
\ea
\ee
From this action we observe that only for certain values of $v$ are either $\varphi$ or $\phi^{\alpha \hat \alpha}$ massless. We will see the same feature arising for the superpartners of these fields in the next section.

\subsubsection*{Fermions}

The theory includes a set of fermions $\rho^{\underline{\hat{m}}}$ whose 5d action is given by
\be
S_\rho=\frac{1}{32\pi^2r}\int d^5x\ \sqrt{g} \Tr\left(i \rho_{\underline{\hat{m}}}\slashed{\mathcal{D}} \rho^{\underline{\hat{m}}}+\rho_{\underline{\hat{m}}}(M_\rho)^{\underline{\hat{m}\hat{n}}}\rho_{\underline{\hat{n}}}\right)
\ee
with
\be
\ba
\mathcal{D}_{a'}\rho^{\underline{\hat{m}}}&=\left(D_{a'}+\frac{1}{4}\omega_{a'}\,^{b'c'}\Gamma_{b'c'}\right)\rho^{\underline{\hat{m}}}-\frac{1}{2}{V_{a'}}^{\underline{\hat{m}}}\,_{\underline{\hat{n}}}\rho^{\underline{\hat{n}}}\\
(M_\rho)^{\underline{\hat{m}\hat{n}}}&=\frac{r}{8} G_{a'b'} \Gamma^{a'b'}\Omega^{\underline{\hat{m}\hat{n}}}-\half T_{a'b'}^{\underline{\hat{m}\hat{n}}}\Gamma^{a'b'}\,.
\ea
\ee
Using the spinor decomposition
\be 
\rho^{\alpha'\underline{\hat{m}}}\to \lambda^{\mu} \varepsilon^{\alpha\hat{m}} 
+ i \xi_a^{\mu} ({\sigma^a})^\alpha\,_\beta \varepsilon^{\beta\hat{m}} 
+ \rho^{{\mu}\alpha\hat{\alpha}}\,, 
\ee 
the kinetic terms and mass terms for the fermions take the form
\be
\ba
S_{\rho}&= \frac{r}{128\pi^2}\int d^3x  \sin \theta d \theta d \phi \Tr\left(-4\xi_{a \mu}(\sigma_3)^{\mu}\,_{\nu} D^{a}\lambda^{\nu} - 2 \varepsilon_{abc} \xi^a_\mu (\sigma_3)^{\mu}\,_{\nu} D^b \xi^{c\nu} \right. \\
& \left. \qquad \qquad +\frac{4i}{r} \lambda_{\mu}\slashed{\mathcal{D}}_{S^2}{}^{\mu}\,_{\nu} \lambda^{\nu} +\frac{4i}{r} \xi_{a\mu} \slashed{\mathcal{D}}_{S^2}{}^{\mu}\,_{\nu} \xi^{a \nu} -\frac{4iv}{r}\lambda_{\mu}(\sigma_3)^{\mu}\,_{\nu}\lambda^{\nu} + i \rho_{\mu \alpha \hat \alpha } (\sigma_3)^{\mu}\,_{\nu} (\sigma_a)^{\alpha}\,_{\beta}  D^a \rho^{\nu \beta \hat \alpha } \right.  \\
& \left. \qquad\qquad + \frac{2i}{r} \rho_{\mu \alpha \hat \alpha  } \slashed{\mathcal{D}}_{S^2}{}^{\mu}\,_{\nu} \rho^{\nu \alpha \hat \alpha} - \frac{i (2-v)}{2r}\rho_{\mu \alpha \hat \alpha} (\sigma_3)^{\mu}\,_{\nu} \rho^{\nu \alpha\hat \alpha} \right)\,,
\ea
\ee
where ${\mathcal{D}}_{S^2}$ is the covariant derivative on $S^2$ with respect to the metric and the gauge field $A_x$. The mass terms for the fermions include contributions from the explicit mass term in 5d and the coupling to the R-symmetry gauge field in the 5d covariant derivative. Note that the supersymmetry partners for the gauge fields on $M_3$, corresponding to $\xi_a^{\sigma}$, are always massless, while the other fermions have $v$ dependent mass terms.

\subsubsection*{Interactions}

The final piece of the 5d action are the interaction terms given by
\be
S_{\text{int}}=\frac{1}{32\pi^2 r}\int d^5x \Tr\left(\rho_{\underline{\hat{m}}}[\Phi^{\underline{\hat{m}\hat{n}}},\rho_{\underline{\hat{n}}}]-\frac{1}{4}[\Phi_{\underline{\hat{m}\hat{n}}},\Phi^{\underline{\hat{n}\hat{r}}}][\Phi_{\underline{\hat{r}\hat{s}}},\Phi^{\underline{\hat{s}\hat{m}}}]\right)\,,
\ee
consisting of the standard Yukawa coupling and quartic scalar potential. After decomposing the fields one obtains
\be
\ba
S_{\text{Yuk}} &= \frac{r}{128\pi^2}\int d^3x  \sin \theta d \theta d \phi \Tr\left(-2 \lambda_{\mu}[\varphi, \lambda^{\mu}] -2 \xi_{a \mu}[ \varphi, \xi^{a \mu}] + \rho_{\mu \alpha \hat \alpha}[\varphi, \rho^{\mu \alpha \hat \alpha }] \right. \\
& \left. \qquad \qquad -2\lambda_\mu[\rho_{\alpha\hat{\alpha}}^\mu,\phi^{\alpha\hat{\alpha}}]+2i \xi_{a\mu}[\rho_{\alpha\hat{\alpha}}^\mu,\phi^{\beta\hat{\alpha}}](\sigma^a)^\alpha\,_\beta\right)\\
S_{\text{pot}} &= \frac{r}{128\pi^2}\int d^3x  \sin \theta d \theta d \phi \Tr\left(2[\varphi, \phi_{\alpha \hat \alpha}][\varphi, \phi^{\alpha \hat \alpha}]+ \half [ \phi_{\alpha \hat \alpha}, \phi_{\beta \hat \beta}][\phi^{\alpha \hat \alpha}, \phi^{\beta \hat \beta}]\right)\,.
\ea
\ee

\subsection{Generalization to Curved $M_3$}
\label{app:CurvedM3Sugra}

In defining the theory on general three-manifold the derivatives need to be covariantized with respect to the curvature on $M_3$ and the connection $\ConnV$ of $SU(2)_\ell$ as discussed in section \ref{sec:DeformationTheory}. In section \ref{sec:5donM3xS2} we added the correction terms \eqref{RphiMass} necessary to preserve supersymmetry.
Alternatively, in this section we use the supergravity background to derive the required curvature terms. Let us perform the twist using the R-symmetry gauge field in the supergravity multiplet. In this we keep the ansatz for the other two fields $T_{a'b}^{\underline{\hat m \hat n}}$ and $D^{\underline{\hat m \hat n}}{}_{\underline{\hat r \hat s}}$ the same and introduce the additional prefactors $d_{\text{twist}}$ and $t_{\text{twist}}$
\be \ba 
{V_a}^{\underline{\hat m}}\,_{\underline{\hat n}} &\rightarrow  {V_a}^{\underline{\hat m}}\,_{\underline{\hat n}} + {(V_{\text{twist}})_a}^{\underline{\hat m}}\,_{\underline{\hat n}} \\
d &\rightarrow d + d_{\text{twist}} \\
t &\rightarrow  t + t_{\text{twist}}\,.
\ea \ee
Since the values of $d$ and $t$ have been fixed by the vanishing of the Killing spinor equations, in the following we only consider the contributions involving the new twist prefactors. It turns out that $t_{\text{twist}}$ can be chosen to vanish so we will not consider it further here. Let $\omega_a\,^{bc}$ be the spin connection of $M_3$ such that the first Killing spinor equation for a constant supersymmetry parameter \eqref{SUSYSinglet} now takes the form
\be
\delta\psi_a^{\underline{\hat{m}}}=\frac{1}{4}\omega_a\,^{bc}\Gamma_{bc}\epsilon^{\underline{\hat{m}}}-\frac{1}{2}{(V_{\text{twist}})_a}^{\underline{\hat m}}\,_{\underline{\hat n}}\epsilon^{\underline{\hat{n}}}\,.
\ee
For a general spin connection this is solved for
\be
{(V_{\text{twist}})_a}^{\underline{\hat m}}\,_{\underline{\hat n}}=\omega_a\,^{bc}\varepsilon_{bc\hat{a}}(\Sigma_{\hat{a}})^{\underline{\hat m}}\,_{\underline{\hat n}}\,,
\ee
where the $\Sigma_{\hat{a}}$ are the generators of $SU(2)_r$ defined in \eqref{SU2lrGenerators}.
The second Killing spinor equation becomes
\be
\ba
\delta\chi^{\underline{\hat{m}\hat{n}}}\,_{\underline{\hat{r}}}&=-{R_{ab}}^{[\underline{\hat{m}}}\,_{\underline{\hat{r}}}\Gamma^{ab}\epsilon^{\underline{\hat{n}}]}-\frac{4}{15}D^{\underline{\hat{m}\hat{n}}}\,_{\underline{\hat{r}\hat{s}}}\epsilon^{\underline{\hat{s}}}-(\text{traces})\,,
\ea
\ee
where the curvature of the R-symmetry gauge field is defined in \eqref{VCurv}.
Using the solution for $V_{\text{twist}}$ we can write the curvature as
\be
R_{ab}^{\underline{\hat{m}\hat{n}}}=\mathcal{R}_{ab}\,^{cd}\varepsilon_{cd\hat{a}}(\Sigma_{\hat{a}})^{\underline{\hat{m}\hat{n}}}\,,
\ee
where $\mathcal{R}_{ab}\,^{cd}$ is the Riemann tensor on $M_3$. One finds that the Killing spinor equation can only be solved if $M_3$ is Einstein, i.e. the Ricci tensor is proportional to the metric
\be
\mathcal{R}_{ab}=\frac{\mathcal{R}}{3} g_{ab}\,,
\ee
where $\mathcal{R}_{ab}$ and $\mathcal{R}$ are the Ricci tensor and scalar on $M_3$. The second Killing spinor equation is then solved by
\be
d_{\text{twist}}=\frac{3\mathcal{R}}{16}\,.
\ee
Turning on these background fields introduces an additional mass term for the scalars given by
\be
(M^2_{\Phi})_{\text{curv}}\Phi^{\underline{\hat{m}\hat{n}}}=\frac{\mathcal{R}}{5}\Phi^{\underline{\hat{m}\hat{n}}}+\frac{1}{15}D^{\underline{\hat{m}\hat{n}}}\,_{\underline{\hat{r}\hat{s}}}\Phi^{\underline{\hat{r}\hat{s}}}\,,
\ee
which reduces to
\be
\ba
(M^2_{\varphi})_{\text{curv}}&=0\\
(M^2_{\phi})_{\text{curv}}&=\frac{\mathcal{R}}{4}\,.
\ea
\ee
This result agrees with the additional curvature term derived in section \ref{sec:5donM3xS2}, and in order to generalize the action to curved $M_3$ this mass term has to be added. It is not clear to the authors where the additional condition, that $M_3$ is Einstein, comes from in the supergravity approach. However, since the twist by hand does not require this constraint it should not be necessary. Similarly, adding an $SU(2)_\ell$ connection, which amounts to turning on the R-symmetry gauge field in the direction ${V_a}^{\hat \alpha}\,_{\hat \beta}$, forces this connection to be flat which should not be necessary in general. 

\subsection{Reduction on $S^2$}

The next step is to reduce the action \eqref{5dActionM3S2} on the $S^2$, keeping only the massless modes on $M_3$. To this end we expand the 5d fields in terms of eigenvectors of the relevant differential operators.

\subsubsection{Spherical Harmonics and Eigenspinors on $S^2$}
\label{app:HarmonicsonS2}

For the Laplacian $\Delta_{S^2}=\mathcal{D}_x\mathcal{D}^x$ the eigenvectors are given by the spherical harmonics $Y_k^m$, where $k$ is a non-negative integer and $|m|\leq k$ counts the multiplicity. They fulfill
\be
\Delta_{S^2} Y_k^m=-k(k+1) Y_k^m
\ee
and are normalized as
\be
\int_{S^2} \sin{\theta}d\theta d\phi\ Y_k^m\ Y_{k'}^{m'}=4\pi (-1)^m\delta_{k,k'}\delta_{m,-m'}\,.
\ee
The modified Dirac operator $(\sigma_3 \slashed{\mathcal{D}}_{S^2})^\mu\,_\nu$, with $\slashed{\mathcal{D}}_{S^2}$ as in \eqref{DiracOperatorS2}, has eigenvalues $\pm n$, for $n$ a positive integer. This can be seen by noticing that $(\sigma_3 \slashed{\mathcal{D}}_{S^2})^2=-(\slashed{\mathcal{D}}_{S^2})^2$ and acknowledging that the usual Dirac operator on the unit sphere has eigenvalues $\pm in$, see \cite{Abrikosov:2002jr}. Let us call the corresponding eigenspinors $\Theta_{n,j}^\mu$, where the subscript $j$ is the eigenvalue under $i\p_\phi$ leading to a degeneracy of $2n$. They thus fulfill
\be
(\sigma_3 \slashed{\mathcal{D}}_{S^2})^\mu\,_\nu \Theta_{n,j}^\nu=n \Theta_{n,j}^\mu\,.
\ee
Concretely, we can write
\be\label{EigenspinorsS2}
\ba
\Theta_{n,j}^\mu=\frac{C(|n|,|j|)}{\sqrt{r}}e^{-ij\phi}\left(\begin{array}{r} (1-x)^{\half |j-\half|} (1+x)^{\half |j+\half|} P_{|n|-|j|-\half}^{|j-\half|,|j+\half|}(x) \\ \text{sgn}(nj) (1-x)^{\half |j+\half|} (1+x)^{\half |j-\half|} P_{|n|-|j|-\half}^{|j+\half|,|j-\half|}(x)\end{array}\right)\,,
\ea
\ee
where $x=\cos\theta$ and the $P_n^{\alpha,\beta}$ are Jacobi polynomials. The additional $r$-dependence is included such that the fermions have the canonical scaling dimension $[\lambda]=1$ in three dimensions. The normalization $C(|n|,|j|)$ is chosen such that
\be \label{NormalizationEigenspinors}
\int_{S^2}\sin{\theta}d\theta d\phi \ {\Theta_{n,j}}_\mu \Theta^\mu_{n',j'}= \text{sgn}(n j)\ \frac{4\pi}{r}\  \delta_{n,-n'}\ \delta_{j,-j'}\,.
\ee

\subsubsection{Spectrum of Massless Fields}
\label{app:MasslessSpectrum}
Depending on the choice of $v$, different fields become massless in 3d. In this section we determine the values of $v$ for which the bosons and fermions become massless.
Let us first look at the the bosonic fields. These are expanded in terms of the spherical harmonics, e.g. for the singlet $\varphi$
\be
\varphi=\sum_{k=0}^\infty \sum_{m=-k}^k \varphi^{(k,m)} Y_k^m (\theta,\phi)\,.
\ee
From the 5d equations of motion we find the conditions for the bosons to be massless in 3d  reduces to
\be \label{vConditionvarphi}
v(k_\varphi)=\left\{
\begin{array}{l}
k_\varphi+1\\
-k_\varphi
\end{array}\right.\,, \quad 
v(k_\phi)=\pm \left(4k_\phi+2\right)\,.
\ee

We can repeat the same logic for the fermions. Expanding them in terms of the eigenspinors \eqref{EigenspinorsS2} of the modified Dirac operator the massless conditions for the fermions are given by
\be \label{vConditionFermions}
v(n_\lambda)=n_\lambda\,, \qquad
v(n_\rho)=-4n_\rho+2\,,
\ee
whereas $\xi$ always receives a mass.
Since there is no closed form for the 3d theory for general $v$, we have to fix $v$ first and then carry out the reduction. In section \ref{sec:RealCS} it was argued that for the value $v=2$ we obtain real Chern-Simons-Dirac theory. In the following we will carry out the reduction for the values $v=0,1$ in more detail and obtain real Chern-Simons theory.

\subsubsection{Real Chern-Simons Theory: $v=0,1$}
\label{app:v=0,1}

In the case $v=1$ the massless field content consists of the gauge field $A_b$ as well as a scalar field $\varphi$, which is constant on the $S^2$, and two fermions $\lambda^\pm$. To obtain the correct action we also have to integrate out the massive fields $\xi_a^\pm$ and $\tilde{\lambda}^\pm$. In terms of the eigenspinors $\Theta^\mu_{n,j}$ of the Dirac operator as in \eqref{EigenspinorsS2} these fields are defined by
\be
\ba
\lambda^\mu&=\lambda^\pm \Theta_{1,\pm\half}^\mu+\tilde{\lambda}^\pm \Theta_{-1,\pm\half}^\mu\\
\xi_a^\mu&=\xi_a^\pm \Theta_{1,\pm\half}^\mu\,.
\ea
\ee
All the modes coming from the fields $\phi^{\alpha\hat{\alpha}}$ and $\rho^{\mu\alpha\hat{\alpha}}$ are massive and can safely be ignored. The 3d action is then given by
\be
\ba
S^{v=1}&=S_A+S^{v=1}_\Phi+S^{v=1}_{\rho}+S^{v=1}_{\text{int}}\\
S_A&=\frac{r}{8\pi}\int d^3x \Tr\left(F\wedge\star F+\frac{2i}{r} \text{CS}(A)\right)\\
S^{v=1}_\Phi&=\frac{r}{8\pi}\int d^3x \Tr\left(\mathcal{D}_a\varphi \mathcal{D}^a\varphi\right)\\
S^{v=1}_{\rho}&=\frac{1}{8\pi}\int d^3x \Tr\left(-\xi_a^+ \mathcal{D}^a\lambda^- +\xi_a^- \mathcal{D}^a\lambda^+ -\frac{2i}{r}\tilde{\lambda}^+\tilde{\lambda}^- +\frac{2i}{r}\xi_a^+\xi^{a-}\right)\\
S^{v=1}_{\text{int}}&=\frac{1}{8\pi}\int d^3x \Tr\left(\tilde{\lambda}^+[\varphi,\lambda^-] - \tilde{\lambda}^-[\varphi,\lambda^+]\right)\,,
\ea
\ee
so that the massive fields are integrated out to
\be
\xi_a^\pm=-\frac{ir}{2}\mathcal{D}_a \lambda^\pm\,,\qquad \tilde{\lambda}^\pm=-\frac{ir}{2}[\varphi,\lambda^\pm]\,,
\ee
to obtain the action
\be
S^{v=1}=S_A+\frac{r}{8\pi}\int d^3x \Tr\left(\mathcal{D}_a\varphi \mathcal{D}^a\varphi + \frac{i}{2} \mathcal{D}_a \lambda^+ \mathcal{D}^a \lambda^- -\frac{i}{2} [\varphi,\lambda^+][\varphi,\lambda^-]\right)\,.
\ee
A very similar action has appeared in the case for $\mathcal{N}=2$ supersymmetry in \cite{Cordova:2013cea}. There it was shown that the fields with kinetic terms scaling with $r$ can be interpreted as ghosts. These can be integrated out to gauge fix the complex Chern-Simons action. In the case at hand we obtain exactly half the field content of \cite{Cordova:2013cea}. We can thus again interpret $\varphi$ and $\lambda^\pm$ as gauge parameters. Consequently, we can safely take the limit $r\to0$, without fixing a gauge, and obtain real Chern-Simons theory
\be
S^{v=1}=\frac{i}{4\pi}\int d^3x \Tr\left(\text{CS}(A)\right)\,.
\ee

The story is slightly different for the case $v=0$. As the Dirac operator on the $S^2$ does not admit any zero-modes there are no massless fermions according to \eqref{vConditionFermions}. The remaining field content is unaltered compared to $v=1$. Thus, we can immediately write down the action
\be
S^{v=0}=\frac{r}{8\pi}\int d^3x \Tr\left(F\wedge\star F+\frac{2i}{r} \text{CS}(A)+\mathcal{D}_a\varphi \mathcal{D}^a\varphi\right)\,.
\ee
Again taking the $r\rightarrow 0$ limit results in
\be
S^{v=0}=\frac{i}{4\pi}\int d^3x \Tr\left(\text{CS}(A)\right)\,.
\ee



\begin{thebibliography}{10}

\bibitem{Alday:2009aq}
L.~F. Alday, D.~Gaiotto and Y.~Tachikawa, \emph{{Liouville Correlation
  Functions from Four-dimensional Gauge Theories}},
  \href{http://dx.doi.org/10.1007/s11005-010-0369-5}{\emph{Lett. Math. Phys.}
  {\bf 91} (2010) 167--197}, [\href{https://arxiv.org/abs/0906.3219}{{\tt
  0906.3219}}].

\bibitem{Wyllard:2009hg}
N.~Wyllard, \emph{{A(N-1) conformal Toda field theory correlation functions
  from conformal N = 2 SU(N) quiver gauge theories}},
  \href{http://dx.doi.org/10.1088/1126-6708/2009/11/002}{\emph{JHEP} {\bf 11}
  (2009) 002}, [\href{https://arxiv.org/abs/0907.2189}{{\tt 0907.2189}}].

\bibitem{Gaiotto:2009we}
D.~Gaiotto, \emph{{N=2 dualities}},
  \href{http://dx.doi.org/10.1007/JHEP08(2012)034}{\emph{JHEP} {\bf 08} (2012)
  034}, [\href{https://arxiv.org/abs/0904.2715}{{\tt 0904.2715}}].

\bibitem{Dimofte:2011ju}
T.~Dimofte, D.~Gaiotto and S.~Gukov, \emph{{Gauge Theories Labelled by
  Three-Manifolds}},
  \href{http://dx.doi.org/10.1007/s00220-013-1863-2}{\emph{Commun. Math. Phys.}
  {\bf 325} (2014) 367--419}, [\href{https://arxiv.org/abs/1108.4389}{{\tt
  1108.4389}}].

\bibitem{Terashima:2011qi}
Y.~Terashima and M.~Yamazaki, \emph{{SL(2,R) Chern-Simons, Liouville, and Gauge
  Theory on Duality Walls}},
  \href{http://dx.doi.org/10.1007/JHEP08(2011)135}{\emph{JHEP} {\bf 08} (2011)
  135}, [\href{https://arxiv.org/abs/1103.5748}{{\tt 1103.5748}}].

\bibitem{Assel:2016lad}
B.~Assel, S.~Schafer-Nameki and J.-M. Wong, \emph{{M5-branes on $S^2 x M_4$:
  Nahm's Equations and 4d Topological Sigma-models}},
  \href{https://arxiv.org/abs/1604.03606}{{\tt 1604.03606}}.

\bibitem{Vafa:1994tf}
C.~Vafa and E.~Witten, \emph{{A Strong coupling test of S duality}},
  \href{http://dx.doi.org/10.1016/0550-3213(94)90097-3}{\emph{Nucl.Phys.} {\bf
  B431} (1994) 3--77}, [\href{https://arxiv.org/abs/hep-th/9408074}{{\tt
  hep-th/9408074}}].

\bibitem{Gadde:2013sca}
A.~Gadde, S.~Gukov and P.~Putrov, \emph{{Fivebranes and 4-manifolds}},
  \href{https://arxiv.org/abs/1306.4320}{{\tt 1306.4320}}.

\bibitem{Nekrasov:2002qd}
N.~A. Nekrasov, \emph{{Seiberg-Witten prepotential from instanton counting}},
  \href{http://dx.doi.org/10.4310/ATMP.2003.v7.n5.a4}{\emph{Adv. Theor. Math.
  Phys.} {\bf 7} (2003) 831--864},
  [\href{https://arxiv.org/abs/hep-th/0206161}{{\tt hep-th/0206161}}].

\bibitem{Pestun:2007rz}
V.~Pestun, \emph{{Localization of gauge theory on a four-sphere and
  supersymmetric Wilson loops}},
  \href{http://dx.doi.org/10.1007/s00220-012-1485-0}{\emph{Commun.Math.Phys.}
  {\bf 313} (2012) 71--129}, [\href{https://arxiv.org/abs/0712.2824}{{\tt
  0712.2824}}].

\bibitem{Teschner:2016yzf}
J.~Teschner, ed., \emph{{New Dualities of Supersymmetric Gauge Theories}}.
\newblock Mathematical Physics Studies. Springer, Cham, Switzerland, 2016,
  \href{http://dx.doi.org/10.1007/978-3-319-18769-3}{10.1007/978-3-319-18769-3}.

\bibitem{Mitev:2017jqj}
V.~Mitev and E.~Pomoni, \emph{{2D CFT blocks for the 4D class $\mathcal{S}_k$
  theories}}, \href{http://dx.doi.org/10.1007/JHEP08(2017)009}{\emph{JHEP} {\bf
  08} (2017) 009}, [\href{https://arxiv.org/abs/1703.00736}{{\tt 1703.00736}}].

\bibitem{Bah:2012dg}
I.~Bah, C.~Beem, N.~Bobev and B.~Wecht, \emph{{Four-Dimensional SCFTs from
  M5-Branes}}, \href{http://dx.doi.org/10.1007/JHEP06(2012)005}{\emph{JHEP}
  {\bf 06} (2012) 005}, [\href{https://arxiv.org/abs/1203.0303}{{\tt
  1203.0303}}].

\bibitem{Gaiotto:2015usa}
D.~Gaiotto and S.~S. Razamat, \emph{{$ \mathcal{N}=1 $ theories of class $
  {\mathcal{S}}_k $}},
  \href{http://dx.doi.org/10.1007/JHEP07(2015)073}{\emph{JHEP} {\bf 07} (2015)
  073}, [\href{https://arxiv.org/abs/1503.05159}{{\tt 1503.05159}}].

\bibitem{Cordova:2013cea}
C.~Cordova and D.~L. Jafferis, \emph{{Complex Chern-Simons from M5-branes on
  the Squashed Three-Sphere}},  \href{https://arxiv.org/abs/1305.2891}{{\tt
  1305.2891}}.

\bibitem{Cordova:2016cmu}
C.~Cordova and D.~L. Jafferis, \emph{{Toda Theory From Six Dimensions}},
  \href{https://arxiv.org/abs/1605.03997}{{\tt 1605.03997}}.

\bibitem{Witten:1999ds}
E.~Witten, \emph{{Supersymmetric index of three-dimensional gauge theory}},
  \href{https://arxiv.org/abs/hep-th/9903005}{{\tt hep-th/9903005}}.

\bibitem{McLean}
R.~C. McLean, \emph{Deformations of calibrated submanifolds},
  \href{http://dx.doi.org/10.4310/CAG.1998.v6.n4.a4}{\emph{Comm. Anal. Geom.}
  {\bf 6} (1998) 705--747}.

\bibitem{Joyce}
D.~Joyce, \emph{Compact Manifolds with Special Holonomy}.
\newblock Oxford mathematical monographs. Oxford University Press, 2000.

\bibitem{Gerchkovitz:2014gta}
E.~Gerchkovitz, J.~Gomis and Z.~Komargodski, \emph{{Sphere Partition Functions
  and the Zamolodchikov Metric}},
  \href{http://dx.doi.org/10.1007/JHEP11(2014)001}{\emph{JHEP} {\bf 11} (2014)
  001}, [\href{https://arxiv.org/abs/1405.7271}{{\tt 1405.7271}}].

\bibitem{Kapustin:2009kz}
A.~Kapustin, B.~Willett and I.~Yaakov, \emph{{Exact Results for Wilson Loops in
  Superconformal Chern-Simons Theories with Matter}},
  \href{http://dx.doi.org/10.1007/JHEP03(2010)089}{\emph{JHEP} {\bf 1003}
  (2010) 089}, [\href{https://arxiv.org/abs/0909.4559}{{\tt 0909.4559}}].

\bibitem{kronheimer2007monopoles}
P.~Kronheimer and T.~Mrowka, \emph{Monopoles and three-manifolds}, vol.~10 of
  \emph{New Mathematical Monographs}.
\newblock Cambridge University Press, 2007,
  \href{http://dx.doi.org/10.1017/CBO9780511543111}{10.1017/CBO9780511543111}.

\bibitem{Witten:1988hf}
E.~Witten, \emph{{Quantum Field Theory and the Jones Polynomial}},
  \href{http://dx.doi.org/10.1007/BF01217730}{\emph{Commun. Math. Phys.} {\bf
  121} (1989) 351--399}.

\bibitem{MR1091619}
N.~Reshetikhin and V.~G. Turaev, \emph{Invariants of {$3$}-manifolds via link
  polynomials and quantum groups},
  \href{http://dx.doi.org/10.1007/BF01239527}{\emph{Invent. Math.} {\bf 103}
  (1991) 547--597}.

\bibitem{Acharya:2001dz}
B.~S. Acharya and C.~Vafa, \emph{{On domain walls of N=1 supersymmetric
  Yang-Mills in four-dimensions}},
  \href{https://arxiv.org/abs/hep-th/0103011}{{\tt hep-th/0103011}}.

\bibitem{Kovalev:2001zr}
A.~Kovalev, \emph{{Twisted connected sums and special Riemannian holonomy}},
  \href{https://arxiv.org/abs/math/0012189}{{\tt math/0012189}}.

\bibitem{CHNP1}
A.~Corti, M.~Haskins, J.~Nordstr{\"o}m and T.~Pacini, \emph{Asymptotically
  cylindrical {C}alabi-{Y}au 3-folds from weak {F}ano 3-folds},
  \href{http://dx.doi.org/10.2140/gt.2013.17.1955}{\emph{Geom. Topol.} {\bf 17}
  (2013) 1955--2059}.

\bibitem{Corti:2012kd}
A.~Corti, M.~Haskins, J.~Nordstr{\"o}m and T.~Pacini,
  \emph{{$\mathrm{G}_{2}$-manifolds and associative submanifolds via semi-Fano
  $3$-folds}}, \href{http://dx.doi.org/10.1215/00127094-3120743}{\emph{Duke
  Math. J.} {\bf 164} (2015) 1971--2092},
  [\href{https://arxiv.org/abs/1207.4470}{{\tt 1207.4470}}].

\bibitem{Braun:2017uku}
A.~P. Braun and S.~Schafer-Nameki, \emph{{Compact, Singular G2-Holonomy
  Manifolds and M/Heterotic/F-Theory Duality}},
  \href{https://arxiv.org/abs/1708.07215}{{\tt 1708.07215}}.

\bibitem{Acharya:2000mu}
B.~S. Acharya, J.~P. Gauntlett and N.~Kim, \emph{{Five-branes wrapped on
  associative three cycles}},
  \href{http://dx.doi.org/10.1103/PhysRevD.63.106003}{\emph{Phys. Rev.} {\bf
  D63} (2001) 106003}, [\href{https://arxiv.org/abs/hep-th/0011190}{{\tt
  hep-th/0011190}}].

\bibitem{BryantSalamon}
R.~L. Bryant and S.~M. Salamon, \emph{On the construction of some complete
  metrics with exceptional holonomy},
  \href{http://dx.doi.org/10.1215/S0012-7094-89-05839-0}{\emph{Duke Math. J.}
  {\bf 58} (1989) 829--850}.

\bibitem{Benini:2016hjo}
F.~Benini and A.~Zaffaroni, \emph{{Supersymmetric partition functions on
  Riemann surfaces}}, {\emph{Proc. Symp. Pure Math.} {\bf 96} (2017) 13--46},
  [\href{https://arxiv.org/abs/1605.06120}{{\tt 1605.06120}}].

\bibitem{Closset:2016arn}
C.~Closset and H.~Kim, \emph{{Comments on twisted indices in 3d supersymmetric
  gauge theories}},
  \href{http://dx.doi.org/10.1007/JHEP08(2016)059}{\emph{JHEP} {\bf 08} (2016)
  059}, [\href{https://arxiv.org/abs/1605.06531}{{\tt 1605.06531}}].

\bibitem{Dimofte:2010tz}
T.~Dimofte, S.~Gukov and L.~Hollands, \emph{{Vortex Counting and Lagrangian
  3-manifolds}}, \href{http://dx.doi.org/10.1007/s11005-011-0531-8}{\emph{Lett.
  Math. Phys.} {\bf 98} (2011) 225--287},
  [\href{https://arxiv.org/abs/1006.0977}{{\tt 1006.0977}}].

\bibitem{Acharya:2004qe}
B.~S. Acharya and S.~Gukov, \emph{{M theory and singularities of exceptional
  holonomy manifolds}},
  \href{http://dx.doi.org/10.1016/j.physrep.2003.10.017}{\emph{Phys. Rept.}
  {\bf 392} (2004) 121--189}, [\href{https://arxiv.org/abs/hep-th/0409191}{{\tt
  hep-th/0409191}}].

\bibitem{Harvey:1982xk}
R.~Harvey and H.~B. Lawson, Jr., \emph{{Calibrated geometries}},
  \href{http://dx.doi.org/10.1007/BF02392726}{\emph{Acta Math.} {\bf 148}
  (1982) 47}.

\bibitem{MR0358873}
N.~Hitchin, \emph{Harmonic spinors},
  \href{http://dx.doi.org/10.1016/0001-8708(74)90021-8}{\emph{Advances in
  Math.} {\bf 14} (1974) 1--55}.

\bibitem{joyce2003riemannian}
D.~Joyce, \emph{Riemannian holonomy groups and calibrated geometry},  in
  \emph{Calabi-Yau Manifolds and Related Geometries}, pp.~1--68.
\newblock Springer, 2003.

\bibitem{MR0156292}
A.~Lichnerowicz, \emph{Spineurs harmoniques}, {\emph{C. R. Acad. Sci. Paris}
  {\bf 257} (1963) 7--9}.

\bibitem{MR1421872}
C.~B\"ar, \emph{Metrics with harmonic spinors},
  \href{http://dx.doi.org/10.1007/BF02246994}{\emph{Geom. Funct. Anal.} {\bf 6}
  (1996) 899--942}.

\bibitem{Joyce:2016fij}
D.~Joyce, \emph{{Conjectures on counting associative 3-folds in
  $G_2$-manifolds}},  \href{https://arxiv.org/abs/1610.09836}{{\tt
  1610.09836}}.

\bibitem{Witten:1994cg}
E.~Witten, \emph{{Monopoles and four manifolds}},
  \href{http://dx.doi.org/10.4310/MRL.1994.v1.n6.a13}{\emph{Math. Res. Lett.}
  {\bf 1} (1994) 769--796}, [\href{https://arxiv.org/abs/hep-th/9411102}{{\tt
  hep-th/9411102}}].

\bibitem{Seiberg:1994rs}
N.~Seiberg and E.~Witten, \emph{{Electric - magnetic duality, monopole
  condensation, and confinement in N=2 supersymmetric Yang-Mills theory}},
  \href{http://dx.doi.org/10.1016/0550-3213(94)90124-4,
  10.1016/0550-3213(94)00449-8}{\emph{Nucl. Phys.} {\bf B426} (1994) 19--52},
  [\href{https://arxiv.org/abs/hep-th/9407087}{{\tt hep-th/9407087}}].

\bibitem{Seiberg:1994aj}
N.~Seiberg and E.~Witten, \emph{{Monopoles, duality and chiral symmetry
  breaking in N=2 supersymmetric QCD}},
  \href{http://dx.doi.org/10.1016/0550-3213(94)90214-3}{\emph{Nucl. Phys.} {\bf
  B431} (1994) 484--550}, [\href{https://arxiv.org/abs/hep-th/9408099}{{\tt
  hep-th/9408099}}].

\bibitem{donaldson1983}
S.~K. Donaldson, \emph{An application of gauge theory to four-dimensional
  topology}, \href{http://dx.doi.org/10.4310/jdg/1214437665}{\emph{J.
  Differential Geom.} {\bf 18} (1983) 279--315}.

\bibitem{dedushenko2017vertex}
M.~Dedushenko, S.~Gukov and P.~Putrov, \emph{Vertex algebras and 4-manifold
  invariants}, {\emph{arXiv preprint arXiv:1705.01645} (2017) },
  [\href{https://arxiv.org/abs/1705.01645}{{\tt 1705.01645}}].

\bibitem{kronheimer1996embedded}
P.~Kronheimer, \emph{Embedded surfaces and gauge theory in three and four
  dimensions}, {\emph{Surveys in differential geometry} {\bf 3} (1996)
  243--298}.

\bibitem{MR3432158}
A.~Haydys and T.~Walpuski, \emph{A compactness theorem for the
  {S}eiberg-{W}itten equation with multiple spinors in dimension three},
  {\emph{Geom. Funct. Anal.} {\bf 25} (2015) 1799--1821}.

\bibitem{akbulut2004associative}
S.~Akbulut and S.~Salur, \emph{Associative submanifolds of a g2 manifold},
  {\emph{arXiv preprint math/0412032} (2004) }.

\bibitem{akbulut2008deformations}
S.~Akbulut and S.~Salur, \emph{Deformations in {$G_2$} manifolds},
  \href{http://dx.doi.org/10.1016/j.aim.2007.09.009}{\emph{Adv. Math.} {\bf
  217} (2008) 2130--2140}.

\bibitem{akbulut2008calibrated}
S.~Akbulut and S.~Salur, \emph{Calibrated manifolds and gauge theory},
  \href{http://dx.doi.org/10.1515/CRELLE.2008.094}{\emph{J. Reine Angew. Math.}
  {\bf 625} (2008) 187--214}.

\bibitem{deBoer:2006bp}
J.~de~Boer, P.~de~Medeiros, S.~El-Showk and A.~Sinkovics, \emph{{Open G(2)
  strings}}, \href{http://dx.doi.org/10.1088/1126-6708/2008/02/012}{\emph{JHEP}
  {\bf 02} (2008) 012}, [\href{https://arxiv.org/abs/hep-th/0611080}{{\tt
  hep-th/0611080}}].

\bibitem{Blau:1996bx}
M.~Blau and G.~Thompson, \emph{{Aspects of N(T) $\geq$ 2 topological gauge
  theories and D-branes}},
  \href{http://dx.doi.org/10.1016/S0550-3213(97)00161-2}{\emph{Nucl. Phys.}
  {\bf B492} (1997) 545--590},
  [\href{https://arxiv.org/abs/hep-th/9612143}{{\tt hep-th/9612143}}].

\bibitem{Howe:1997fb}
P.~S. Howe, E.~Sezgin and P.~C. West, \emph{{Covariant field equations of the M
  theory five-brane}},
  \href{http://dx.doi.org/10.1016/S0370-2693(97)00257-8}{\emph{Phys. Lett.}
  {\bf B399} (1997) 49--59}, [\href{https://arxiv.org/abs/hep-th/9702008}{{\tt
  hep-th/9702008}}].

\bibitem{Witten:1982df}
E.~Witten, \emph{{Constraints on Supersymmetry Breaking}},
  \href{http://dx.doi.org/10.1016/0550-3213(82)90071-2}{\emph{Nucl. Phys.} {\bf
  B202} (1982) 253}.

\bibitem{Braun:2018fdp}
A.~P. Braun, M.~Del~Zotto, J.~Halverson, M.~Larfors, D.~R. Morrison and
  S.~Schafer-Nameki, \emph{{Infinitely Many M2-instanton Corrections to
  M-theory on $G_2$-manifolds}},  \href{https://arxiv.org/abs/1803.02343}{{\tt
  1803.02343}}.

\bibitem{Gaiotto:2009hg}
D.~Gaiotto, G.~W. Moore and A.~Neitzke, \emph{{Wall-crossing, Hitchin Systems,
  and the WKB Approximation}},  \href{https://arxiv.org/abs/0907.3987}{{\tt
  0907.3987}}.

\bibitem{Witten:1986bf}
E.~Witten, \emph{{Elliptic Genera and Quantum Field Theory}},
  \href{http://dx.doi.org/10.1007/BF01208956}{\emph{Commun. Math. Phys.} {\bf
  109} (1987) 525}.

\bibitem{Gopakumar:1997dv}
R.~Gopakumar and C.~Vafa, \emph{{Branes and fundamental groups}}, {\emph{Adv.
  Theor. Math. Phys.} {\bf 2} (1998) 399--411},
  [\href{https://arxiv.org/abs/hep-th/9712048}{{\tt hep-th/9712048}}].

\bibitem{Witten:1989sx}
E.~Witten, \emph{{Topology Changing Amplitudes in (2+1)-Dimensional Gravity}},
  \href{http://dx.doi.org/10.1016/0550-3213(89)90591-9}{\emph{Nucl. Phys.} {\bf
  B323} (1989) 113--140}.

\bibitem{Birmingham:1989is}
D.~Birmingham, M.~Blau and G.~Thompson, \emph{{Geometry and Quantization of
  Topological Gauge Theories}},
  \href{http://dx.doi.org/10.1142/S0217751X90002014}{\emph{Int. J. Mod. Phys.}
  {\bf A5} (1990) 4721--4752}.

\bibitem{Geyer:2001yc}
B.~Geyer and D.~Mulsch, \emph{{N(T) = 4 equivariant extension of the 3-D
  topological model of Blau and Thompson}},
  \href{http://dx.doi.org/10.1016/S0550-3213(01)00461-8}{\emph{Nucl. Phys.}
  {\bf B616} (2001) 476--494},
  [\href{https://arxiv.org/abs/hep-th/0108042}{{\tt hep-th/0108042}}].

\bibitem{Dijkgraaf:1996tz}
R.~Dijkgraaf and G.~W. Moore, \emph{{Balanced topological field theories}},
  \href{http://dx.doi.org/10.1007/s002200050097}{\emph{Commun. Math. Phys.}
  {\bf 185} (1997) 411--440}, [\href{https://arxiv.org/abs/hep-th/9608169}{{\tt
  hep-th/9608169}}].

\bibitem{PidstrigachTyurin}
V.~{Pidstrigach} and A.~{Tyurin}, \emph{{Localisation of the Donaldson's
  invariants along Seiberg-Witten classes}},  in \emph{eprint
  arXiv:dg-ga/9507004}, July, 1995.

\bibitem{MR1405956}
C.~Okonek and A.~Teleman, \emph{Quaternionic monopoles}, {\emph{Comm. Math.
  Phys.} {\bf 180} (1996) 363--388}.

\bibitem{blau1993n}
M.~Blau and G.~Thompson, \emph{{N=2} topological gauge theory, the {E}uler
  characteristic of moduli spaces, and the {C}asson invariant},
  \href{http://dx.doi.org/10.1007/BF02097057}{\emph{Comm. Math. Phys.} {\bf
  152} (1993) 41--71}, [\href{https://arxiv.org/abs/hep-th/9112012}{{\tt
  hep-th/9112012}}].

\bibitem{Witten:1990bs}
E.~Witten, \emph{{Introduction to cohomological field theories}},
  \href{http://dx.doi.org/10.1142/S0217751X91001350}{\emph{Int. J. Mod. Phys.}
  {\bf A6} (1991) 2775--2792}.

\bibitem{Bergshoeff:1999db}
E.~Bergshoeff, E.~Sezgin and A.~Van~Proeyen, \emph{{(2,0) tensor multiplets and
  conformal supergravity in D = 6}},
  \href{http://dx.doi.org/10.1088/0264-9381/16/10/311}{\emph{Class.Quant.Grav.}
  {\bf 16} (1999) 3193--3206},
  [\href{https://arxiv.org/abs/hep-th/9904085}{{\tt hep-th/9904085}}].

\bibitem{Cordova:2013bea}
C.~Cordova and D.~L. Jafferis, \emph{{Five-Dimensional Maximally Supersymmetric
  Yang-Mills in Supergravity Backgrounds}},
  \href{https://arxiv.org/abs/1305.2886}{{\tt 1305.2886}}.

\bibitem{Kugo:2000hn}
T.~Kugo and K.~Ohashi, \emph{{Supergravity tensor calculus in 5-D from 6-D}},
  \href{http://dx.doi.org/10.1143/PTP.104.835}{\emph{Prog. Theor. Phys.} {\bf
  104} (2000) 835--865}, [\href{https://arxiv.org/abs/hep-ph/0006231}{{\tt
  hep-ph/0006231}}].

\bibitem{Gustavsson:2011af}
A.~Gustavsson, \emph{{M5 brane on $R^{1,2} \times S^3$}},
  \href{http://dx.doi.org/10.1007/JHEP01(2012)057}{\emph{JHEP} {\bf 01} (2012)
  057}, [\href{https://arxiv.org/abs/1111.5392}{{\tt 1111.5392}}].

\bibitem{2017arXiv170106061M}
A.~J. {Moreno} and H.~N.~S. {Earp}, \emph{{The Weitzenb\"ock formula for the
  Fueter-Dirac operator}},  \href{https://arxiv.org/abs/1701.06061}{{\tt
  1701.06061}}.

\bibitem{FreedGompf}
D.~S. Freed and R.~E. Gompf, \emph{Computer calculation of {W}itten's
  {$3$}-manifold invariant}, {\emph{Comm. Math. Phys.} {\bf 141} (1991)
  79--117}.

\bibitem{Jeffrey:1992tk}
L.~C. Jeffrey, \emph{{Chern-Simons-Witten invariants of lens spaces and torus
  bundles, and the semiclassical approximation}},
  \href{http://dx.doi.org/10.1007/BF02097243}{\emph{Commun. Math. Phys.} {\bf
  147} (1992) 563--604}.

\bibitem{LawrenceRozansky}
R.~Lawrence and L.~Rozansky, \emph{Witten-{R}eshetikhin-{T}uraev invariants of
  {S}eifert manifolds},
  \href{http://dx.doi.org/10.1007/s002200050678}{\emph{Comm. Math. Phys.} {\bf
  205} (1999) 287--314}.

\bibitem{Marino:2002fk}
M.~Marino, \emph{{Chern-Simons theory, matrix integrals, and perturbative three
  manifold invariants}},
  \href{http://dx.doi.org/10.1007/s00220-004-1194-4}{\emph{Commun. Math. Phys.}
  {\bf 253} (2004) 25--49}, [\href{https://arxiv.org/abs/hep-th/0207096}{{\tt
  hep-th/0207096}}].

\bibitem{ClossetKimWillett}
C.~Closset, H.~Kim and B.~Willett, \emph{{"Three-dimensional N = 2
  supersymmetric gauge theories on Seifert manifolds"}, to appear}, .

\bibitem{Aharony:2012nh}
O.~Aharony, G.~Gur-Ari and R.~Yacoby, \emph{{Correlation Functions of Large N
  Chern-Simons-Matter Theories and Bosonization in Three Dimensions}},
  \href{http://dx.doi.org/10.1007/JHEP12(2012)028}{\emph{JHEP} {\bf 12} (2012)
  028}, [\href{https://arxiv.org/abs/1207.4593}{{\tt 1207.4593}}].

\bibitem{Jain:2013gza}
S.~Jain, S.~Minwalla and S.~Yokoyama, \emph{{Chern Simons duality with a
  fundamental boson and fermion}},
  \href{http://dx.doi.org/10.1007/JHEP11(2013)037}{\emph{JHEP} {\bf 11} (2013)
  037}, [\href{https://arxiv.org/abs/1305.7235}{{\tt 1305.7235}}].

\bibitem{Seiberg:2016gmd}
N.~Seiberg, T.~Senthil, C.~Wang and E.~Witten, \emph{{A Duality Web in 2+1
  Dimensions and Condensed Matter Physics}},
  \href{http://dx.doi.org/10.1016/j.aop.2016.08.007}{\emph{Annals Phys.} {\bf
  374} (2016) 395--433}, [\href{https://arxiv.org/abs/1606.01989}{{\tt
  1606.01989}}].

\bibitem{Karch:2016sxi}
A.~Karch and D.~Tong, \emph{{Particle-Vortex Duality from 3d Bosonization}},
  \href{http://dx.doi.org/10.1103/PhysRevX.6.031043}{\emph{Phys. Rev.} {\bf X6}
  (2016) 031043}, [\href{https://arxiv.org/abs/1606.01893}{{\tt 1606.01893}}].

\bibitem{Armoni:2009vv}
A.~Armoni, A.~Giveon, D.~Israel and V.~Niarchos, \emph{{Brane Dynamics and 3D
  Seiberg Duality on the Domain Walls of 4D N=1 SYM}},
  \href{http://dx.doi.org/10.1088/1126-6708/2009/07/061}{\emph{JHEP} {\bf 07}
  (2009) 061}, [\href{https://arxiv.org/abs/0905.3195}{{\tt 0905.3195}}].

\bibitem{Gomis:2017ixy}
J.~Gomis, Z.~Komargodski and N.~Seiberg, \emph{{Phases Of Adjoint QCD$_3$ And
  Dualities}},  \href{https://arxiv.org/abs/1710.03258}{{\tt 1710.03258}}.

\bibitem{Bashmakov:2018wts}
V.~Bashmakov, J.~Gomis, Z.~Komargodski and A.~Sharon, \emph{{Phases of ${\cal
  N}=1$ Theories in 2+1 Dimensions}},
  \href{https://arxiv.org/abs/1802.10130}{{\tt 1802.10130}}.

\bibitem{Benini:2018umh}
F.~Benini and S.~Benvenuti, \emph{{$\mathcal{N}{=}1$ dualities in 2+1
  dimensions}},  \href{https://arxiv.org/abs/1803.01784}{{\tt 1803.01784}}.

\bibitem{Braun:2018joh}
A.~P. Braun and S.~Schafer-Nameki, \emph{{Spin(7)-Manifolds as Generalized
  Connected Sums and 3d N=1 Theories}},
  \href{https://arxiv.org/abs/1803.10755}{{\tt 1803.10755}}.

\bibitem{blau2000relationship}
M.~Blau and G.~Thompson, \emph{On the relationship between the
  {R}ozansky-{W}itten and the 3-dimensional {S}eiberg-{W}itten invariants},
  \href{http://dx.doi.org/10.4310/ATMP.2001.v5.n3.a3}{\emph{Adv. Theor. Math.
  Phys.} {\bf 5} (2001) 483--498},
  [\href{https://arxiv.org/abs/hep-th/0006244}{{\tt hep-th/0006244}}].

\bibitem{Rozansky:1996bq}
L.~Rozansky and E.~Witten, \emph{{HyperKahler geometry and invariants of three
  manifolds}}, \href{http://dx.doi.org/10.1007/s000290050016}{\emph{Selecta
  Math.} {\bf 3} (1997) 401--458},
  [\href{https://arxiv.org/abs/hep-th/9612216}{{\tt hep-th/9612216}}].

\bibitem{Kaloper:1993fg}
N.~Kaloper, \emph{{Topological mass generation in three-dimensional string
  theory}}, \href{http://dx.doi.org/10.1016/0370-2693(94)90817-6}{\emph{Phys.
  Lett.} {\bf B320} (1994) 16--20},
  [\href{https://arxiv.org/abs/hep-th/9310011}{{\tt hep-th/9310011}}].

\bibitem{Schonfeld:1980kb}
J.~F. Schonfeld, \emph{{A Mass Term for Three-Dimensional Gauge Fields}},
  \href{http://dx.doi.org/10.1016/0550-3213(81)90369-2}{\emph{Nucl. Phys.} {\bf
  B185} (1981) 157--171}.

\bibitem{Cecotti:2011iy}
S.~Cecotti, C.~Cordova and C.~Vafa, \emph{{Braids, Walls, and Mirrors}},
  \href{https://arxiv.org/abs/1110.2115}{{\tt 1110.2115}}.

\bibitem{Gates:1983nr}
S.~J. Gates, M.~T. Grisaru, M.~Rocek and W.~Siegel, \emph{{Superspace Or One
  Thousand and One Lessons in Supersymmetry}}, {\emph{Front. Phys.} {\bf 58}
  (1983) 1--548}, [\href{https://arxiv.org/abs/hep-th/0108200}{{\tt
  hep-th/0108200}}].

\bibitem{Abrikosov:2002jr}
A.~A. Abrikosov, Jr., \emph{{Dirac operator on the Riemann sphere}},
  \href{https://arxiv.org/abs/hep-th/0212134}{{\tt hep-th/0212134}}.

\end{thebibliography}
\bibliographystyle{JHEP}

\providecommand{\href}[2]{#2}\begingroup\raggedright\endgroup


\end{document}